\documentclass[a4paper,12pt]{article}
\pdfoutput=1 

\usepackage{jheppub} 

\usepackage[T1]{fontenc}
\usepackage[italian,english]{babel}
\usepackage{hyperref}
\usepackage{ifpdf}
\usepackage{subfigure}
\usepackage{amssymb}
\usepackage{amsfonts}
\usepackage{epsf}
\usepackage{rotating}
\usepackage{graphicx}
\usepackage{amsmath}
\usepackage{fancyhdr}
\usepackage{lineno}
\usepackage{dutchcal}
\usepackage{babel}
\usepackage{graphics}
\usepackage{pstricks}
\usepackage{color}
\usepackage{multirow}

\newcommand{\nn}{\nonumber}
\newcommand{\lsim}{\mathrel{\mathop{\kern 0pt \rlap
  {\raise.2ex\hbox{$<$}}}
  \lower.9ex\hbox{\kern-.190em $\sim$}}}
\newcommand{\gsim}{\mathrel{\mathop{\kern 0pt \rlap
  {\raise.2ex\hbox{$>$}}}
  \lower.9ex\hbox{\kern-.190em $\sim$}}}

\newcommand{\be}{\begin{equation}}
\newcommand{\ee}{\end{equation}}
\newcommand{\bea}{\begin{eqnarray}}
\newcommand{\eea}{\end{eqnarray}}

\def\etmiss{\not\!\!{E_T}}

\title{\boldmath General Analysis of the Charged Higgs Sector of the 
$Y=0$ Triplet-Singlet Extension of the MSSM at the LHC}

\author[a]{Priyotosh Bandyopadhyay}

\author[a,b]{Claudio Corian\`o}
\author[a]{Antonio Costantini}

\affiliation[a]{Dipartimento di Matematica e Fisica "Ennio De Giorgi", Universit\`a del Salento and INFN-Lecce, \\ Via Arnesano, 73100 Lecce, Italy}
\affiliation[b]{STAG Research Centre and Mathematical Sciences,\\ University of Southampton, Southampton SO17 1BJ, UK}

\emailAdd{priyotosh.bandyopadhyay@le.infn.it}
\emailAdd{claudio.coriano@le.infn.it}
\emailAdd{antonio.costantini@le.infn.it}

\abstract{We investigate the extended Higgs sectors, specially the charged Higgs sector, in a supersymmetric  $Y=0$ $SU(2)$ triplet and a Standard Model (SM) gauge singlet extension of SM. We show that 
in this model, the allowed data for the Higgs boson interaction eigenstates tend to group into separate blocks for a $SU(2)$ triplet, doublet and singlet. A typical mass spectrum has  a doublet type Standard Model like Higgs of 125 GeV, a triplet-like light charged Higgs boson and a very light singlet-like pseudoscalar with the rest relatively decoupled.  
Later we investigate the different decay processes allowed in a charged Higgs boson of this model. Specifically, we search for new decay modes of the charged Higgs bosons in order to distinguish between Higgs fields belonging to $SU(2)$ doublet and triplet representations and also to show the existence of a light pseudoscalar which belong to the singlet representation. The different production modes for the light charged Higgs boson have been discussed, including the limiting case of $|\lambda_T| \simeq 0$. We also propose few final state modes carrying the distinctive signatures of this model which could be investigated at LHC and future colliders.  The signatures of singlet and/or triplet can be explored with an earlier reach of $120$ fb$^{-1}$ for some final states at the LHC with 14 TeV of center of mass energy.}

\keywords{\footnotesize Extended Higgs sector, Supersymmetric Phenomenology, Charged Higgs boson}

\begin{document}
\maketitle
\flushbottom

\section{Introduction}
The recently discovered Higgs boson with a mass around $125$ GeV has confirmed the presence of at least one CP-even scalar responsible for the mechanism of electro-weak symmetry breaking (EWSB), in agreement with the Standard Model prediction \cite{CMS, CMS2, ATLAS}. The existence of an extended Higgs sector and its possible contribution to the EWSB mechanism, however, has not been ruled out. 
In fact, even with its success, the Standard Model is not a complete theory of the fundamental interactions. This point of view is supported by various limitations of the theory,  the unsolved gauge hierarchy problem and the mounting evidence in favour of dark matter, which does not find any justification within the model, being just two among several. 

Supersymmetric extensions of SM, even if disfavoured in their minimal formulations, such as in constrained minimal supersymmetric extension of SM (MSSM) scenarios, address the two issues mentioned above in a natural way. Specifically, the introduction of a conserved $R$-parity guarantees that the lightest supersymmetric particle (LSP) takes the role of a dark matter (DM) component \cite{rparity}. 

{ In the MSSM we have two Higgs doublets giving masses to up- and down-type quarks 
respectively.  After EWSB we have two  CP-even light neutral Higgs bosons among which one can be the discovered Higgs around 125 GeV, a CP-odd neutral Higgs boson and a charged Higgs boson pair. 
Observation of a charged Higgs boson will be a obvious proof of the existence of another Higgs doublet which is necessary in the context of supersymmetry. 

Searches for the extended Higgs sector by looking for charged Higgs boson at the LHC are not new. In fact, both the CMS and ATLAS collaborations have investigated scenarios with charged Higgs bosons, even under the assumption of these being lighter than the top quark ($m_{H^\pm}\leq m_t$). In this case, the channel in question has been the $pp\to t\bar{t}$ production channel, with one of the top decaying into $b H^\pm$. In the opposite case of a charged Higgs heavier than the top ($m_{H^\pm}\geq m_t$), the most studied channels have been the $bg \to tH^\pm$ and $pp \to  tb H^\pm$, with the charged Higgs decaying into $\tau \nu_\tau$ \cite{ChCMS, ChATLAS}. We recall that both doublet type charged and neutral Higgs bosons couple to fermions with Yukawa interactions which are proportional to the mixing angle of the up- and down-type $SU(2)$ doublets. 

The extension of the MSSM with a  SM gauge singlet, i.e. the NMSSM \cite{Ellwanger}, has a scalar which does not couple to fermions or gauge bosons thus changes the search phenomenology. Similar extensions are possible with only $SU(2)$ triplet superfields with $Y=0 \pm 2 $ hypercharges  \cite{pbas1, pbas2, DiChiara, pbas3, EspinosaQuiros}. In the case of $Y=0$, the neutral part of the triplet scalar does not couple to $Z$ boson
and does not contribute to $Z$ mass, whereas non-zero hypercharge triplets  contribute both in $W^\pm$
and $Z$ mass.

The supersymmetric extensions of the Higgs sectors with $Z_3$ symmetry have the common feature of a light pseudoscalar in the spectrum, known as R axion in the literature. Such feature is common to NMSSM with $Z_3$ symmetry \cite{Ellwanger} and also to extensions with singlet and triplet(s) with appropriate hypercharges 
\cite{TNMSSM1, TNMSSM2, tnssm, tnssma}.

In this article we consider an extension of the MSSM with $SU(2)$ triplet superfield of $Y=0$ hypercharge and SM gauge singlet superfield, named as TNMSSM \cite{TNMSSM1, TNMSSM2}, with $Z_3$ symmetry. The main motivation to work with $Y=0$ triplet is that it is the simplest triplet extension in supersymmetric context, where the triplet only contribute in $W^\pm$ mass. For a model with non-zero hypercharges we need at least two triplets and also we get constrained from both $W^\pm$ and $Z$ masses \cite{tnssma}.  The light pseudoscalar in this model is mostly singlet and hence does not have any coupling to fermions or gauge bosons.  For this reason such light pseudoscalar is still allowed by the earlier LEP \cite{LEPb} data and current LHC data\cite{CMS, CMS2, ATLAS}. Similarly the triplet type Higgs bosons also do not couple to fermions  \cite{pbas1, pbas2, DiChiara, pbas3} which still allows a light triplet-like charged Higgs in charged Higgs searches \cite{ChCMS, ChATLAS} and such Higgs bosons have to looked for in different production as well as decay modes.}

General features of this model have been presented in \cite{TNMSSM1}, while a more detailed investigation of the hidden pseudoscalar has been discussed by the authors in \cite{TNMSSM2}. Existence of the light pseudoscalar makes the phenomenology of the Higgs sector very rich for both the neutral and the charged sectors, along with other signatures. In the TNMSSM, we have three physically charged Higgs bosons $h^\pm_{1,2,3}$, two of which are triplet type in the gauge basis. The neutral part of the Higgs sector has four  CP-even ($h_{1,2,3,4}$) and three CP-odd sectors ($a_{1,2,3}$) states. In the gauge basis two of CP-even states are doublet-like one of which should be the discovered Higgs around $125$ GeV, one triplet type and one singlet type.  For the CP-odd states, there are one doublet type, one triplet type and one singlet type. Often it is the singlet-like pseudoscalar which becomes very light, which makes the phenomenology very interesting. The mass spectrum often splits into several regions with distinctively doublet/triplet blocks. The goal of our analysis will be to address the main features of this complete spectrum, characterizing its main signatures in the complex environment of a hadron collider.  

Our work is organized as follows. In section~\ref{model} we review the model very briefly. 
We present a scan over the parameter space of the model in the light of recent LHC data and discuss
 the Higgs boson mass hierarchy in section~\ref{scans} . The structure of the charged Higgs bosons are detailed in section~\ref{chcdcy}. The new and modified charged Higgs decay modes consider in section~\ref{chdcys}. In section~\ref{ch1dcy} various decay branching fractions are shown for all the three charged Higgs bosons with the allowed data points, while the several production modes at the LHC are contained in section~\ref{ch1prod}.  Finally we discuss in section~\ref{pheno} the prospect for future searches of triplet and extra doublet Higgs bosons at the LHC and possible ways to distinguish scalar states belonging to such different representations of $SU(2)$ and in section~\ref{dis} we conclude. 
\section{The Model}\label{model}
The superpotential of the TNMSSM, $W_{TNMSSM}$, contains a SU(2) triplet $\hat{T}$ of zero hypercharge ($Y=0$)  together with a SM gauge singlet ${\hat S}$ added to the superpotential of the MSSM.


The triplet superfield and the two Higgs doublets are then expressed as 
\begin{equation}\label{spf}
 \hat T=\begin{pmatrix}
       \sqrt{\frac{1}{2}}\hat T^0 & \hat T_2^+ \cr
      \hat T_1^- & -\sqrt{\frac{1}{2}}\hat T^0
       \end{pmatrix},\, \hat{H}_u= \begin{pmatrix}
      \hat H_u^+  \cr
       \hat H^0_u
       \end{pmatrix},\, \hat{H}_d= \begin{pmatrix}
      \hat H_d^0  \cr
       \hat H^-_d
       \end{pmatrix}.
 \end{equation}
In the previous expression $\hat T^0$ is a complex neutral superfield, while  $\hat T_1^-$ and $\hat T_2^+$ are the charged Higgs superfields. 

The two terms of the superpotential are combined in the form 

 \begin{equation}
 W_{TNMSSM}=W_{MSSM} + W_{TS},
 \end{equation}
with
\begin{equation}
W_{MSSM}= y_t \hat U \hat H_u\!\cdot\! \hat Q - y_b \hat D \hat H_d\!\cdot\! \hat Q - y_\tau \hat E \hat H_d\!\cdot\! \hat L\ ,
\label{spm}
 \end{equation}
 being the superpotential of the MSSM, while 
 \begin{equation}
W_{TS}=\lambda_T  \hat H_d \cdot \hat T  \hat H_u\, + \, \lambda_S \hat S  \hat H_d \cdot  \hat H_u\,+ \frac{\kappa}{3}\hat S^3\,+\,\lambda_{TS} \hat S  \textrm{Tr}[\hat T^2]
\label{spt}
 \end{equation}
accounts for the extended scalar sector which includes a triplet and a singlet superfields.
 The MSSM Higgs doublets are the only superfields which couple to the fermion multiplet via Yukawa coupling, as in Eq.~(\ref{spm}). After supersymmetry breaking the theory is also characterized by a discrete $Z_3$ symmetry. The soft breaking terms in the scalar potential are given by

 \bea\nn
V_{soft}& =&m^2_{H_u}|H_u|^2\, +\, m^2_{H_d}|H_d|^2\, +\, m^2_{S}|S|^2\, +\, m^2_{T}|T|^2\nn\\
&&+\, m^2_{Q}|Q|^2 + m^2_{U}|U|^2\,+\,m^2_{D}|D|^2 \nn\\
&&+(A_S S H_d\cdot H_u\, +\, A_{\kappa} S^3\, +\, A_T H_d\cdot T\cdot H_u \nn\\ 
 &&+\, A_{TS} S Tr(T^2)+\, A_U U H_U\cdot Q\, +\, \, A_D D H_D\cdot Q \nn\\
 &&+ h.c.),
\label{softp}
 \eea
while the D-terms take the form 
 \begin{equation}
 V_D=\frac{1}{2}\sum_k g^2_k ({ \phi^\dagger_i t^a_{ij} \phi_j})^2 .
 \label{dterm}
 \end{equation}
 As in our previous study, also in this case we assume that all the coefficients involved in the Higgs sector are real in order to preserve CP invariance. The breaking of the $SU(2)_L\times U(1)_Y$ electroweak symmetry is then obtained by giving real vacuum expectation values (VEVs) to the neutral components of the Higgs field
 \be
 <H^0_i>=\frac{v_i}{\sqrt{2}},\, <S>=\frac{v_S}{\sqrt{2}}, \,  <T^0>=\frac{v_T}{\sqrt{2}},\,\,i=u,d
 \ee
 which give mass to the $W^\pm$ and $Z$ bosons
 \bea\label{gbmass}
 m^2_W=\frac{1}{4}g^2_L(v^2 + 4v^2_T),&\ m^2_Z=\frac{1}{4}(g^2_L \, +\, g^2_Y)v^2,\nn\\
  v^2=(v^2_u\, +\, v^2_d), 
&\tan\beta=\frac{v_u}{v_d}. 
\eea
The presence of $\hat{S}$ and $\hat{T}$ in the superpotential allows a $\mu$-term of the form $ \mu_D=\frac{\lambda_S}{\sqrt 2} v_S+ \frac{\lambda_T}{2} v_T$. We also recall that the triplet VEV $v_T$ is strongly  constrained by the global fit on the measurement of the $\rho$ parameter \cite{rho}
 \be
 \rho =1.0004^{+0.0003}_{-0.0004} ,
 \ee 
 which restricts its value to $v_T \leq 5$ GeV. The non-zero triplet contribution to the $W^\pm$ mass leads to a deviation of the $\rho$ parameter
 \be
 \rho= 1+ 4\frac{v^2_T}{v^2} .
 \ee
As in \cite{TNMSSM1}, in our current numerical analysis we have chosen $v_T =3$ GeV. { The detailed minimisation conditions both at tree-level as well at one-loop are given in \cite{TNMSSM1}. We also present the tree-level expressions for the neutral and charged Higgs mass matrices in the Appendix.}

\section{A scan over the parameter space and the LHC selection criteria}\label{scans}

The main goal of our previous works and of our current one is to search for a suitable region of parameter space, in the form of specific benchmark points,  which could allow one or more hidden Higgs particles, compatible with the current LHC limits.

As already pointed out before \cite{TNMSSM1,TNMSSM2}, there are four CP-even neutral ($h_1,h_2,h_3,h_4$), three CP-odd neutral ($a_1,a_2, a_3$) and three charged Higgs bosons ($h_1^\pm,h_2^\pm,h_3^\pm$). In general the interaction eigenstates are obtained via a mixing of the two Higgs doublets, the triplet and the singlet scalar. However, the singlet does not contribute to the charged Higgs bosons, which are mixed states generated only by the $SU(2)$ doublets and triplets.
The rotation from gauge eigenstates to the interaction eigenstates are
\bea\label{chmix}
h_i= \mathcal{R}^S_{ij} H_j\nn\\
a_i= \mathcal{R}^P_{ij} A_j\\
h^\pm_i= \mathcal{R}^C_{ij} H^\pm_j\nn
\eea
where the eigenstates on the left-hand side are interaction eigenstates whereas the eigenstates on th right-hand side are gauge eigensates. Explicitly we have $h_i=(h_1,h_2,h_3,h_4)$, $H_i=(H^0_{u,r},H^0_{d,r},S_r,T^0_r)$, $a_i=(a_0,a_1,a_2,a_3)$, $A_i=(H^0_{u,i},H^0_{d,i},S_i,T^0_i)$, $h_i^\pm=(h_0^\pm,h_1^\pm,h_2^\pm,h_3^\pm)$ and $H_i^+=(H_u^+,T_2^+,H_d^{-*},T_1^{-*})$. { Using these definitions we can write the doublet and triplet fraction for the scalar and pseudoscalar Higgs bosons as
\bea
h_i|_{D}=(\mathcal{R}^S_{i,1})^2+(\mathcal{R}^S_{i,2})^2, \,\, a_i|_{D}=(\mathcal{R}^P_{i,1})^2+(\mathcal{R}^P_{i,2})^2
\eea
\bea
h_i|_{S}=(\mathcal{R}^S_{i3})^2, \,\, a_i|_{S}=(\mathcal{R}^P_{i3})^2
\eea
\bea
h_i|_T=(\mathcal{R}^S_{i4})^2, \,\, a_i|_T=(\mathcal{R}^P_{i4})^2
\eea
and the triplet and doublet fraction of the charged Higgs bosons as
\bea
h_i^\pm|_D=(\mathcal{R}^C_{i1})^2+(\mathcal{R}^C_{i3})^2, \,\, h_i^\pm|_T=(\mathcal{R}^C_{i2})^2+(\mathcal{R}^C_{i4})^2 .
\eea
We call a scalar(pseudoscalar) Higgs boson doublet-like if $h_i|_D(a_i|_D)\geq\,90\%$, singlet-like if $h_i|_S(a_i|_S)\geq\,90\%$ and triplet-like if $h_i|_T(a_i|_T)\geq\,90\%$. Similarly a charged Higgs boson will be doublet-like if $h_i^\pm|_D\geq\,90\%$ or triplet-like if $h_i^\pm|_D\geq\,90\%$.}

If the discovered Higgs is the lightest CP-even boson, $h_1\equiv h_{125}$, then $h_1$ must be doublet-like and the lightest CP-odd and charged Higgses must be triplet/singlet-like, in order to evade the experimental constraint from LEP \cite{LEPb} for the pseudoscalar and charged Higgses.
{ LEP searched for the Higgs boson via the $e^+ e^- \to Z h$ and $e^+e^- \to h_1h_2$ channels (in models with multiple Higgs bosons) and their fermionic decay modes ($h \to \bar{b}b,\bar\tau \tau$ and $Z \to \ell\ell$). The higher centre of mass energy at LEP II (210 GeV) allowed to set a lower bound of 114.5 on the SM-like Higgs boson and of 93 GeV for the MSSM-like Higgs boson in the maximal mixing scenario \cite{LEPb}. Interestingly, neither the triplet nor the singlet type Higgs boson couple to Z or to leptons (see
Eq.~\ref{spt}), and we checked explicitly to ensure that the demand of $\geq 90\%$ singlet and/or triplet is sufficient 
for the light pseudoscalar to be allowed by LEP data. We also checked explicitly to see that the LHC allowed parameter space for the light pseudoscalar and the details can be found out in \cite{TNMSSM2}. Later we also discuss how  the criteria of $\geq 90\%$ singlet/triplet is enough to fulfill the constraints coming from the B-observables.}
Similar constraints on the structure of the Higgses must be imposed if $h_2\equiv h_{125}$. To scan the parameter space we have used a code written by us, in which we have randomly selected $1.35\times10^6$ points that realize the EWSB mechanism at tree-level. In particular, we have performed the scan using the following criteria for the couplings and the soft parameters
\bea\label{scan}
&|\lambda_{T, S, TS}| \leq 1, \, |\kappa|\leq 3, \, |v_s|\leq 1 \, \rm{TeV}, \, 1\leq \tan{\beta}\leq 10,\nn\\
&|A_{T, S, TS, U, D}|\leq 1\, \rm{GeV},\,|A_\kappa|\leq 3\, \rm{GeV},\\
&65\leq|M_{1, 2}|\leq10^3\,\rm{GeV},\,  3\times10^2\leq m_{Q_3, \bar{u}_3, \bar{d}_3}\leq10^3\,\rm{GeV}.\nn
\eea
We have selected those points which have one of the four Higgs bosons with a one-loop mass of $\sim125$ GeV { with one-loop minimization conditions} and, out of the $1.35\times10^6$ points, over $10^5$ of them pass this constraint. On this set of Higgs candidates we have imposed the constraints on the structure of the lightest CP-even, CP-odd and charged Higgses. The number of points with $h_1\equiv h_{125}$ doublet-like and $a_1$ singlet-like is about 70 \% but we have just one point with $h_1\equiv h_{125}$ which is doublet-like and $a_1$ triplet-like. If we add the requirement on the lightest charged Higgs to be triplet-like, we find that the number of points with $h_1\equiv h_{125}$ doublet-like, $a_1$ singlet-like and $h_1^\pm$ triplet-like is 26 \%. The case of $h_2\equiv h_{125}$ doublet-like allows more possibilities, because in this case we have also to check the structure of $h_1$. However we find 75 points only when $h_1$ is triplet-like, $h_2\equiv h_{125}$ is doublet-like and $a_1$ is singlet-like. This selection is insensitive to the charged Higgs selection, i.e. we still have 75 points with $h_1$ triplet-like, $h_2\equiv h_{125}$ doublet-like, $a_1$ singlet-like and $h_1^\pm$ triplet-like.\\
The LHC constraints have been imposed on those points with $h_1\equiv h_{125}$, because they provide a better statistics. For these points we demand that
\bea\label{LHCdata}
&\mu_{WW^*}=0.83\pm0.21\,\,\mu_{ZZ^*}=1.00\pm0.29\\
&\mu_{\gamma\gamma}=1.12\pm0.24\nn
\eea
at 1$\sigma$ of confidence level \cite{CMS2}. The LHC selection give us 12223 points out of the 26776 points that have  $h_1\equiv h_{125}$ doublet-like, $a_1$ singlet-like and $h_1^\pm$ triplet-like.

\begin{figure}[thb]
\begin{center}
\mbox{\hskip -10pt\subfigure[]{\includegraphics[width=0.45\linewidth]{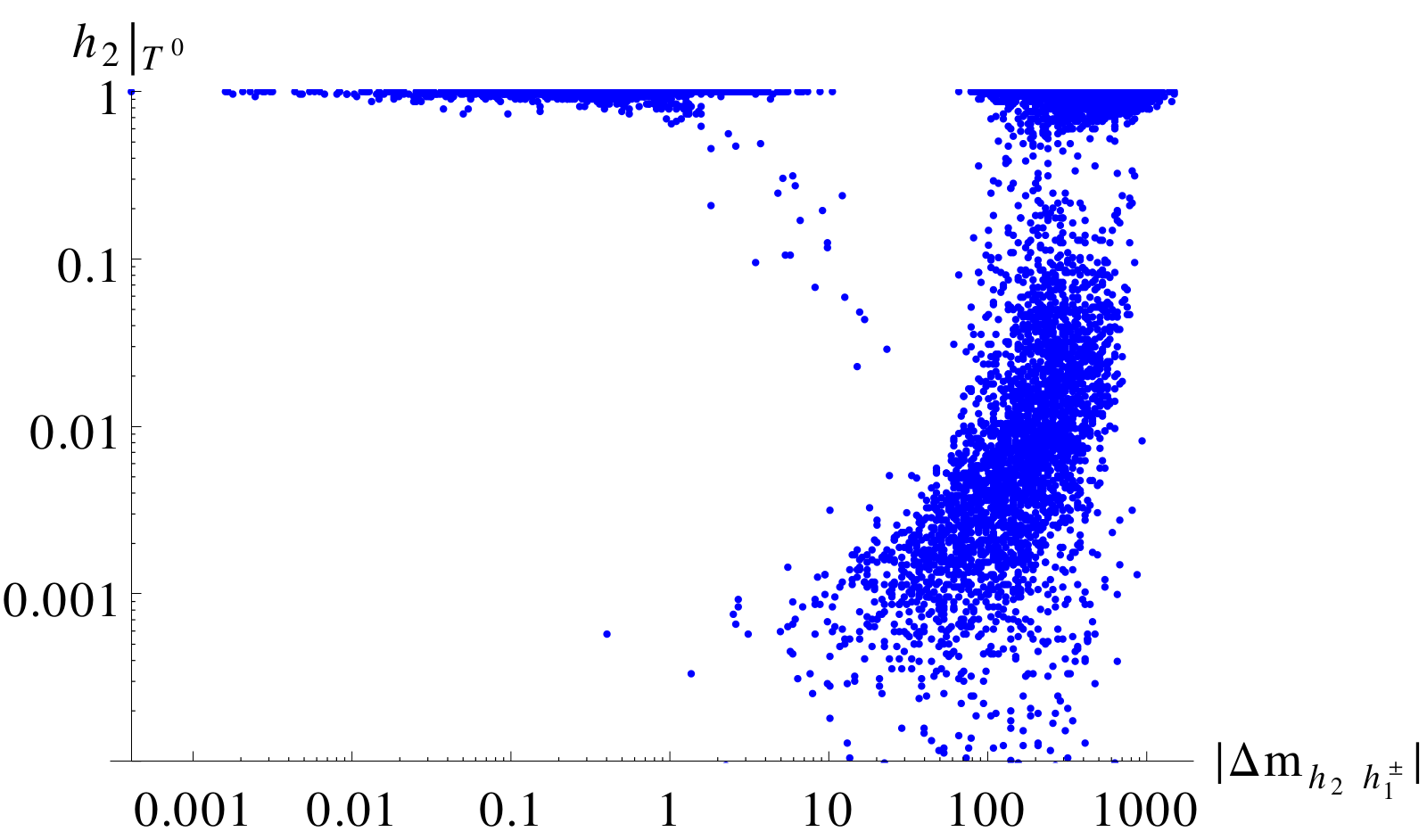}}
\subfigure[]{\includegraphics[width=0.45\linewidth]{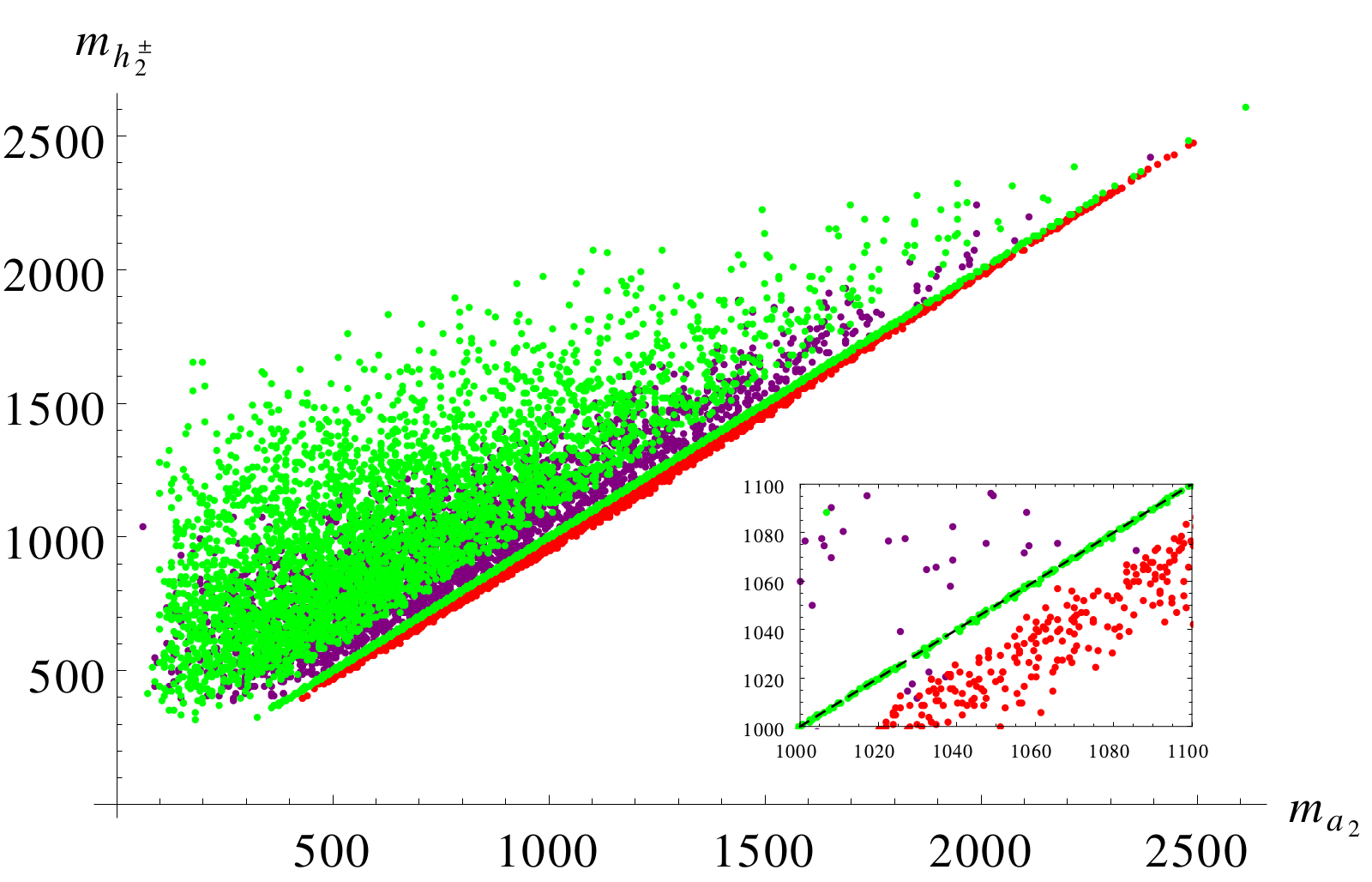}}}
\mbox{\hskip -10pt\subfigure[]{\includegraphics[width=0.45\linewidth]{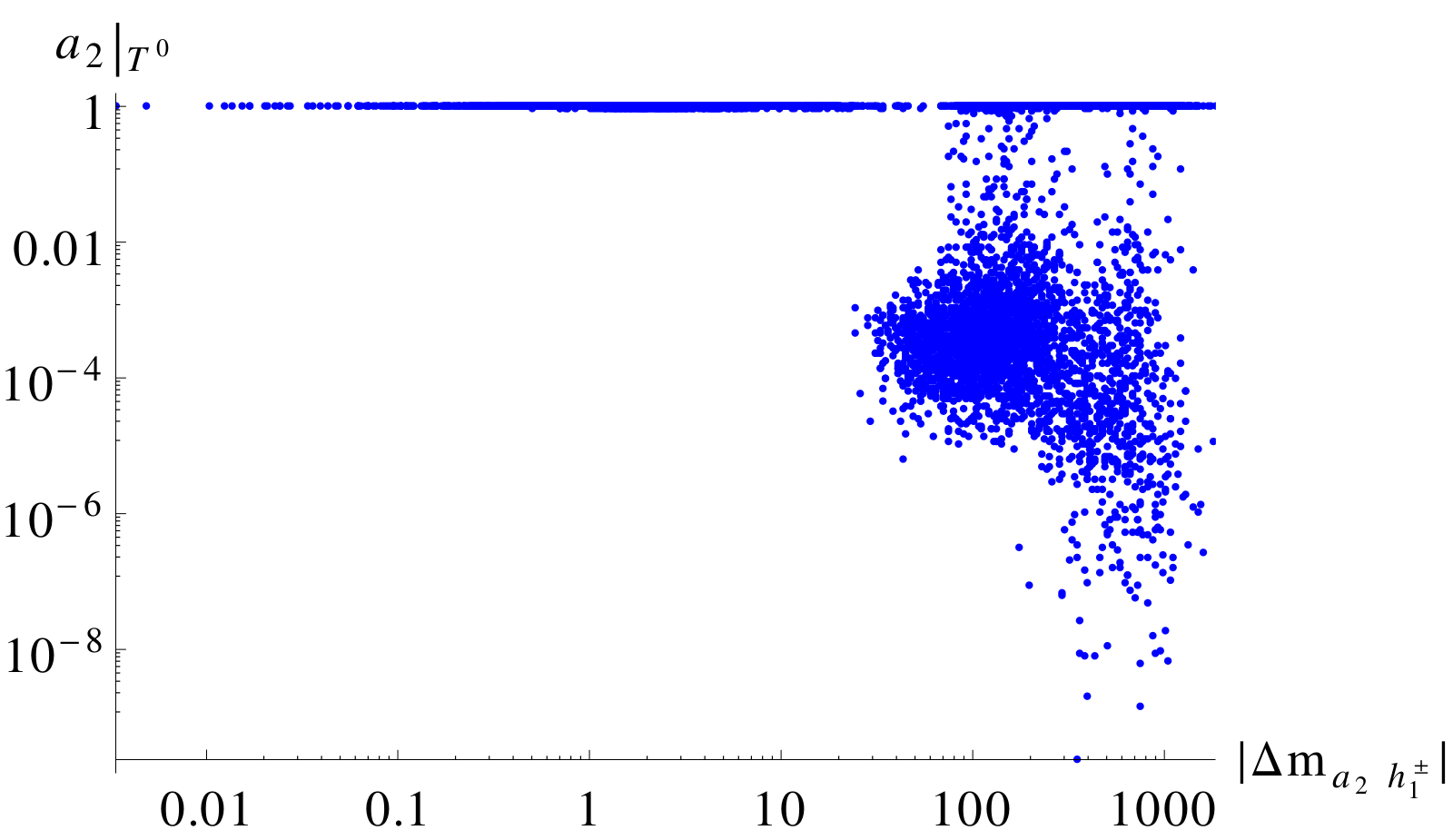}}
\subfigure[]{\includegraphics[width=0.45\linewidth]{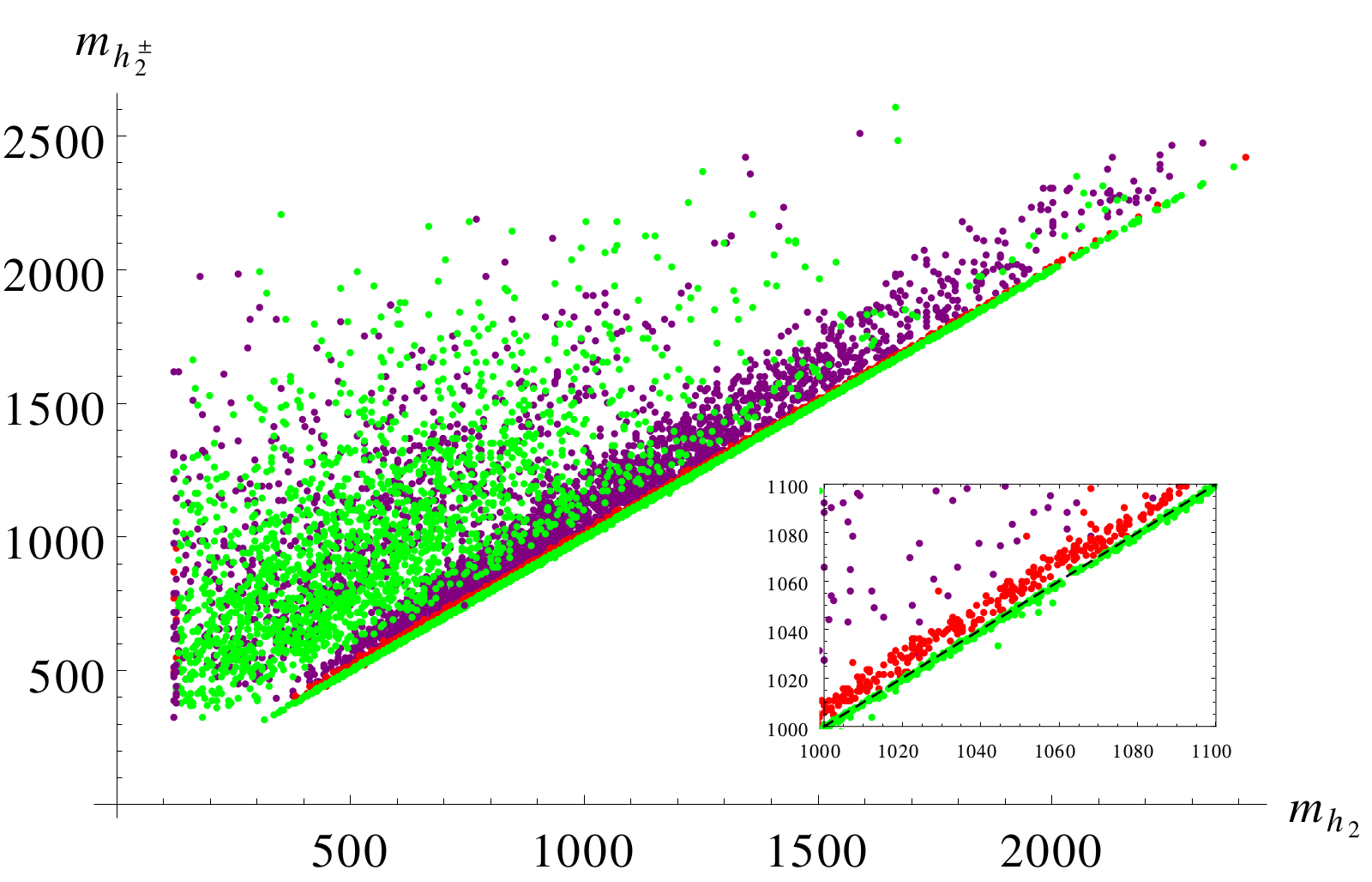}}
}
\caption{We show the fraction of triplets of $h_2$ (a) and $a_2$ (c) as a function of the mass difference $|\Delta m_{h_2/a_2\, h^\pm_1}|$ between $h_2/a_2$ and $h^\pm_1$ respectively. We plot the mass correlation between $a_2$ and $h_2^\pm$ (b) and between $h_2$ and $h_2^\pm$ (d). These exhaust the possible hierarchies for the triplet eigenstates. We mark in red the points with both $a_2$ and $h_2^\pm$ doublet-type, in purple the points with $a_2$ triplet-type and $h_2^\pm$ doublet-type or viceversa, and in green the points with both $a_2$ and $h_2^\pm$ triplet-like.}\label{ch1h2a2}
\end{center}
\end{figure}
{ Apart from the LEP \cite{LEPb} and LHC  \cite{CMS2} constraints, we also ensure the validity of the constrains coming from the $B$-observables. For this particular reason we claim the light pseudoscalar $a_1$ to be $\geq 90\%$ singlet-type and the light charged Higgs $h^\pm_1$ to be $90\%$ triplet-type. 
A very light scalar or pseudoscalar, with a mass around $1-10$ GeV, gets strong bounds from bottomonium decay to $a_1\gamma$ \cite{bottomonium1}. The decay rate for $\Upsilon \to a_1 \gamma$ can be approximated as follows 
\be
\mathcal{Br}(\Upsilon \to a_1 \gamma)=\mathcal{Br}(\Upsilon \to a_1 \gamma)_{SM}\times g^2_{a_1 b\bar{b}},
\ee
where $g_{a_1 b\bar{b}}$ is the reduced down-type Yukawa coupling with respect to SM \cite{bottomonium}. We checked explicitly that the requirement of more than $90\%$ singlet type $a_1$ and low $\tan{\beta}$ ensure that we are in the region of validity. 

Another important constraint for a light pseudoscalar comes from $\mathcal{Br}(B_s \to \mu \mu)$ which can be summerised as follows \cite{bottomonium}
\be
\mathcal{Br}(B_s \to \mu \mu)\simeq \frac{2\tau_{B_s}M^5_{B_s}f^2_{B_s}}{64\pi}|C|^2( \mathcal{R}^P_{12})^4,
\ee
with 
\bea
C=\frac{G_F\alpha}{\sqrt2\pi}V_{tb}V^*_{ts}\frac{\tan^3\beta}{4\sin^2\theta_w}\frac{m_\mu m_t |\mu_r|}{m_W^2(m^2_{a_1}-m^2_{B_s})}\frac{\sin2\theta_{\tilde t}}{2}\Delta f_3\nn\\
\eea
where $\Delta f_3=f_3(x_2)-f_3(x_1)$, $x_i=m^2_{\tilde t_i}/|\mu_r|^2$, $f_3(x)=x\ln x/(1-x)$, $\theta_{\tilde t}$ is the stop mixing angle and $\mathcal{R}^P_{12}$ is the rotation angle, defined in Eq.~\ref{chmix}, which gives the coupling with the down type Higgs ($H_d$) with leptons and down type quarks.  The demand of mostly singlet $a_1$ ($\geq 90\%$) on the data set ensures that we are well below the current upper limit \cite{lhcb}. 

Another constraint that affects the models with extra Higgs boson, specially the charged Higgs bosons, comes from the rare decay of $B\to X_s \gamma$. The charged Higgs bosons which are doublet in nature couple to quarks via Yukawa couplings and contribute to the rare decay of $B\to X_s \gamma$. Similar contributions also come from the charginos which couple to the quarks, namely doublet-type Higgsinos and Wino. However when we have charged Higgs or charginos which are triplet in nature they do not couple to the fermions and thus do not contribute in such decays \cite{pbas1,pbas2}. If the light charged Higgs bosons are triplet in nature the dominant Wilson coefficients $F_{7,8}$ are suppressed by the charged Higgs rotation angles $\mathcal{R}^C_{11,13}$ as defined in Eq.~\ref{chmix}. The demand of the light charged Higgs boson mostly triplet $\geq 90\%$ enable us to avoid the constraint from $\mathcal{Br}(B\to X_s \gamma)$ \cite{pbas1,pbas2}.

}

\begin{figure}[thb]
\begin{center}
\mbox{\subfigure[]{
\includegraphics[width=0.6\linewidth]{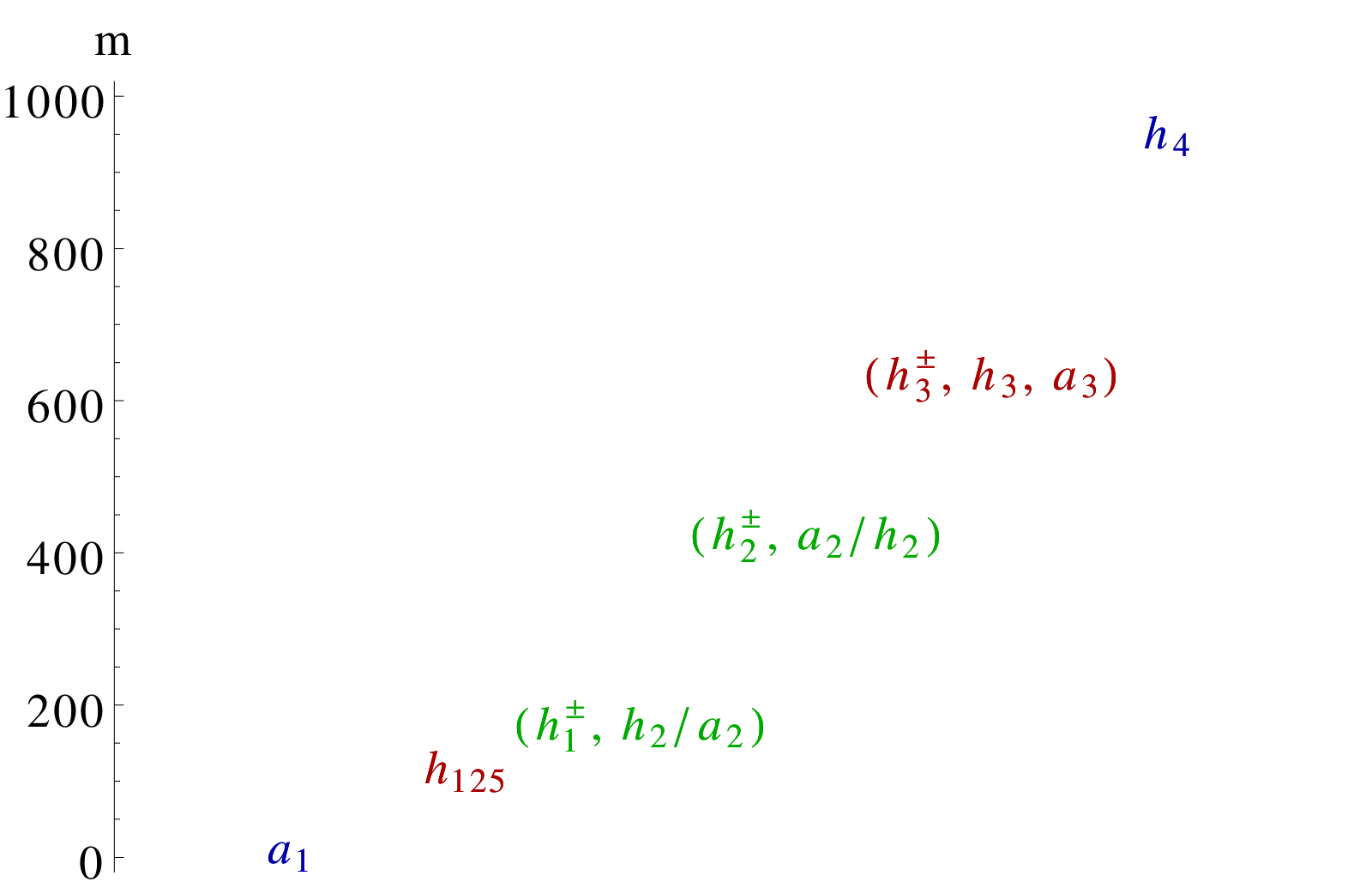}}}
\caption{A typical mass hierarchy of the scalar sector, with the singlet in blue, the doublets in red 
and the triplet Higgs bosons in green colour. The eigenstates of the triplet sector with $a_2/h_2$ or  
$h_2/a_2$ are alternative: if $h_1^\pm$ pairs with the neutral $h_2$, then $h_2^\pm$ is mass degenerate with the pseudoscalar $a_2$ (and viceversa).
}\label{cartoon}
\end{center}
\end{figure}
In Figure \ref{ch1h2a2}(a) we plot the triplet fraction of $h_2$ in function of the mass splitting between $h_2$ and $h_1^\pm$. The lightest charged Higgs is selected to be triplet-like ($\geq 90\%$). It is evident that in the case of mass degeneracy between  $h_2$ and $h_1^\pm$ the triplet-like structure of $h_1^\pm$ is imposed also on $h_2$. In Figure \ref{ch1h2a2}(b) we plot the mass correlation between $a_2$ and $h_2^\pm$. We use the following color code: we mark in red the points with both $a_2$ and $h_2^\pm$ doublet-type, in purple the points with $a_2$ triplet-type and $h_2^\pm$ doublet-type or viceversa, and in green the points with both $a_2$ and $h_2^\pm$ triplet-like. In the inset the dashed line indicates a configuration of mass degeneracy. It is evident that the mass degeneracy between $a_2$ and $h_2^\pm$ implies that both of them are triplet-like. As we depict in Figure \ref{cartoon}, there could be an exchange between $a_2$ and $h_2$ in the triplet pairs, shown in green. For this reason we illustrate also the other possible hierarchy path in Figure \ref{ch1h2a2}(c) and \ref{ch1h2a2}(d). As one may notice, the two sets of plots are qualitatively similar, although there is a quantitative difference between the red points of Figures \ref{ch1h2a2}(b) and \ref{ch1h2a2}(d). The points in the latter are closer than the former to the line of mass degeneracy.
\begin{figure}[thb]
\begin{center}
\mbox{\hskip -10pt\subfigure[]{\includegraphics[width=0.55\linewidth]{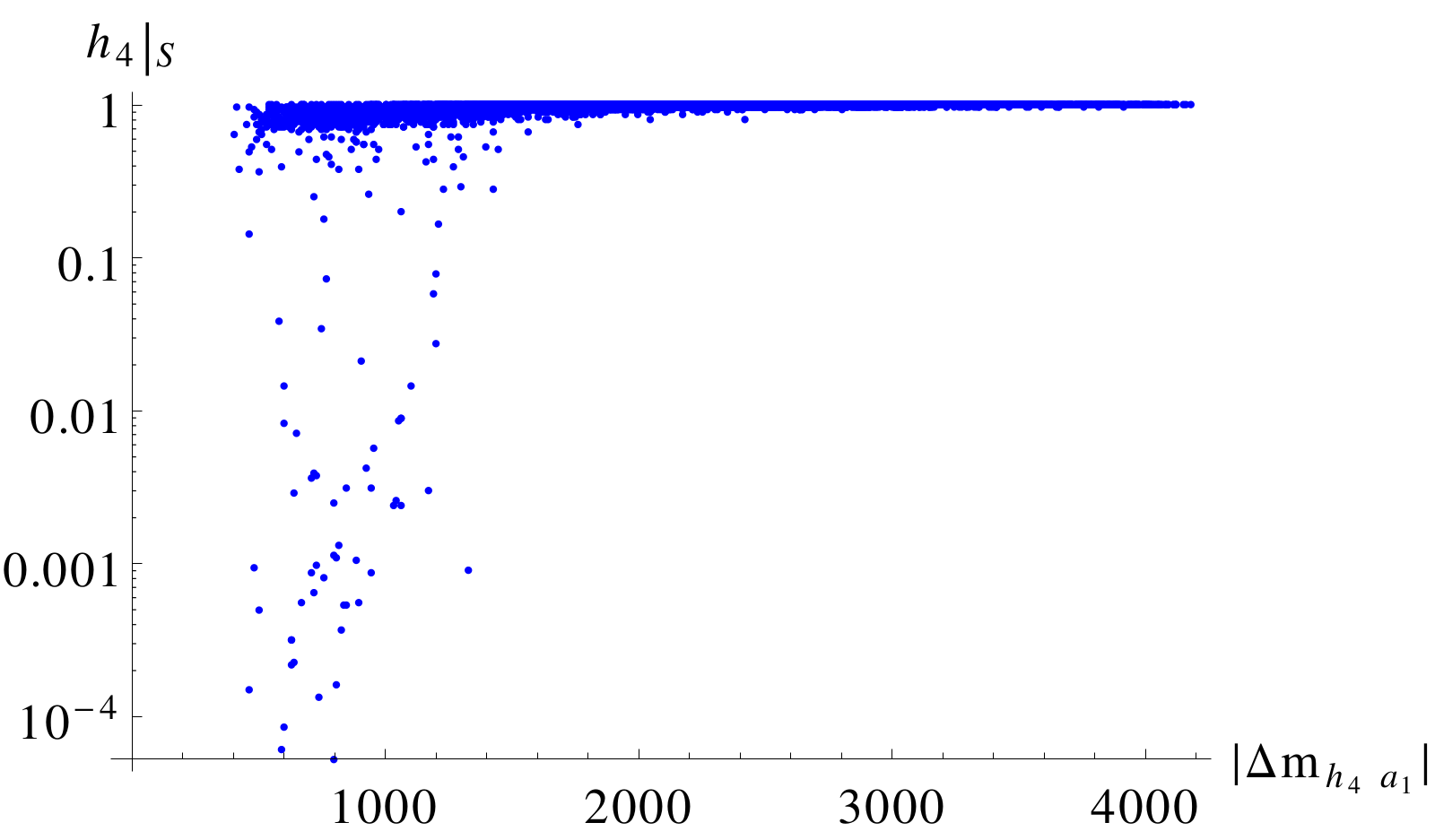}}
\subfigure[]{\includegraphics[width=0.55\linewidth]{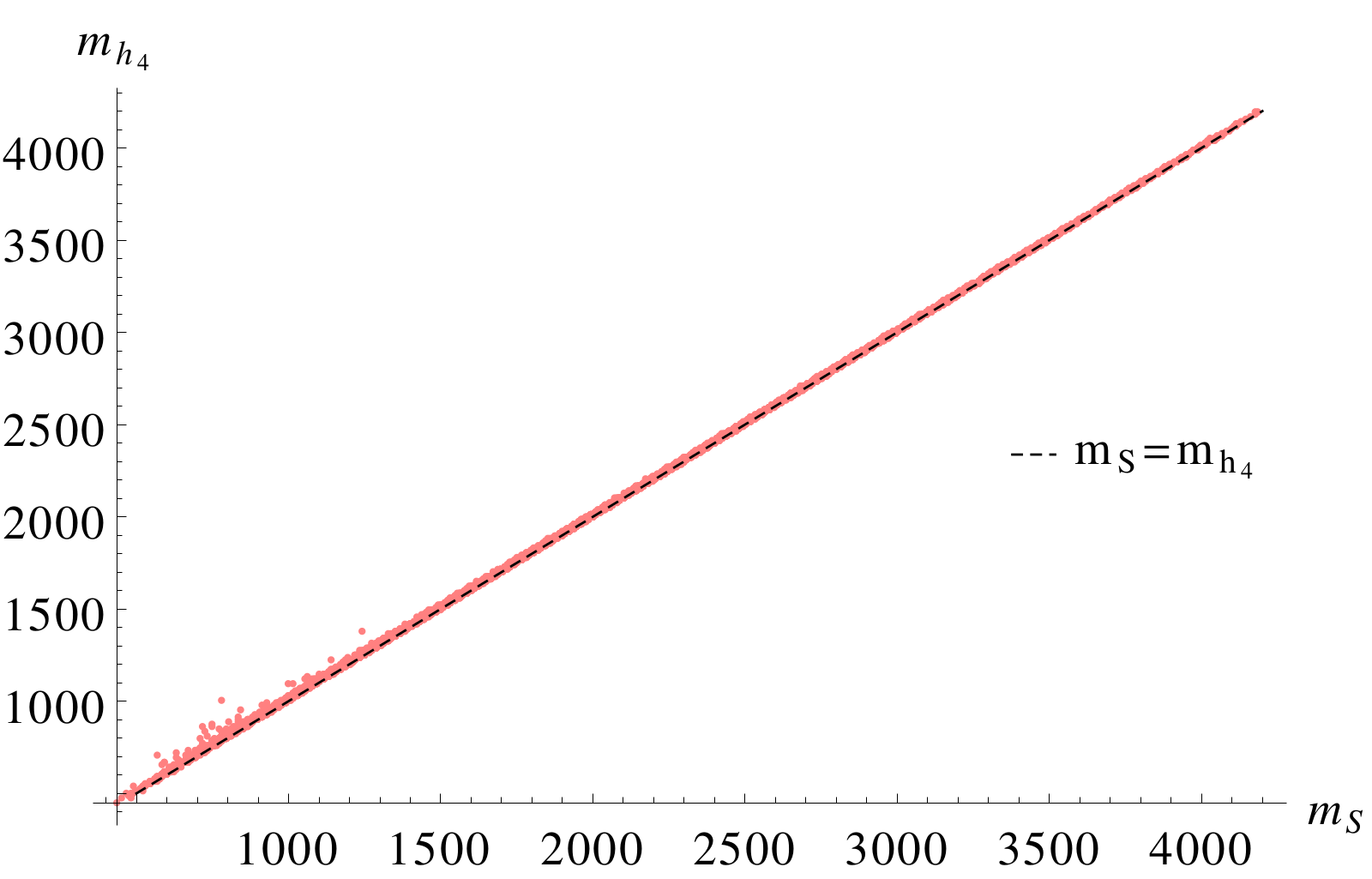}}
}
\caption{We show the singlet fraction of $h_4$ as a function of mass difference $|\Delta m_{h_4\, a_1}|$ 
between the two states $h_4$ and $a_1$ (a), and the mass correlation between $h_4$ and $m_S$ (b).}\label{h4a1}
\end{center}
\end{figure}
Figure \ref{h4a1}(a)  shows that the more $h_4$ is decoupled,
compared to $a_1$, the more it tends to be in a singlet-like eigenstate. We remind that $a_1$ is a pseudo NG mode and hence it is naturally light. From Figure \ref{h4a1}(b) it is evident that $h_4$ takes the soft mass $m_S$ coming from the singlet.
\begin{figure}[htb]
\begin{center}
\mbox{\hskip -10pt\subfigure[]{\includegraphics[width=0.55\linewidth]{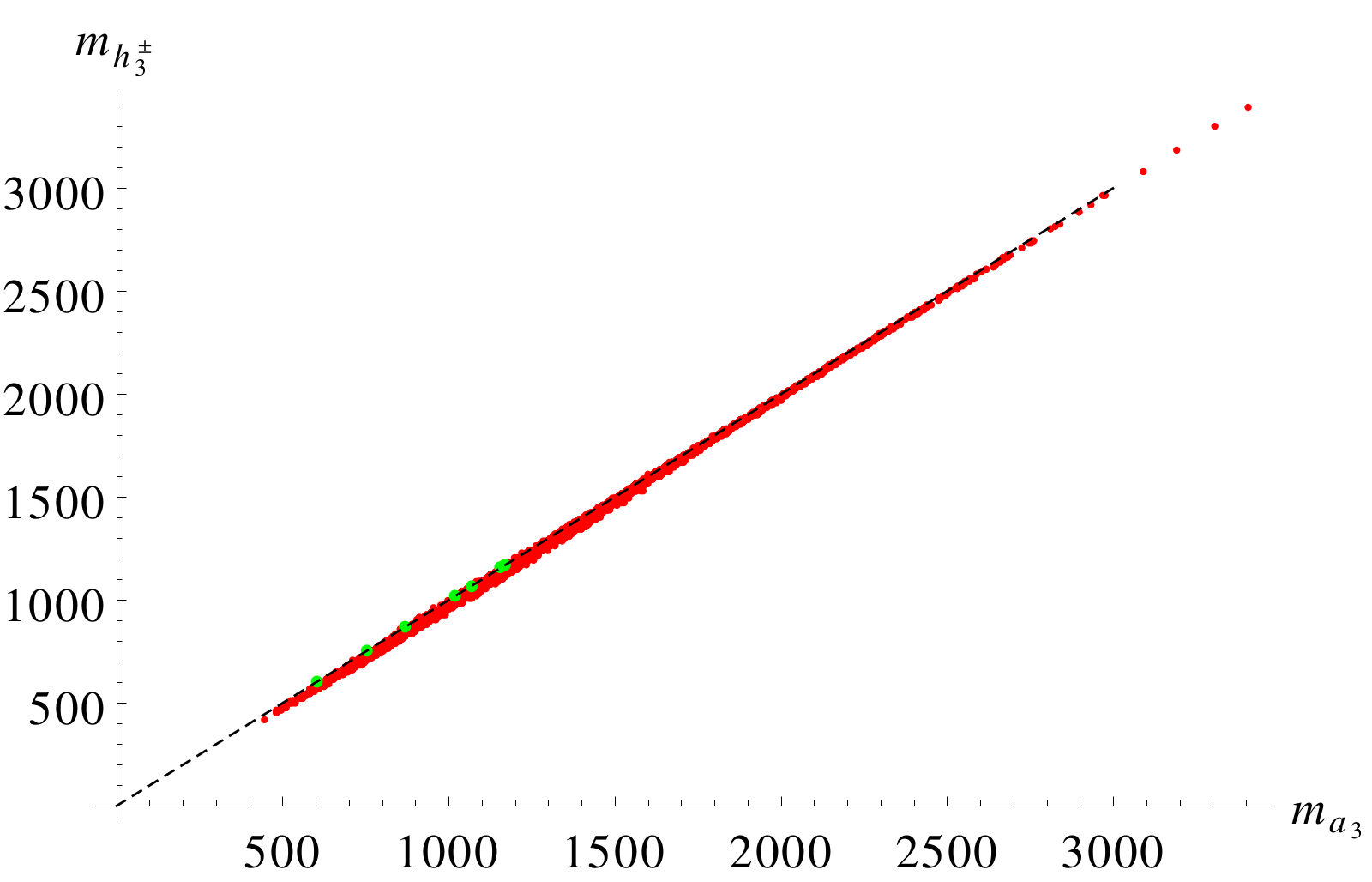}}
\subfigure[]{\includegraphics[width=0.55\linewidth]{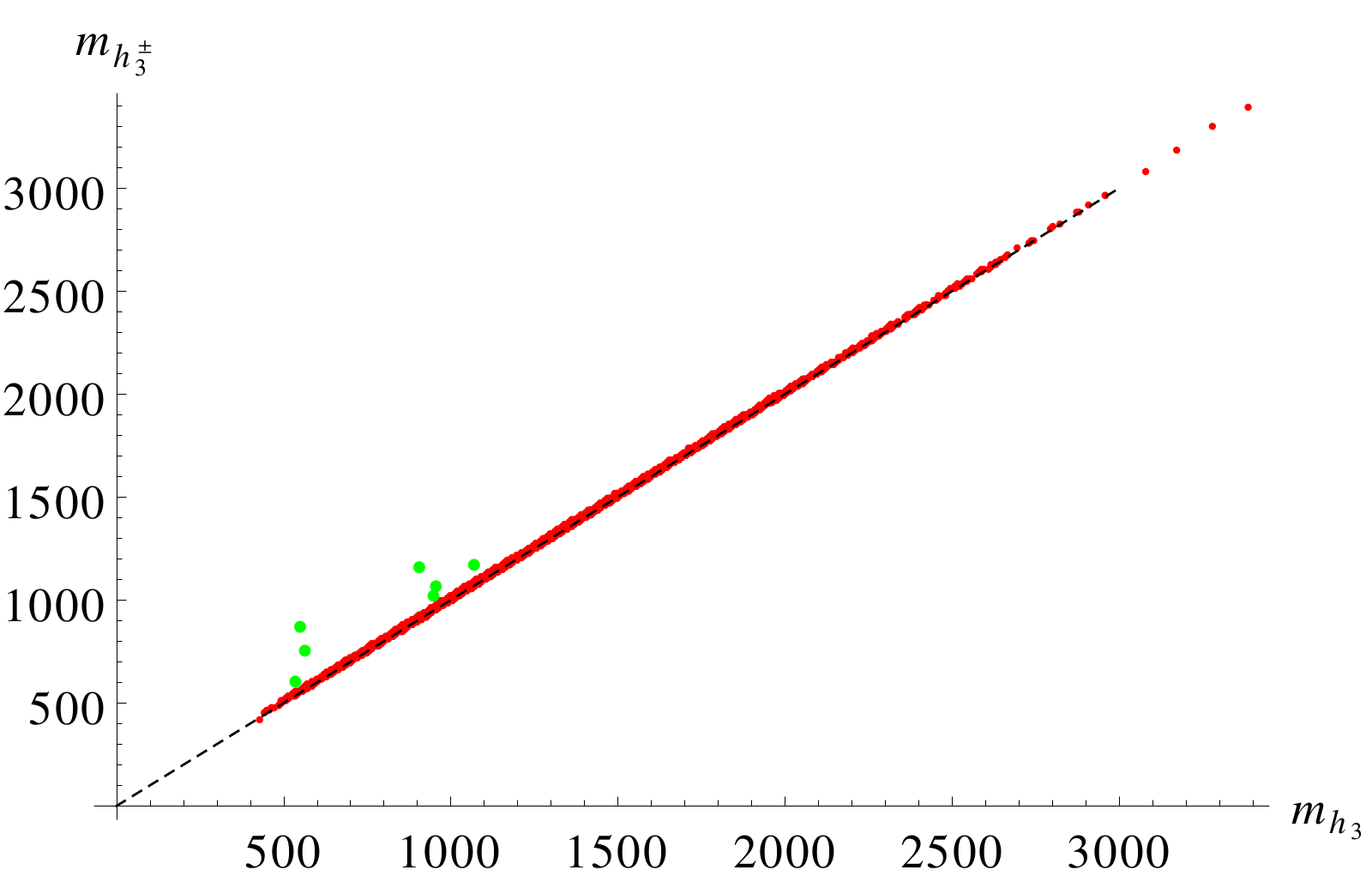}}}
\caption{Scattered plots of the mass correlation between $a_3$ and $h_3^\pm$ (a) and between $h_3$ and $h_3^\pm$ (b). The color code is defined as follows: we mark in red the points where $h_3, a_3, h^\pm_3$ are mostly doublets ($\geq90\%$) and in green the points where they are mostly triplet.}\label{mcrl2}
\end{center}
\end{figure}
Figure~\ref{mcrl2}(a) shows the mass correlations between $h^\pm_3$ and $a_3$, while Figure~\ref{mcrl2}(b)  shows the same correlation but between $h^\pm_3$, $h_3$  where 
all of them are of doublet-type nature and are marked in red. It is easily seen that all the
three doublet-like Higgs bosons $h^\pm_3$, $h_3$ and $a_3$ remain degenerate. 
There are only 7 points which behave like triplets and are shown in green. 
Thus it is evident from the above analysis that eigenstates dominated by the same representation (i.e mostly singlet or mostly triplet) tend to be hierarchically clustered. In this case of a $Z_3$ symmetric Lagrangian, the light pseudoscalar is actually a pseudo NG mode of a continuous $U(1)$ symmetry of the Higgs potential, also known as R-axion \cite{Ellwanger}, and remains very light across the entire allowed parameter space.


Though the interaction eigenstates are a mixture of the gauge eigenstates, there seems to be a pattern for the various representations of the Higgs sector.
A given representation tries to keep their masses in the same block, i.e., the masses of scalar, pseudoscalar and charged components of the triplets will form a different mass block than the doublet Higgs sectors. A typical mass hierarchy is shown in Figure~\ref{cartoon}, where  a light pseudoscalar which is a pseudo-NG boson lays hidden below $100$ GeV and the scalar state $h_4$ takes a heavy mass $\sim m_S$, and is therefore decoupled from the low energy spectrum. There is a CP-even Higgs boson of doublet type around $125$ GeV and  doublet-like heavy Higgs bosons of larger mass ($h^\pm_3, h_3, a_3$), shown in red. Apart from doublet and singlet interaction eigenstates, we have two triplets $T_1$ and $T_2$ which then forms two different sets, ($h^\pm_1, h_2/a_2$)  and ($h^\pm_2, a_2/h_2$) in the mass hierarchy, shown in green colours. Of course this is not the most general situation but it comes from the phenomenological constraints that should be applied to the scanned points in the parameter space. We remind again that these constraints include a scalar Higgs boson with a mass around 125 GeV which satisfy the LHC constraint of Eq.~\ref{LHCdata} and no light doublet-like pseudoscalar or charged Higgs boson. We take care of the latter requesting that the lightest pseudoscalar  as mostly singlet and lightest charged Higgs boson is mostly triplet.

 
\section{Charged Higgs bosons and its structure}\label{chcdcy}
In this section we will describe the feature of the charged Higgs sector, emphasizing the role of the rotation angles in the limit $|\lambda_T|\simeq0$. The charged Higgs bosons are a mixture of two doublet and two triplet fields, as can be seen from Eq.~\ref{chH},
\be\label{chH}
h^\pm_i= \mathcal{R}^C_{i1}H_u^+ +  \mathcal{R}^C_{i2}T_2^+ + \mathcal{R}^C_{i3}H_d^{-*} +  \mathcal{R}^C_{i4}T_1^{-*}
\ee
with $\mathcal{R}^C_{i1, i3}$  and $\mathcal{R}^C_{i2, i4}$ determining the doublet and triplet part respectively.  In general $\mathcal{R}^C_{ij}$ is a function of all the VEVs, $\lambda_{T, TS, S}$ and the $A_i$ parameters and we can write schematically
\bea\label{rc}
\mathcal{R}^C_{ij} = f^C_{ij}\left(v_u, v_d, v_T, v_S, \lambda_T, \lambda_{TS}, \lambda_S, A_i\right).
\eea
The charged Higgs mass matrix which is given in the Appendix (Eq.~\ref{chMM}), shows the similar dependency on the parameters. However, the charged Goldstone mode, expressed in terms of the gauge eigenstates, is a function only of the VEVs and the gauge couplings, as we expect from the Goldstone theorem. 
\bea\label{gstn}
h_0^\pm=\pm N_T \left(\sin\beta H_u^+\, -\cos\beta H_d^{-*}  \,\mp\sqrt2\,\frac{v_T}{v}(T_2^+ +T_1^{-*})\right)\, ,\quad N_T=\frac{1}{\sqrt{1+4\frac{v_T^2}{v^2}}}
\eea
Eq.~\ref{gstn} presents the explicit expression of the charged Goldstone mode and we can see that it is independent of any other kind of couplings or parameters. Among the three kind of VEVs entering in the charged Goldstone mode, the triplet  VEV is very small ($v_T\lesssim 5$ GeV) due to its contribution in the $W^\pm$ boson mass, as already discussed in Eq.~\ref{gbmass}. The triplet VEV, being restricted by the $\rho$ parameter \cite{rho}, makes the charged Goldstone always doublet-type. However among the massive states in the gauge basis, two of them are triplet-like and one is doublet-like. We shall see later that this small triplet contribution to the Goldstone boson protects one of the three physical charged Higgs bosons from becoming absolute triplet-like.

\begin{figure}[thb]
\begin{center}
\includegraphics[width=0.6\linewidth]{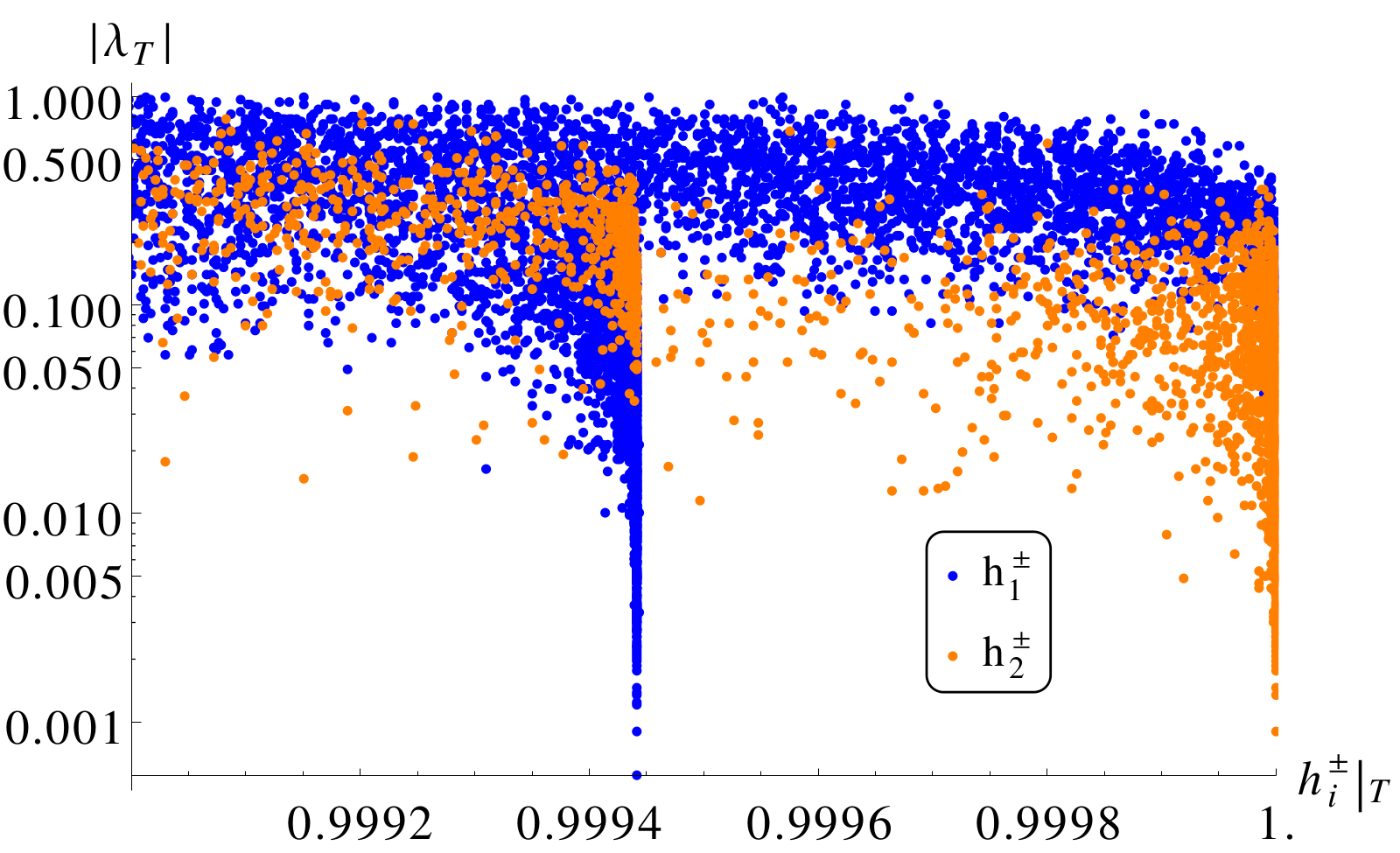}
\includegraphics[width=0.6\linewidth]{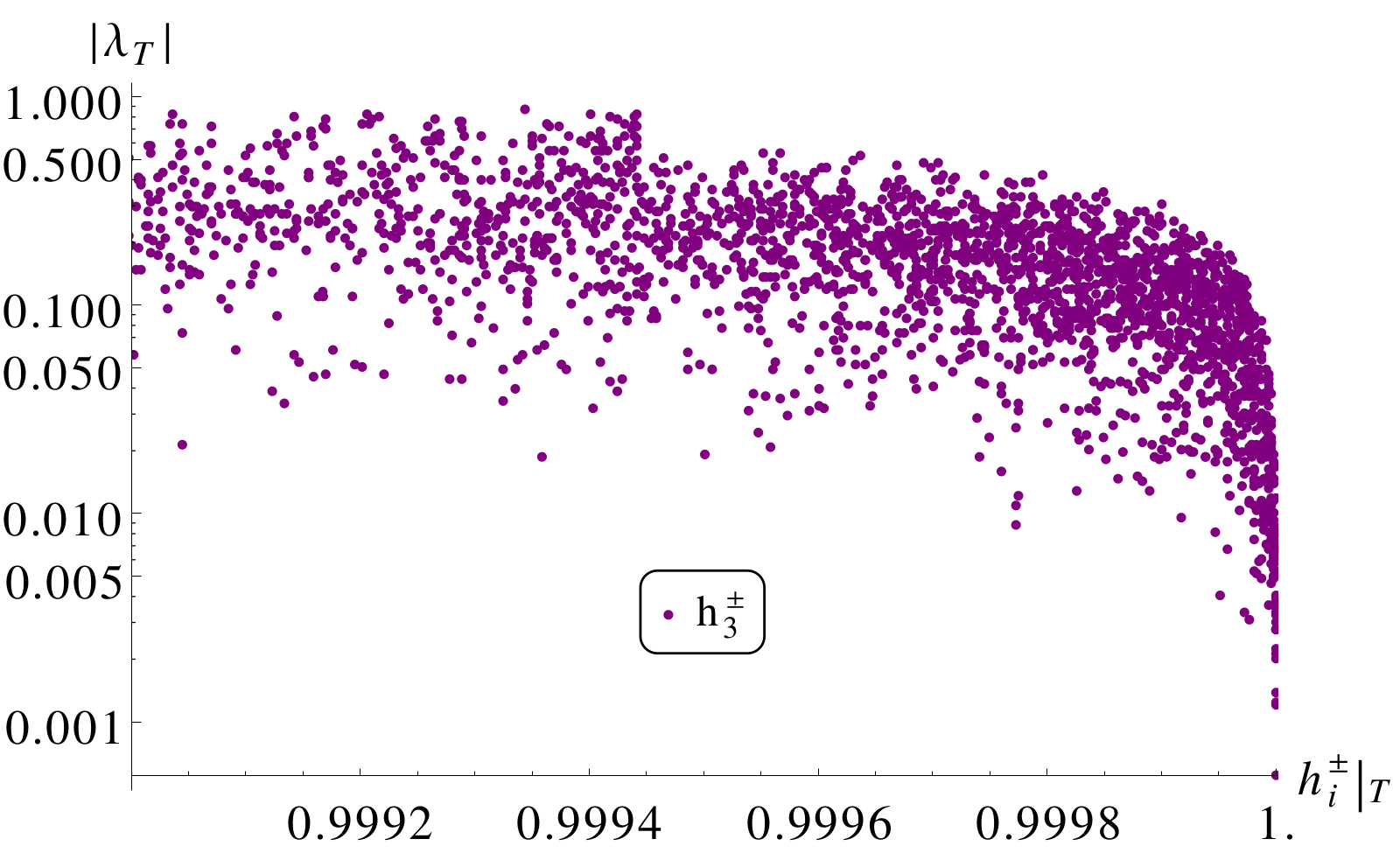}
\caption{Triplet component of the massive charged Higgs bosons versus $\lambda_T$.}\label{chpslmbda}
\end{center}
\end{figure}
In Figure~\ref{chpslmbda} we show the structure of the 
charged Higgs bosons as a function of  $|\lambda_T|$, where we demand 
the lightest charged Higgs massive state to be mostly triplet. One can realize that that for a non-zero $\lambda_T$, their tendency is to mix. However, as we move towards the $|\lambda_T|\simeq 0$ region, one of the charged Higgs boson gives away the $\sim (\frac{v_T}{v})^2$ triplet part to the charged Goldstone and fails to become 100\% triplet (see the blue points in Figure~\ref{chpslmbda}). 
{ In the models where $A_T$ parameter is proportional to $\lambda_T$, the mixing induced by the soft parameter $A_T$ automatically goes to zero in this limit. However the mixing of doublet and triplet  in the charged  Goldstone comes from the corresponding VEVs and it is independent of $\lambda_T$ or $A_T$ as can be seen from Eq. 21. Now all the other massive charged Higgs bosons are orthogonal to the Goldstone boson, which makes the similar mixing in the massive states as well. This mixing goes to zero only when the triplet does not play any role 
in EWSB, i.e. $v_T=0$. However for non-zero $\lambda_T$ and $A_T$ the additional mixings come for the massive eigenstates.}

Anyone of the three massive charged Higgs boson can show this feature but we see it only for $h_1^\pm$ because in the selection criteria we have demanded that $h_1^\pm$ must be triplet-like. Thus for non-zero triplet VEV even with $|\lambda_T|=0$, complete decoupling of doublet and triplet representations is not possible. Therefore by 'decoupling limit' we mean $|\lambda_T|\simeq 0$ here onwards.  In this decoupling limit either the $h^\pm_2$ or the $h^\pm_3$ become completely of triplet-type.  A similar conclusion was shown for the triplet extension of the supersymmetric standard model \cite{EspinosaQuiros}.
\begin{figure}[thb]
\begin{center}
\includegraphics[width=0.98\linewidth]{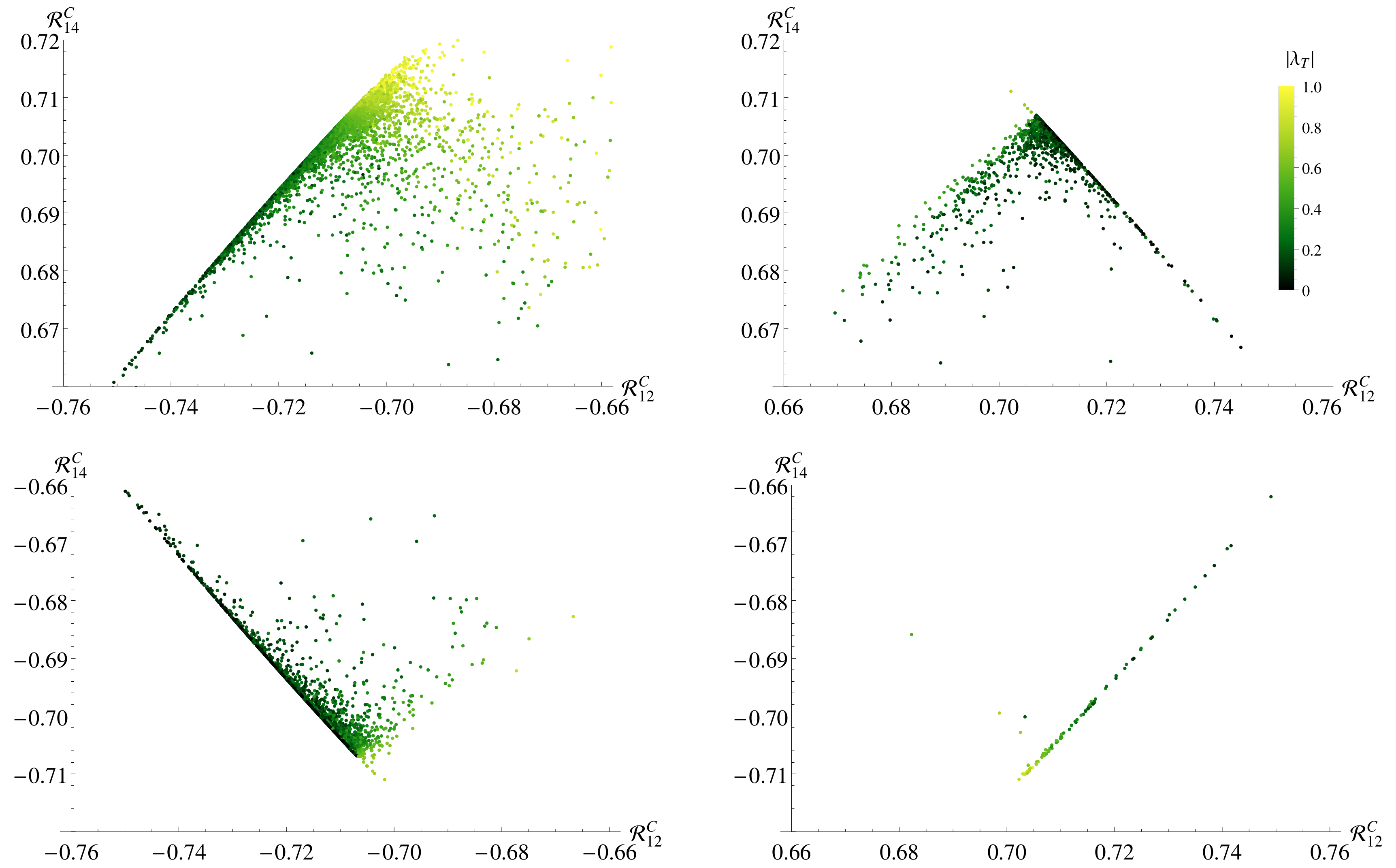}
\caption{Correlations of the rotation angles of the lightest charged Higgs boson $h^\pm_1$ as a function of $\lambda_T$.}\label{ssoslmbda}
\end{center}
\end{figure}

\begin{table}
\begin{center}
\renewcommand{\arraystretch}{1.4}
\begin{tabular}{||c||c|c||}
\hline
&$10^{-2}<|\lambda_T|<1$&$|\lambda_T|<10^-2$\\
\hline
\rm{sign} $\mathcal{R}^C_{12}$ $\mathcal{R}^C_{14}$&+ or -&+\\
\hline
\end{tabular}
\caption{The sign of the product $\mathcal{R}^C_{12}$ $\mathcal{R}^C_{14}$. The sign of the two rotation angles of the lightest charged Higgs boson plays a crucial role in the interactions of a triplet-like charged Higgs boson. In the limit $|\lambda_T|\sim0$ these two rotation angles have the same sign. This feature has important consequences for the interaction, and hence the cross-section, of the lightest charged Higgs boson in various channels.}\label{r2r4s}
\end{center}
\end{table}


The decoupling limit of $|\lambda_T|\sim 0$ not only affects the structure of the charged Higgs bosons, where two of them become triplet-like and one of them doublet-like, but also affects the respective coupling via the corresponding rotation angles. In Figure~\ref{ssoslmbda} we show the rotation matrix elements for the light charged Higgs boson $h^\pm_1$ with respect to $|\lambda_T|$. We can see that when $\lambda_T$ becomes very small the mixing angles in the triplet component of the light charged Higgs boson $h^\pm_1$, $\mathcal{R}^C_{12}$ and $\mathcal{R}^C_{14}$, as defined in Eq.~\ref{chH}, take same signs, unlike the general case. We will see later that the presence of same signs in $\mathcal{R}^C_{12}$ and  $\mathcal{R}^C_{14}$  in the decoupling limit, causes an enhancement of some production channels and decrement for some other ones.

\section{Decays of the charged Higgs bosons}\label{chdcys}

As briefly mentioned above, the phenomenology of the Higgs decay sector of the TNMSSM, as discussed in \cite{TNMSSM1}, is affected by the presence of a light pseudoscalar which induces new decay modes. In this section we consider its impact in the decay of a light charged Higgs boson $h^\pm_1$. Along with the existence of the light pseudoscalar, which opens up the $h^\pm_1 \to a_1 W^\pm$ decay mode, the triplet-like charged Higgs adds new decay modes, not possible otherwise. In particular, a $Y=0$ triplet-like charged Higgs boson gets a new decay mode into $ZW^\pm$ which is a signature of custodial symmetry breaking.  Apart from that, the usual doublet-like decay modes into $\tau\nu$ and $tb$ are present via the mixings with the doublets. 

\subsection{$h_i^\pm  \to W^\pm h_j/a_i$}
The trilinear couplings with charged Higgses, scalar (pseudoscalar) Higgses and $W^\pm$ are given by 
\begin{align}\label{hachW}
g_{h_i^\pm W^\mp h_j}&=\frac{i}{2}g_L\Big(\mathcal R_{j2}^S\mathcal R_{i3}^C-\mathcal R_{j1}^S\mathcal R_{i1}^C+\sqrt2\mathcal R_{j4}^S\left(\mathcal R_{i2}^C+\mathcal R_{i4}^C\right)\Big), \\
g_{h_i^\pm W^\mp a_j}&=\frac{g_L}{2}\Big(\mathcal R_{j1}^P\mathcal R_{i1}^C+\mathcal R_{j2}^P\mathcal R_{i3}^C+\sqrt2\mathcal R_{j4}^P\left(\mathcal R_{i2}^C-\mathcal R_{i4}^C\right)\Big).
\end{align}
Both the triplet and doublet has $SU(2)$ charges so they couple to $W^\pm$ boson. Their coupling in association with neutral Higgs bosons have to be  doublet(triplet) type for doublet(triplet) type charged Higgs bosons. For the phenomenological studies we have considered a doublet-like Higgs boson around $125$ GeV, a light triplet-like charged Higgs boson $\lesssim 200$ GeV and a very light singlet type pseudoscalar $\sim 20$ GeV. Hence the mixing angles become really important. In the next few section we will see how the various rotation angles involved with the charged Higgs bosons and their relative signs determine the strength of the couplings and thus of the decay widths. Eq.~\ref{hachW} shows that for $h_i^\pm  \to W^\pm h_j$ decay the rotation angles $\mathcal R^C_{i2}$ and $\mathcal R^C_{i4}$ come as additive where as for $h_i^\pm  \to W^\pm a_j$ they come as subtractive.

\begin{figure}[thb]
\begin{center}
\includegraphics[width=0.7\linewidth]{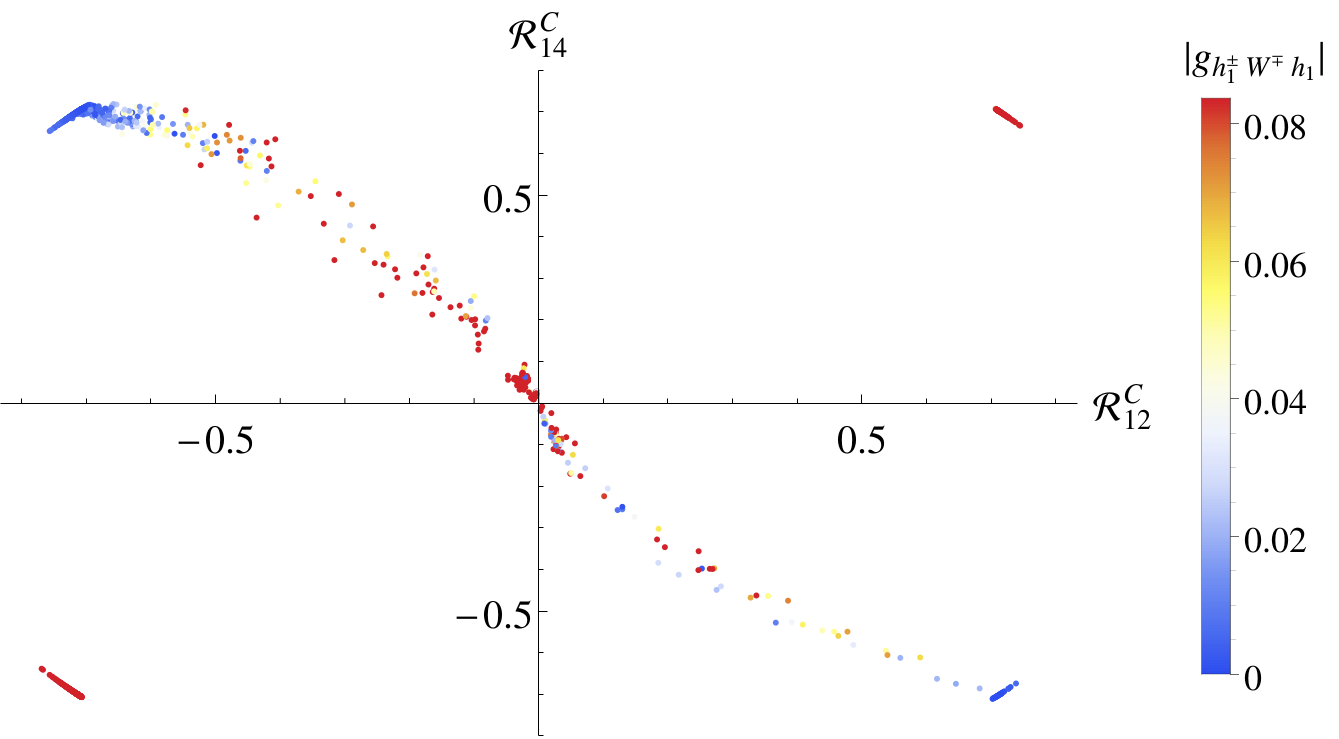}
\caption{Correlation of $g_{h^\pm_1 W^\mp h_1}$ with  $\mathcal{R}^C_{12}$ and $\mathcal{R}^C_{14}$. { The coupling is enhanced when $\mathcal{R}^C_{12}$ and $\mathcal{R}^C_{14}$ are small, i.e. for a doublet-like charged Higgs $h_1^\pm$. The enhancement in the I and III quadrants are related to the same sign of $\mathcal{R}^C_{12}$ and $\mathcal{R}^C_{14}$, cfr. Eq.~\ref{hachW}.}}\label{ghmp1h1W}
\end{center}
\end{figure}

The decay width of a massive charged Higgs boson in a $W$ boson and a scalar (or pseudoscalar) boson is given by
\bea \label{chwah}
\Gamma_{h_i^\pm\rightarrow W^\pm h_j/a_j}&=&\frac{G_F}{8\sqrt2\pi}m^2_{W^\pm}|g_{h_i^\pm W^\mp h_j/a_j}|^2 \,\sqrt{\lambda(1,x_W,x_{h_j/a_j})}\,\lambda(1,y_{h_i^\pm},y_{h_j/a_j})
\eea

where $x_{W,h_j}=\frac{m^2_{W,h_j}}{m^2_{h_i^\pm}}$ and $y_{h_i^\pm,h_j}=\frac{m^2_{h_i^\pm,h_j}}{m^2_{W^\pm}}$ and similarly for $a_j$.
\begin{figure}[thb]
\begin{center}
\includegraphics[width=0.7\linewidth]{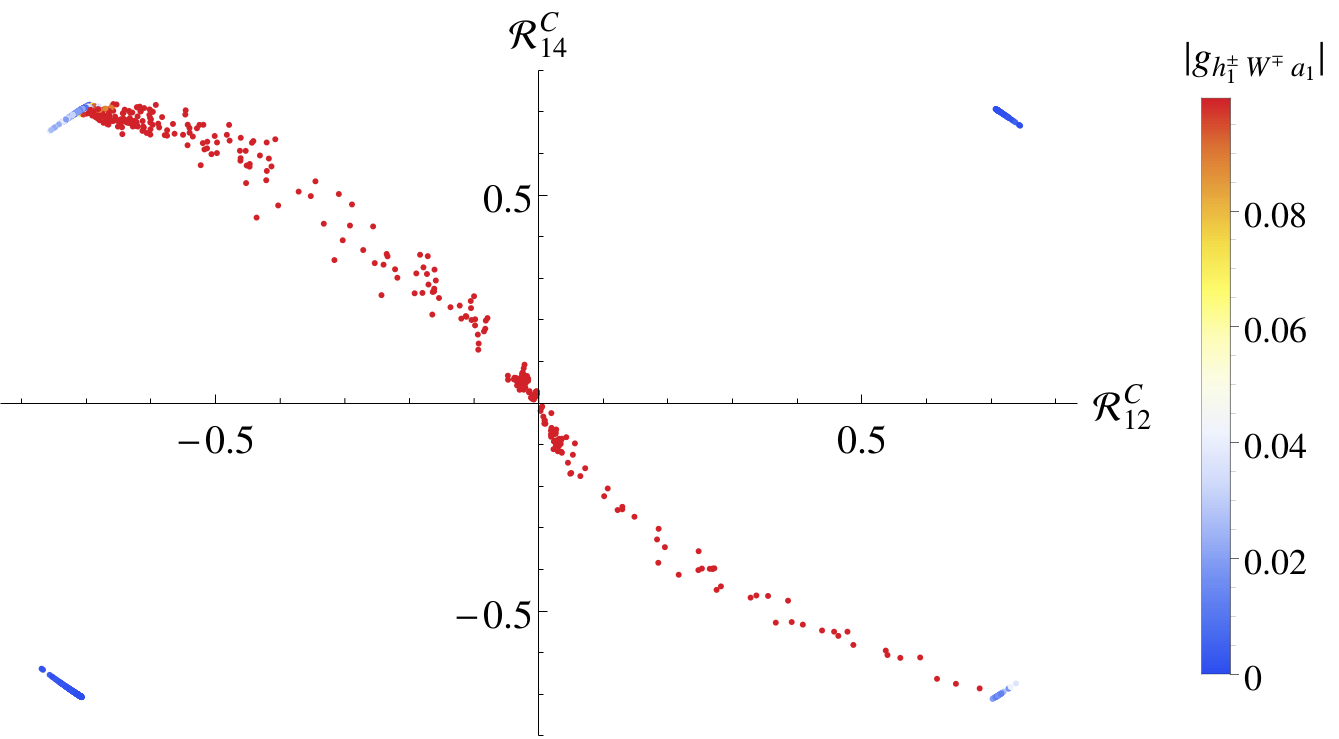}
\caption{Correlation of $g_{h^\pm_1 W^\mp a_1}$ with  $\mathcal{R}^C_{12}$ and $\mathcal{R}^C_{14}$. { For the blue points in II and IV quadrants the low values of the coupling are due to the selection of a singlet-like $a_1$, which means that $\mathcal{R}^P_{13}\sim1$, whereas for the blue points in the I and  III quadrants the low value of $|g_{h^\pm_1 W^\mp a_1}|$ comes from the cancellation between $\mathcal{R}^C_{12}$ and $\mathcal{R}^C_{14}$.}}\label{ghmp1a1W}
\end{center}
\end{figure}
Figure~\ref{ghmp1a1W} shows the dependency of the $g_{h^\pm_1 W^\mp a_1}$ coupling with the triplet components of the lightest charged Higgs eigenstate, i.e., $\mathcal{R}^C_{12}$ and $\mathcal{R}^C_{14}$.  We have seen from Figure~\ref{ssoslmbda} and Table~\ref{r2r4s} the behaviour of $\mathcal{R}^C_{12}$ $\mathcal{R}^C_{14}$ as a function of $\lambda_T$, i.e. that for $\lambda_T \sim 0$ they take same sign. We can see that in the decoupling limit, i.e. for $\lambda_T\sim 0$, the coupling decreases because $\mathcal{R}^C_{12}$ and $\mathcal{R}^C_{14}$ take same sign and they tend to cancel, cfr. Eq.~\ref{hachW}. 
{ A low value of this coupling can come even when the pseudoscalar Higgs boson ($a_j$) is singlet-like, which means that $\mathcal{R}^P_{j3}\sim1$.}
The situation is just opposite in the case of $g_{h^\pm_1 W^\mp h_1}$, as one can see from Figure~\ref{ghmp1h1W}. Here in the decoupling limit the coupling  $g_{h^\pm_1 W^\mp h_1}$ is enhanced. { In Figure~\ref{ghmp1h1W} we can also see some blue points with low $\mathcal{R}^C_{12}$, $\mathcal{R}^C_{14}$. In this case the charged Higgs boson is not triplet-like and the suppression in the coupling is due to the accidental cancellation of $\Big(\mathcal R_{12}^S\mathcal R_{13}^C-\mathcal R_{11}^S\mathcal R_{11}^C\Big)$, cfr. Eq.~\ref{hachW}. This cancellation is of course not related to the limit $\lambda_T\sim 0$.} We see later how it affects the corresponding production processes. 

\subsection{$h_i^\pm  \to W^\pm Z$}
The charged sector of a theory with scalar triplet(s) is very interesting due to the tree-level interactions $h_i^\pm-W^\mp-Z$ for $Y=0, \pm 2$ hypercharge triplets which break the custodial symmetry \cite{pbas3,EspinosaQuiros,tnssm, tnssma}. In the TNMSSM this coupling is given by 
\bea\label{zwch}
g_{h_i^\pm W^\mp Z}&=&-\frac{i}{2}\left(g_L\, g_Y\left(v_u\sin\beta\,\mathcal R^C_{i1}-v_d\cos\beta\,\mathcal R^C_{i3}\right)+\sqrt2\,g_L^2v_T\left(\mathcal R^C_{i2}+\mathcal R^C_{i4}\right)\right),
\eea
where the rotation angles are defined in Eq.~\ref{chmix}. The on-shell decay width is given by
\bea\label{chzw}
\Gamma_{h_i^\pm\rightarrow W^\pm Z}&=&\frac{G_F\,\cos^2\theta_W}{8\sqrt2\pi}m^3_{h_i^\pm}|g_{h_i^\pm W^\mp Z}|^2\,\sqrt{\lambda(1,x_W,x_Z)}\left(8\,x_W\,x_Z+(1-x_W-x_Z)^2\right)
\eea
where $\lambda(x,y,z)=(x-y-z)^2-4\,y\,z$ and $x_{Z,W}=\frac{m^2_{Z,W}}{m^2_{h_i^\pm}}$ \cite{Asakawa}.

\begin{figure}[thb]
\begin{center}
\includegraphics[width=0.7\linewidth]{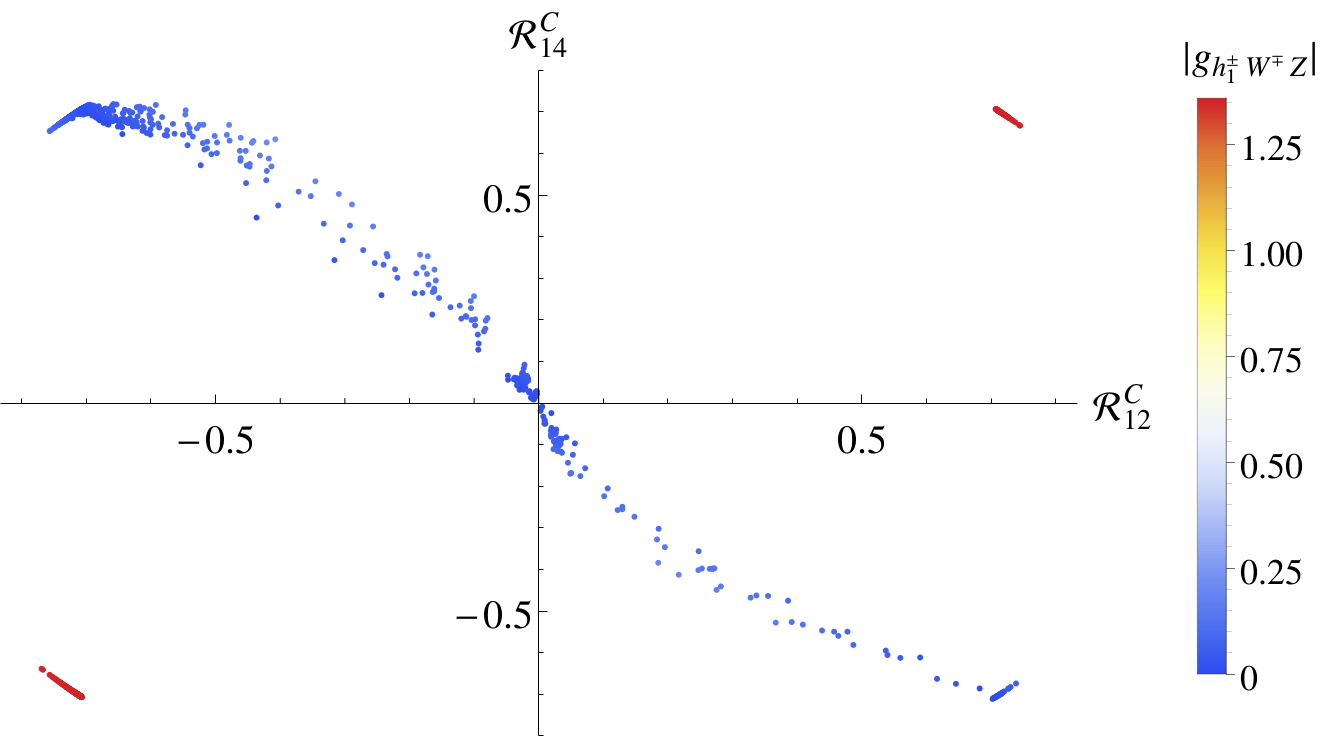}
\caption{Correlation of $g_{h^\pm_1 W^\mp Z}$ with  $\mathcal{R}^C_{12}$ and  $\mathcal{R}^C_{14}$.}\label{ghmp1WZ}
\end{center}
\end{figure}

Figure~\ref{ghmp1WZ} shows the dependency of $g_{h_i^\pm W^\mp Z}$ with respect to $\mathcal{R}^C_{12}$ and $\mathcal{R}^C_{14}$. We see that for 
$\lambda_T \sim 0$ $\mathcal{R}^C_{12}$ and $\mathcal{R}^C_{14}$ take the same sign, and hence the $h_i^\pm-W^\mp-Z$  coupling is enhanced.

\subsection{$h_i^\pm \to t b$}
\begin{figure}[thb]
\begin{center}
\includegraphics[width=0.7\linewidth]{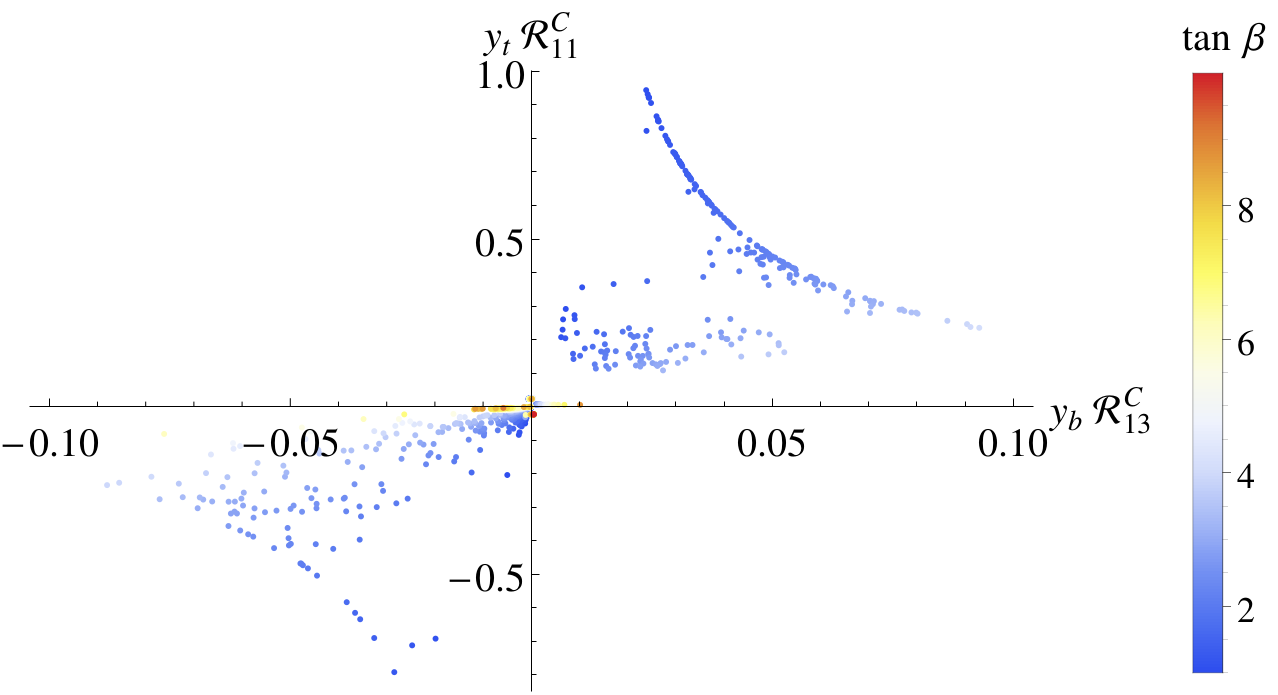}
\caption{Correlation of  $y_t\mathcal{R}^C_{11}$ and $y_b\mathcal{R}^C_{13}$ as a function of $\tan{\beta}$.}\label{ghmp1tb}
\end{center}
\end{figure}

Beside the non-zero $h^\pm_i-W^\mp-Z$ coupling at the tree-level due to custodial symmetry breaking, the charged Higgs bosons can also decay into fermions through the Yukawa interaction given below
\bea
g_{h_i^+ \bar u d}=i\left(y_u\,\mathcal R^C_{i1}\,\mathtt{P_L}+y_d\,\mathcal R^C_{i3}\,\mathtt{P_R}\right)
\eea
governed by doublet part of the charged Higgses. The decay width at leading order is
\begin{align}\label{chtb}
\Gamma_{h_i^\pm\rightarrow u\,d}&=\frac{3}{4}\frac{G_F}{\sqrt2\pi}m_{h_i^\pm}\sqrt{\lambda(1,x_u,x_d)}\Bigg[(1-x_u-x_d)\,\left(\frac{m^2_u}{\sin^2\beta}(\mathcal R^C_{i1})^2+\frac{m_d^2}{\cos^2\beta}(\mathcal R^C_{i3})^2\right)\nn\\
&\hspace{4.5cm}-4\frac{m_u^2m_d^2}{m^2_{h_i^\pm}}\frac{\mathcal R^C_{i1}\mathcal R^C_{i3}}{\sin\beta\cos\beta}\Bigg]
\end{align}
where $x_{u,d}=\frac{m^2_{u,d}}{m^2_{h_i^\pm}}$. The QCD correction to the leading order formula are the same as in the MSSM and are given in \cite{anatomy2}. The decay of the charged Higgs bosons into quarks is then suppressed in the case of triplet-like eigenstates, as one can easily realize from the expression above. In Figure~\ref{ghmp1tb} we show the correlation of the effective Yukawa coupling $(y_u\,\mathcal R^C_{i1}$ and $y_d\,\mathcal R^C_{i3})$ of top and bottom quark respectively as a function of $\tan\beta$. The dominant contribution comes from the top for small $\tan\beta$, as we expected.

\section{Decay branching ratios of the charged Higgs bosons}\label{ch1dcy}
Prepared with the possibilities of new decay modes we finally analyze such scenarios with the data satisfying various theoretical and experimental constraints.  The points here have a CP-even neutral Higgs boson around 125 GeV which satisfies the LHC constraint given in Eq.~\ref{LHCdata}.
To study the decay modes and calculate the branching fractions we have implemented our model in \texttt{SARAH$\_$4.4.6} \cite{sarah} and we have generated the model files for \texttt{CalcHEP$\_$3.6.25} \cite{calchep}.
\begin{figure}[thb]
\begin{center}
\mbox{\subfigure[]{
\includegraphics[width=0.23\linewidth]{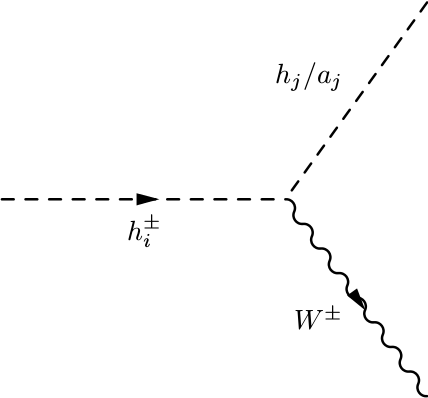}}
\hspace*{.5cm}
\subfigure[]{
\includegraphics[width=0.23\linewidth]{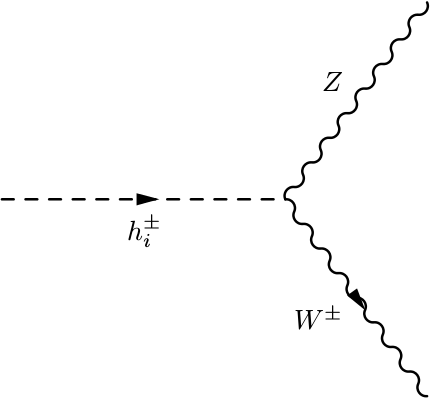}}
\hspace*{.5cm}
\subfigure[]{
\includegraphics[width=0.23\linewidth]{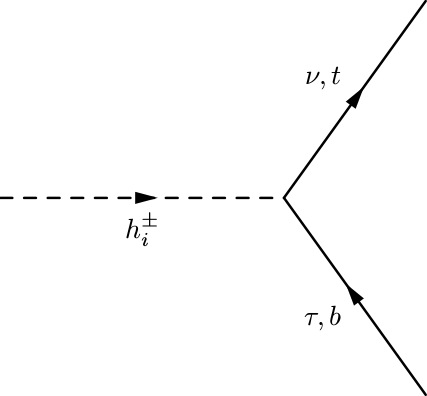}}}
\caption{The new and modified decay channels of the Higgs bosons at the LHC.}\label{higgdcy}
\end{center}
\end{figure}
\begin{figure}[thb]
\begin{center}
\mbox{\subfigure[]{\includegraphics[width=.5\linewidth]{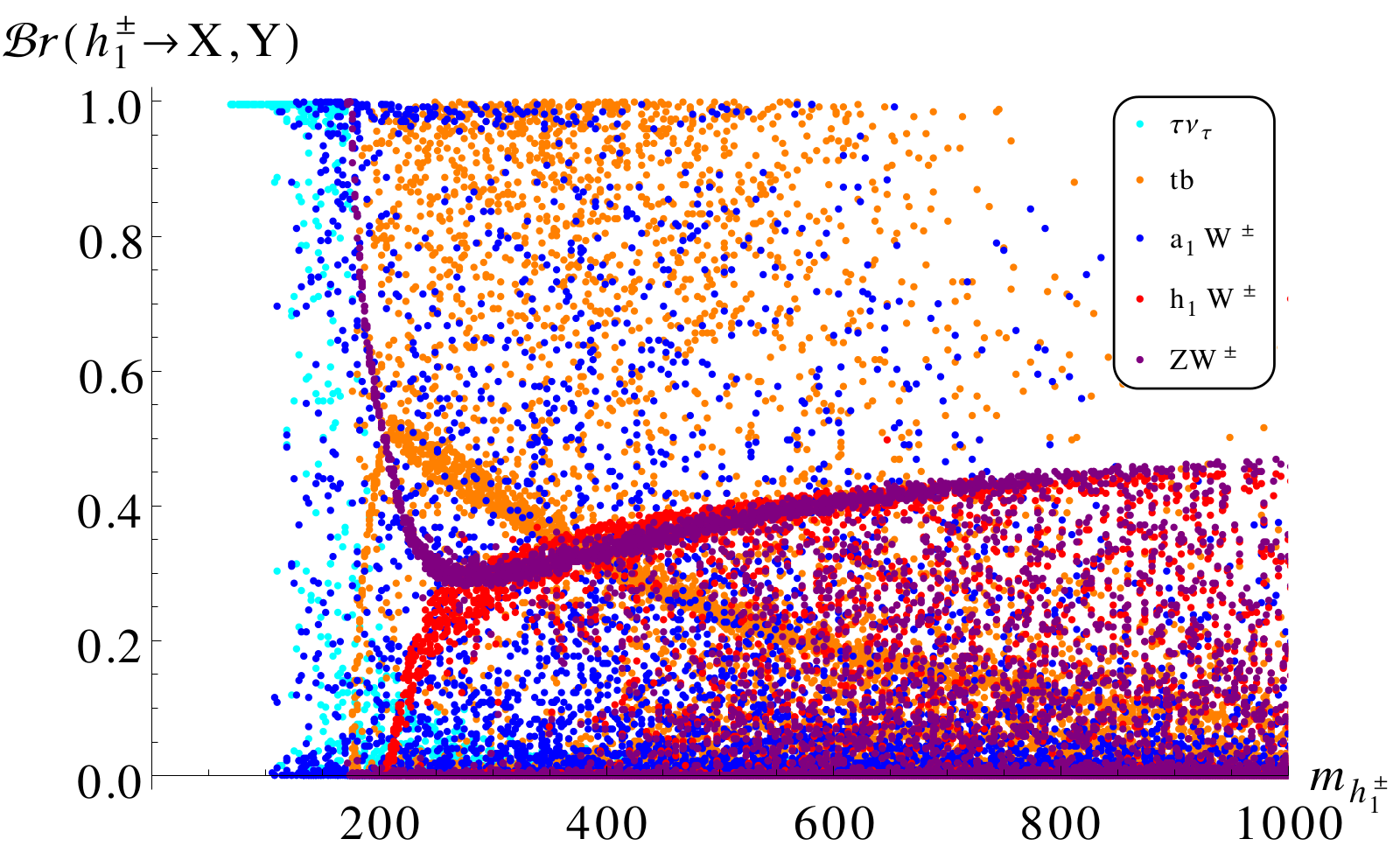}}
\subfigure[]{\includegraphics[width=.5\linewidth]{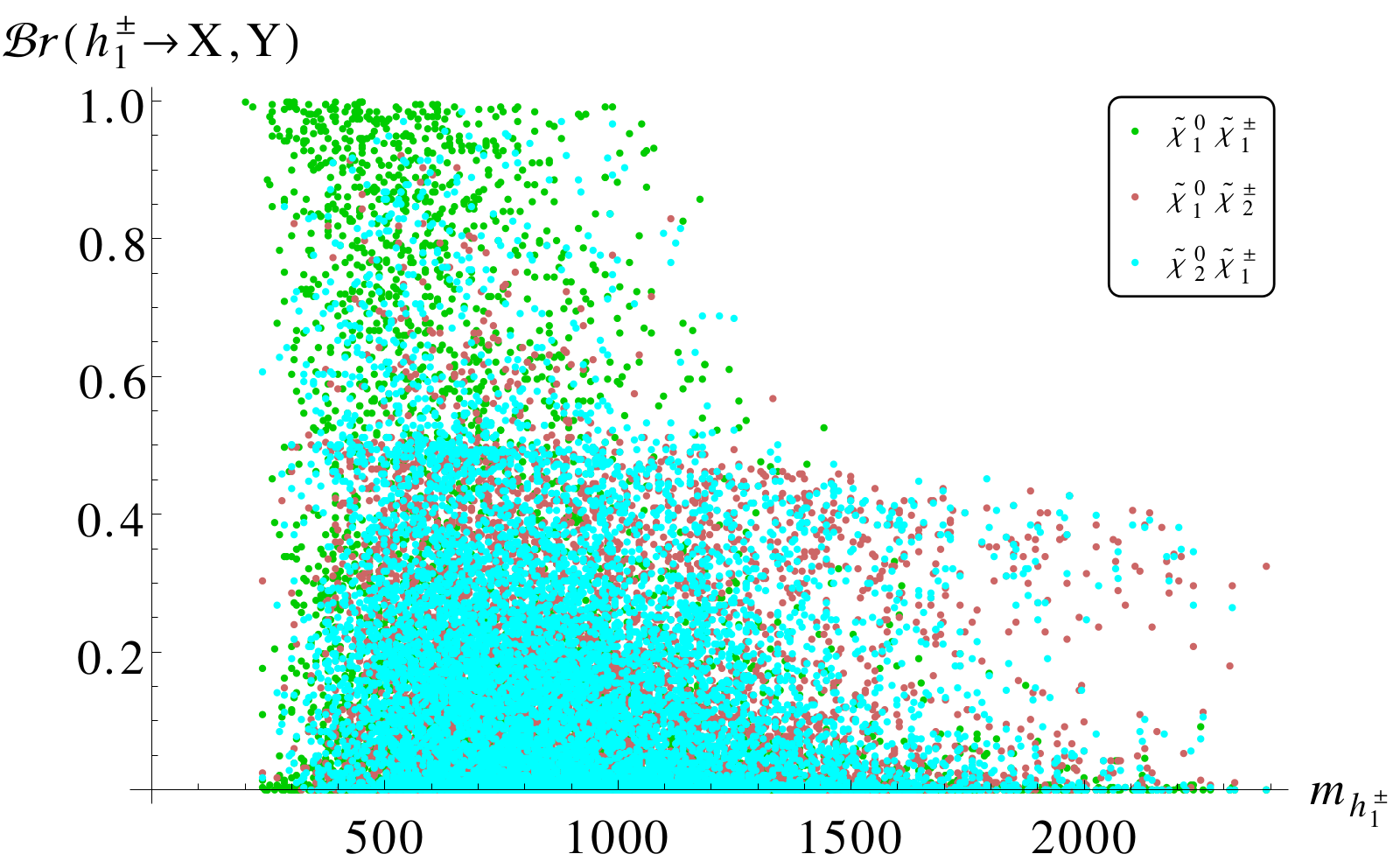}}}
\caption{The branching ratios for the decay of the lightest charged Higgs boson $h^\pm_1$ into non-supersymmetric (a) and supersymmetric modes (b).}\label{ch1br}
\end{center}
\end{figure}

\begin{figure}[bht]
\begin{center}
\includegraphics[width=.5\linewidth]{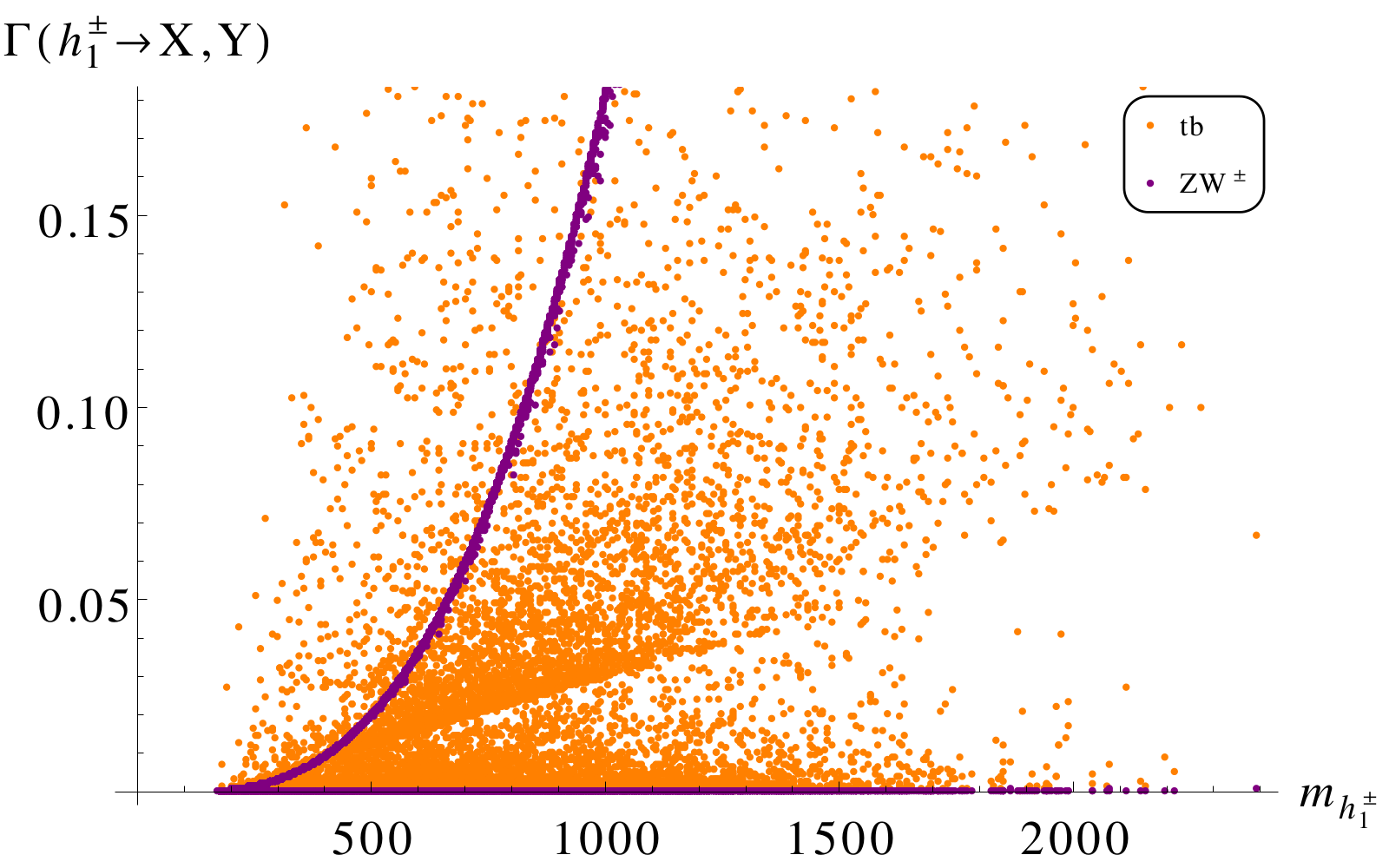}
\caption{The decay widths of the lightest charged Higgs boson $h^\pm_1$ to $tb$ and $ZW^\pm$.}\label{ch1dc}
\end{center}
\end{figure}

Figure~\ref{ch1br}(a) presents the decay branching ratios of the light charged Higgs boson $h^\pm_1$
into non-supersymmetric modes.  This includes the $a_1W^\pm$, $h_1W^\pm$, $ZW^\pm$, $tb$ and $\tau\nu$ channels.
The points in the Figure~\ref{ch1br} include a discovered Higgs boson at $\sim 125$ GeV  and a triplet-like light charged Higgs boson $h_1^\pm$. When $a_1$ is singlet-type, the $a_1W^\pm$ decay mode is suppressed in spite of being kinematically open.  One can notice that, being the $ h^\pm_1$ triplet-like, the decay mode $ZW^\pm$ can be very large, even close to $100\%$. When the $tb$ mode is kinematically open, the $ZW^\pm$ gets an apparent suppression, but it increases again for a charged Higgs bosons of larger mass ($m_{h^\pm_1}\sim 400$ GeV). This takes place because the $h^\pm_i \to ZW^\pm$ decay width is proportional to $m_{h^\pm_i}^3$, unlike the $tb$ one, which is proportional to $m_{h^\pm_i}$ (see Eq.~\ref{chzw} and Eq.~\ref{chtb}). The variation of these two decay widths, as a function of $m_{h^\pm_1}$, are shown in Figure~\ref{ch1dc}.\\
Figure~\ref{ch1br}(b) shows the decays of the lightest charged Higgs boson into the supersymmetric modes, i.e. into charginos $\tilde{\chi}^\pm_i$ and neutralinos $\tilde{\chi}^0_j$, when these modes are are kinematically allowed. We observe that for a charged Higgs boson of a relatively higher mass $m_{h^\pm_i} \gsim 300$ GeV, these modes open up and can have very large branching ratios. 

Apart from the lightest charged Higgs boson, there are two additional charged Higgs bosons, $h^\pm_2$ and $h^\pm_3$. As we have pointed out many times, we have selected data points for which the light charged Higgs boson is triplet-type. Certainly, in the decoupling limit, i.e. when $|\lambda_T|\simeq 0$, either one of $h^\pm_{2,3}$ is triplet-like and the other one is doublet-like. The points that we have generated, which satisfy also the precondition of allowing a $h_{125}$ in the spectrum, have a $h^\pm_2$ as a triplet- and a $h^\pm_3$ as a doublet-like Higgs boson, cfr. Figure~\ref{chpslmbda}.
\begin{figure}[thb]
\begin{center}
\mbox{\subfigure[]{
\includegraphics[width=0.5\linewidth]{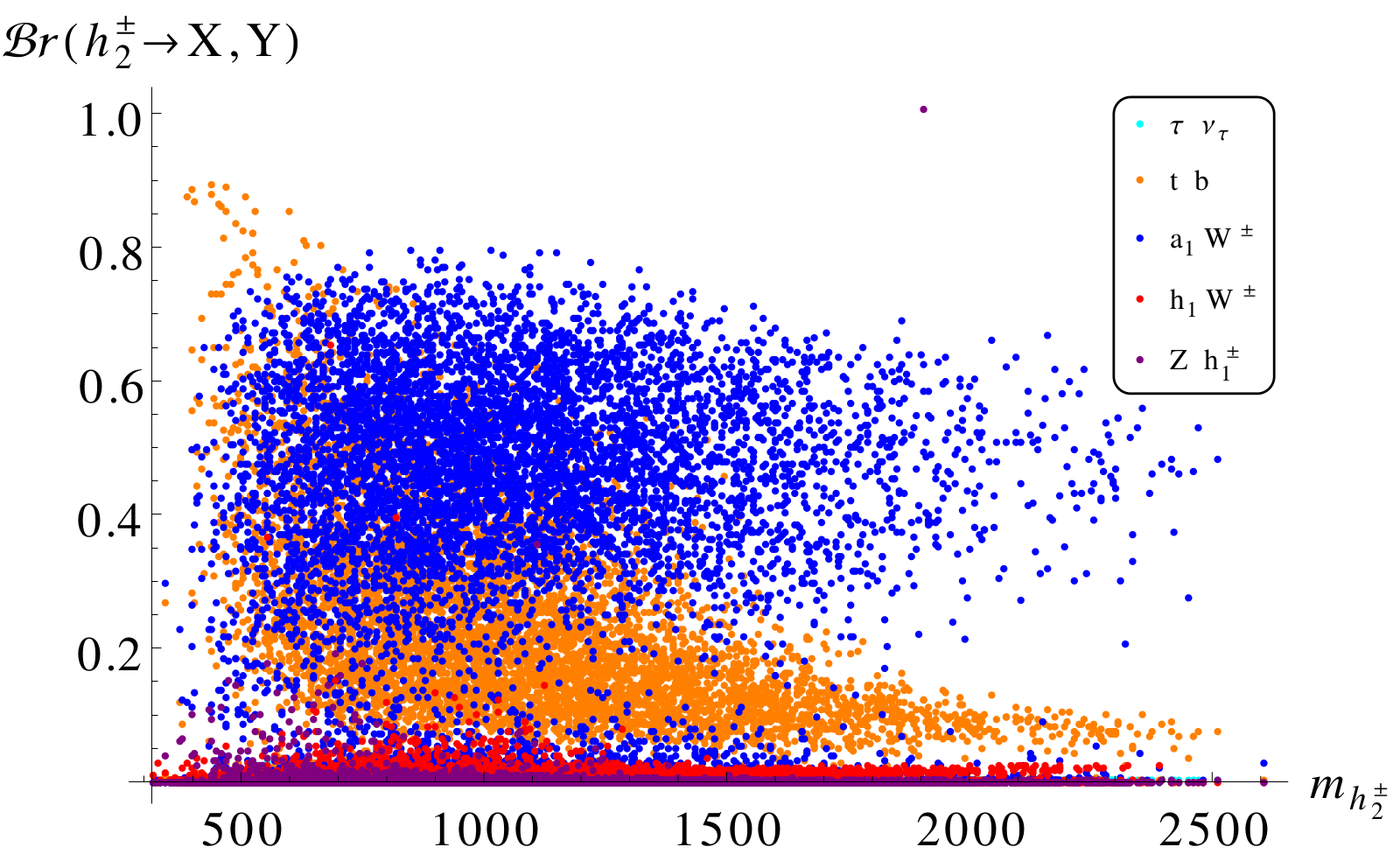}}
\subfigure[]{\includegraphics[width=0.5\linewidth]{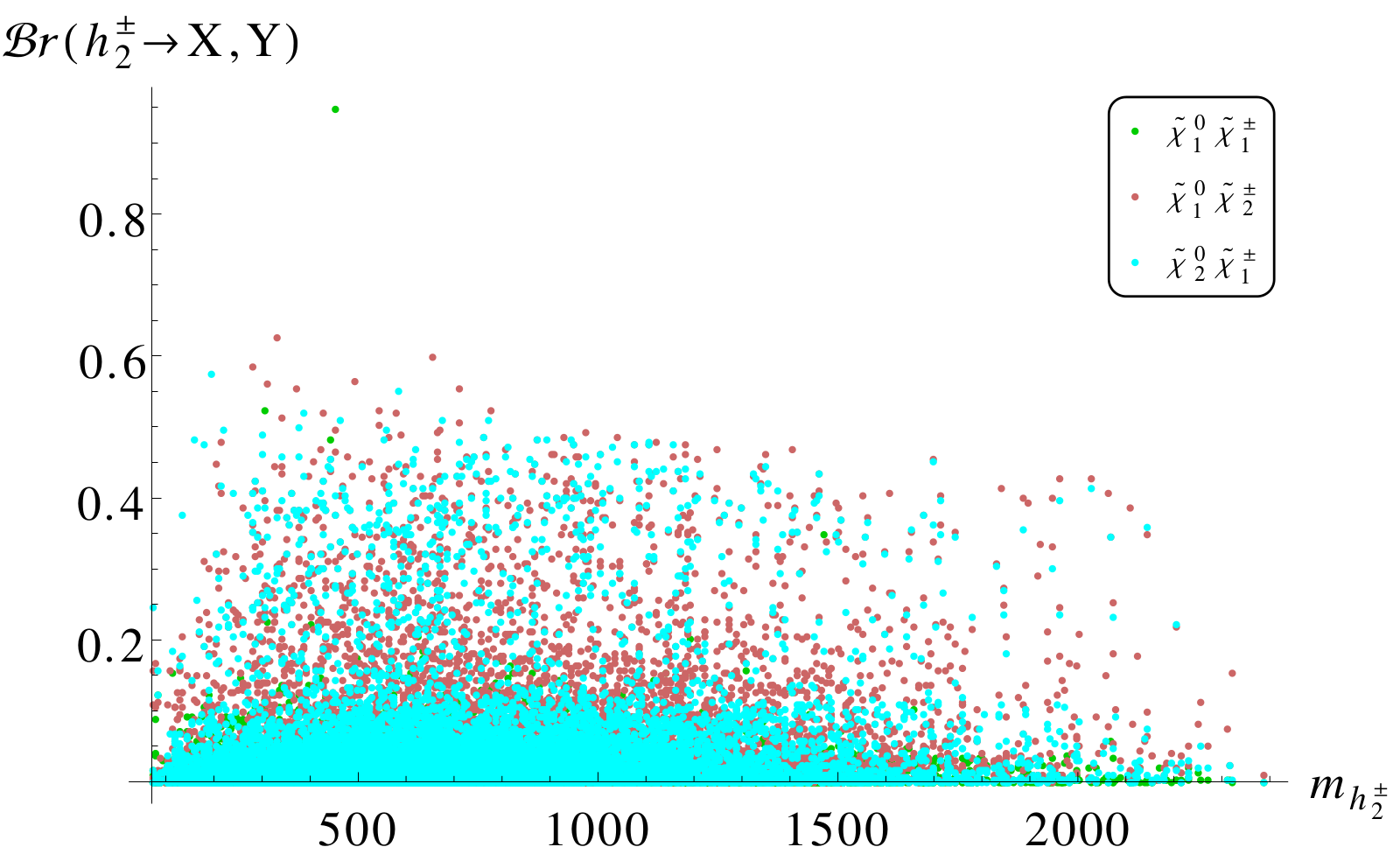}}}
\mbox{\subfigure[]{\includegraphics[width=0.5\linewidth]{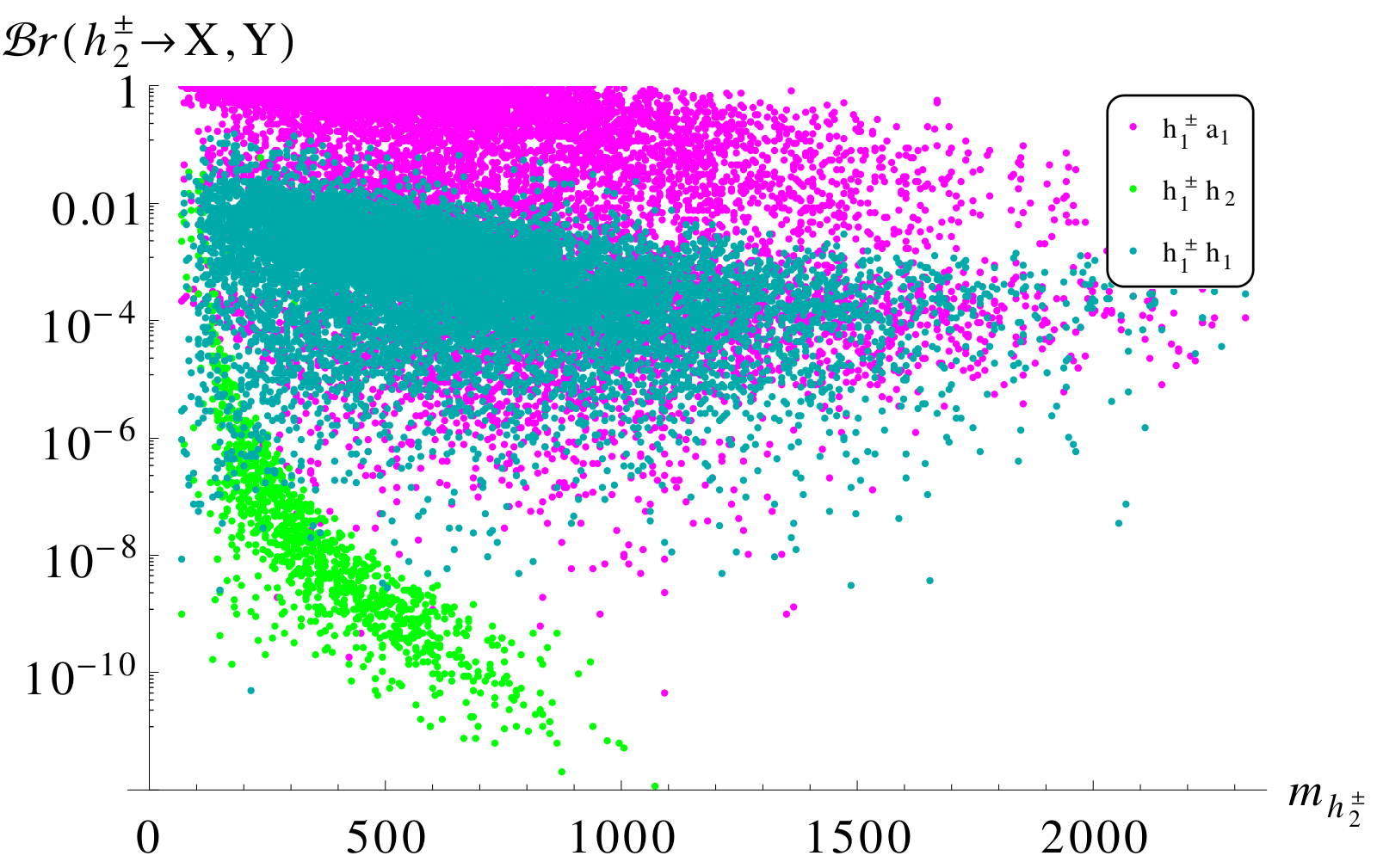}}}
\caption{The branching ratios of the decay of the charged Higgs boson $h^\pm_2$ into non-supersymmetric (a), supersymmetric modes (b) and into Higgs bosons (c).}\label{ch2br}
\end{center}
\end{figure}
In Figure~\ref{ch2br} we present the decay branching ratios of the second charged Higgs boson $h^\pm_2$. Figure~\ref{ch2br}(a) shows the ratios in $\tau\nu$, $tb$, $a_1W^\pm$, 
$h_1 W^\pm$ and $Z h^\pm_1$. As one can observe, $tb$ and $a_1 W^\pm$ are the dominant modes reaching up to $\sim 90\%$ and $\sim80\%$ respectively.  Figure~\ref{ch2br}(b) shows the branching ratios into supersymmetric modes with neutralinos and charginos, which are kinematically allowed. For some benchmark points these modes can have decay ratios as large as $\sim 60\%$. Figure~\ref{ch2br}(c)  shows the ratios for $h^\pm_2$ decaying into two scalars, i.e. to $h_1^\pm h_{1,2}$ and $h^\pm_1 a_1$, with the $h^\pm_1 a_1$ final state being the dominant among all. 
\begin{figure}[thb]
\begin{center}
\mbox{\subfigure[]{
\includegraphics[width=0.44\linewidth]{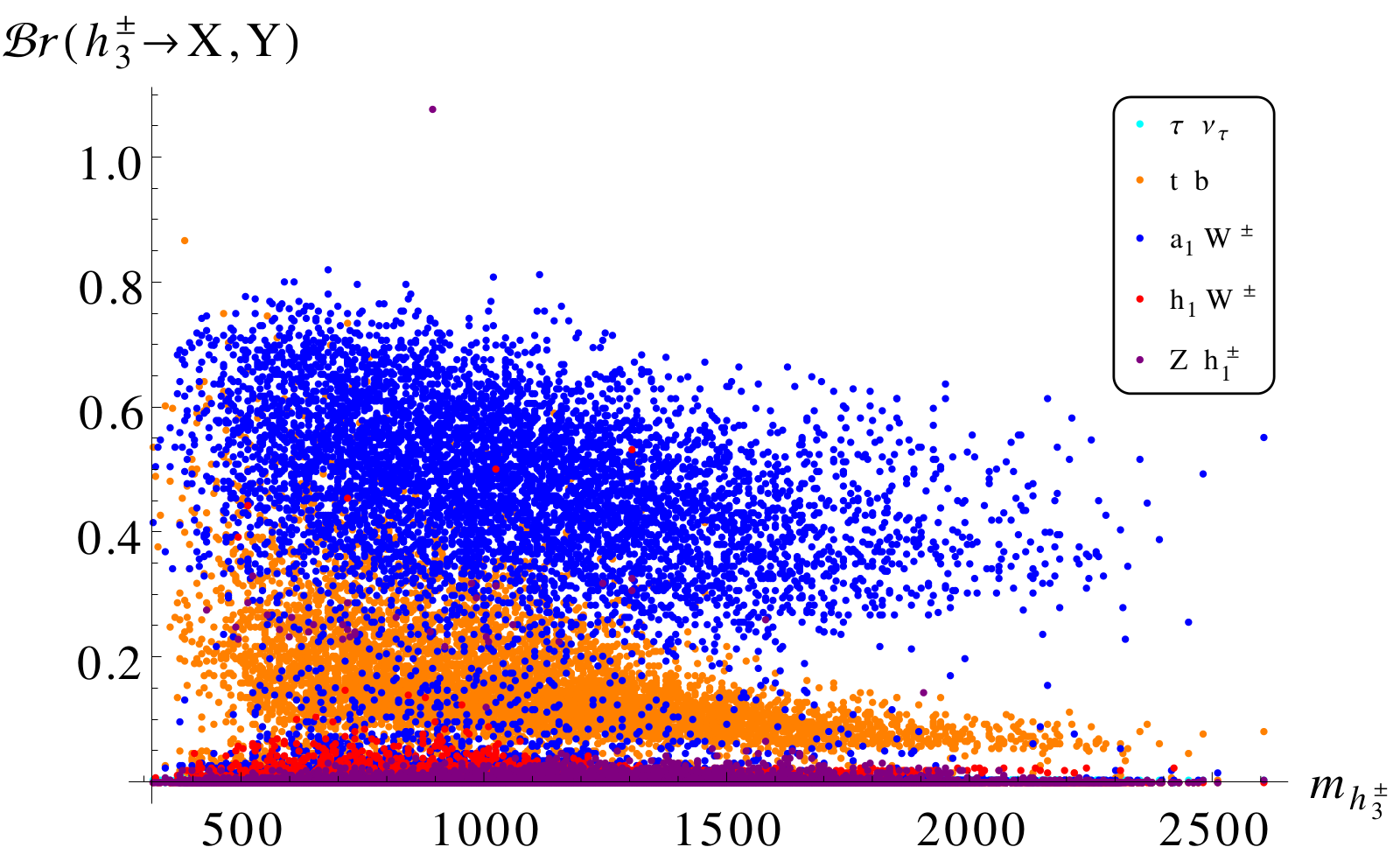}}}
\mbox{\subfigure[]{\includegraphics[width=0.44\linewidth]{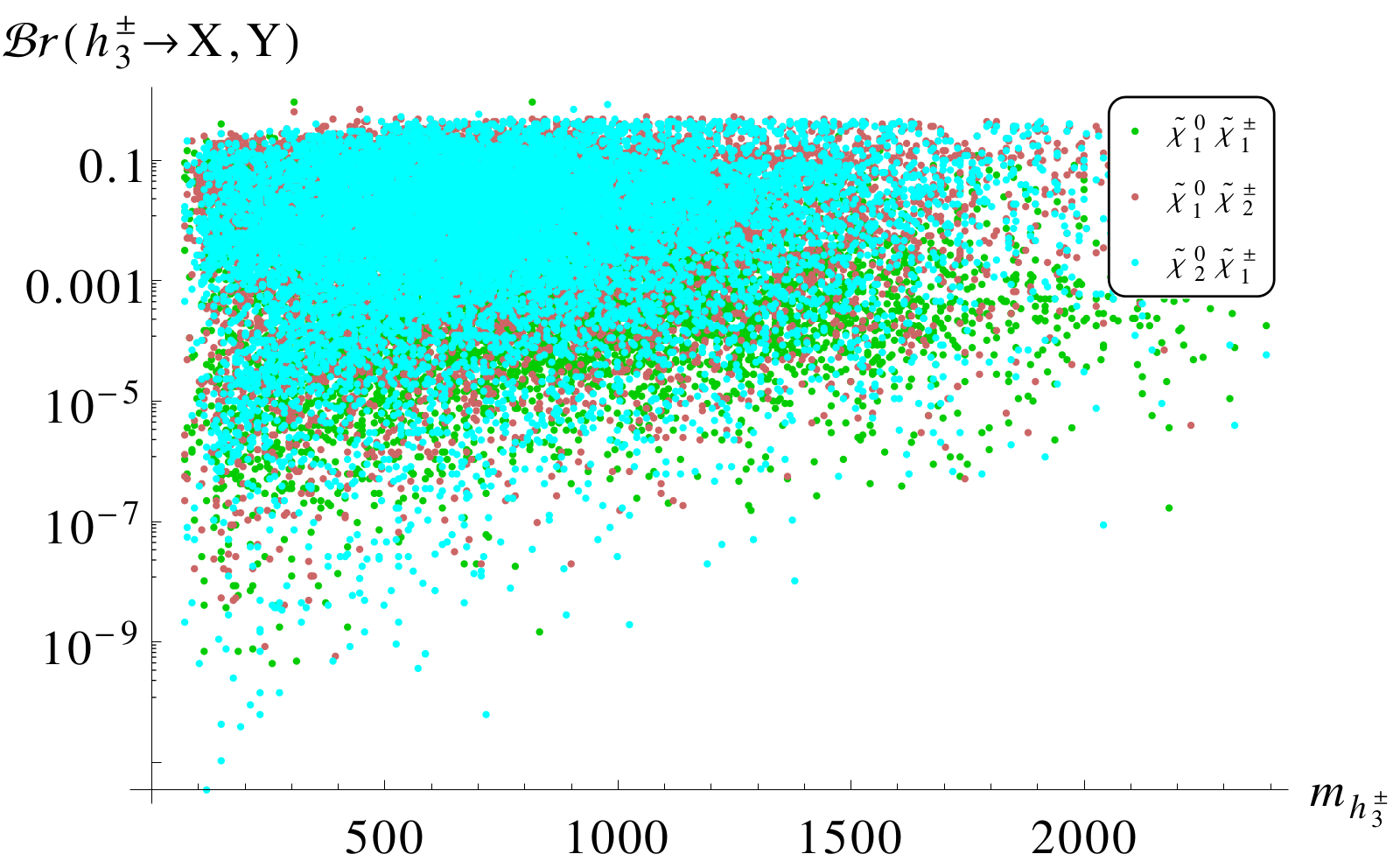}}}
\mbox{\subfigure[]{\includegraphics[width=0.44\linewidth]{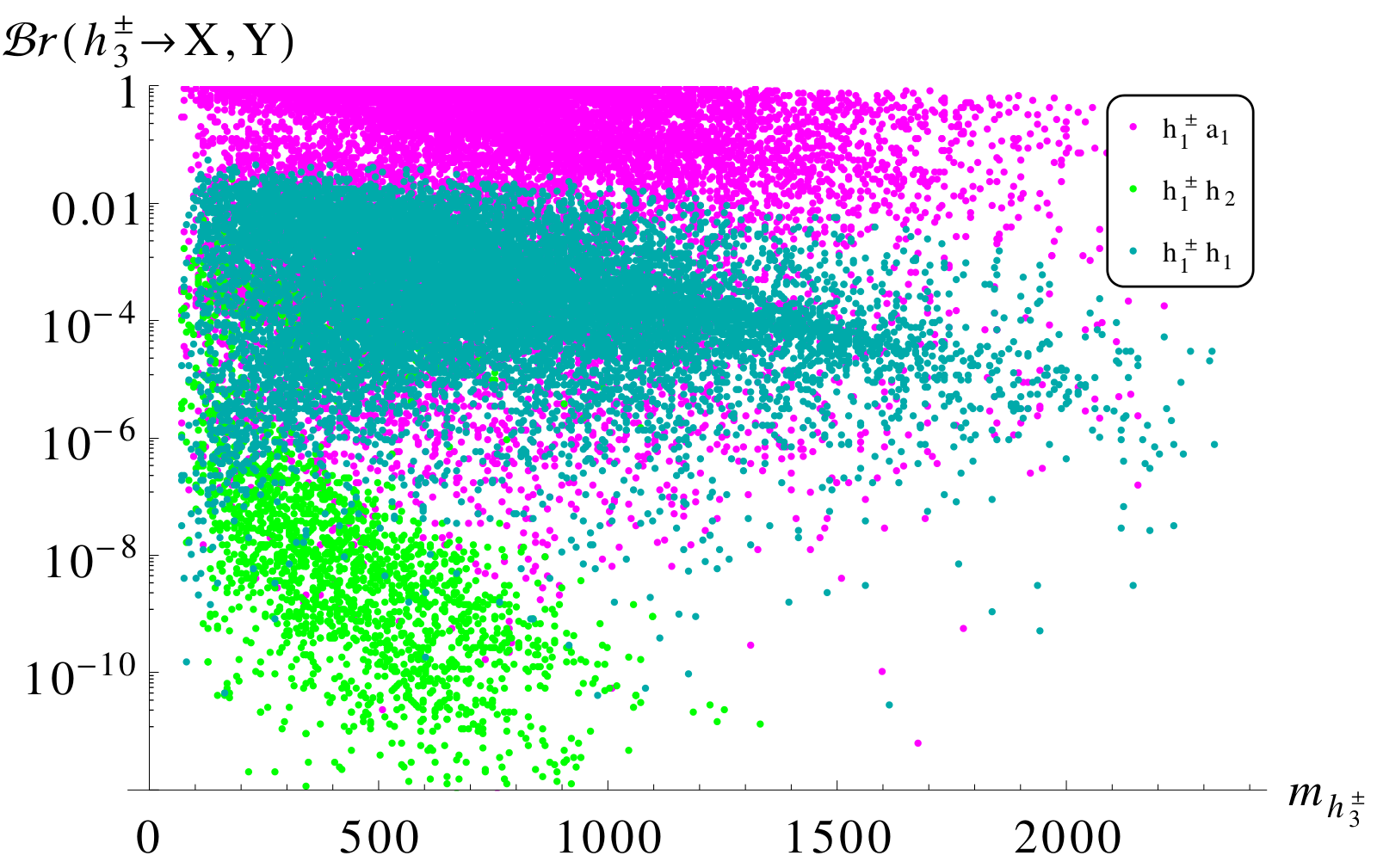}}}
\mbox{\subfigure[]{\includegraphics[width=0.44\linewidth]{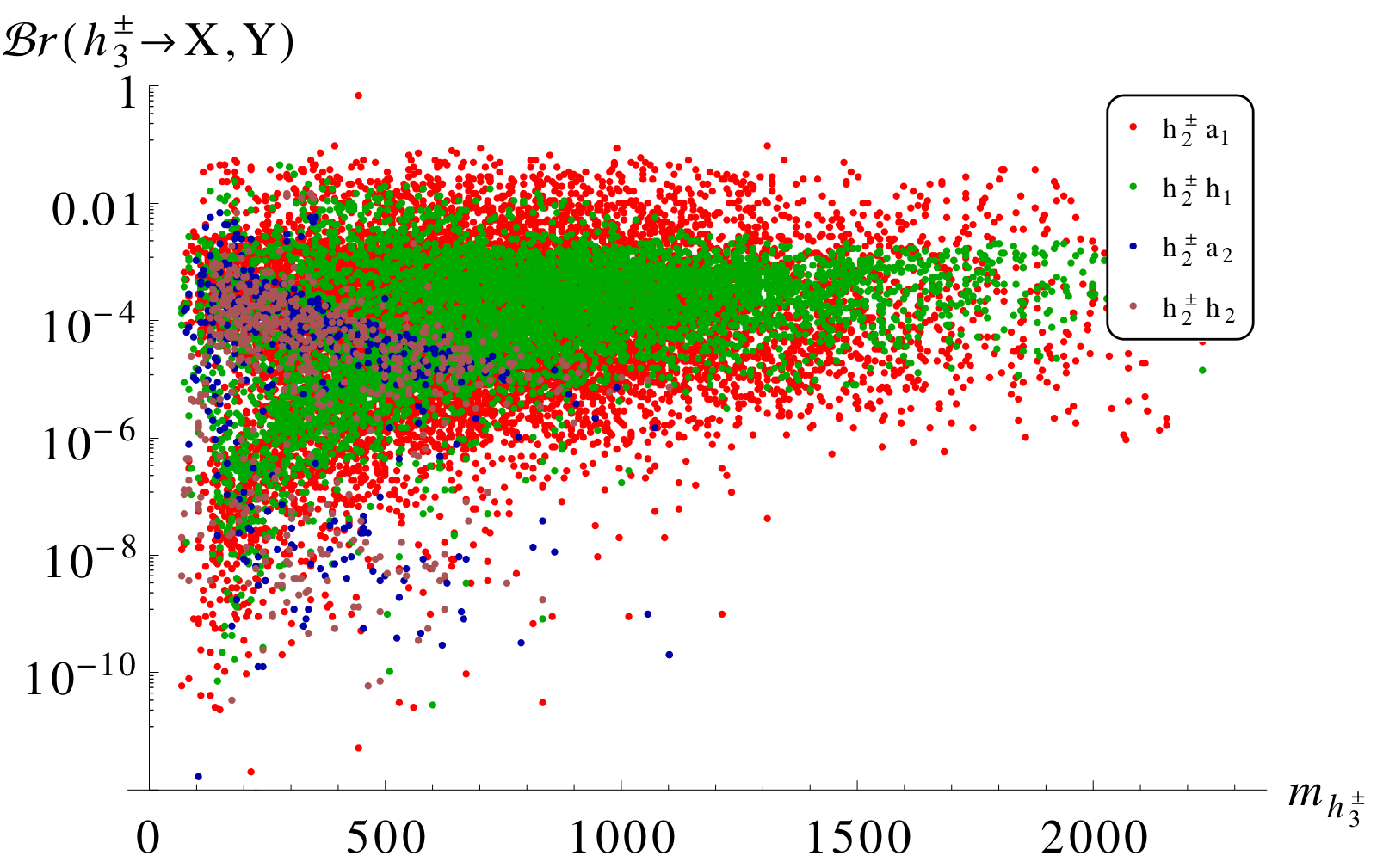}}}
\caption{The branching ratios of the decay of the charged Higgs boson $h^\pm_3$ into non-supersymmetric (a), supersymmetric modes (b), lightest charged Higgs boson $h^\pm_1$ in association with the neutral Higgs bosons (c) and second light charged Higgs boson $h^\pm_2$ in association with the neutral Higgs bosons (d).}\label{ch3br}
\end{center}
\end{figure}

Figure~\ref{ch3br} presents the third charged Higgs boson $h^\pm_3$ decays. From Figure~\ref{ch3br}(a) we can see that for a large parameter space the decay branching fraction to $a_1W^\pm$ is the most relevant mode which can be probed at the LHC. Even though $tb$ mode is kinematically open but not the most dominant one. Figure~\ref{ch3br}(b) shows that $\tilde{\chi}^0_2 \tilde{\chi}^\pm_1$ mode is kinematically open and also one of the most important. Figure~\ref{ch3br}(c) shows the decay branching ratios for the decay modes into the lightest charged Higgs boson in association with the neutral Higgs bosons. It is evident that the $h^\pm_1 a_1$ mode is the most important and one can probe more than one charged Higgs boson and also the light pseudoscalar. In Figure~\ref{ch3br}(d) the branching ratios are shown where the heaviest charged Higgs boson $h^\pm_3$ decays to second lightest charged Higgs boson $h^\pm_2$ in  association with the neutral Higgs bosons. Again the light pseudoscalar mode can have large branching ratios.
\section{Production channels of a light charged Higgs boson}\label{ch1prod}
The triplet nature of the charged Higgs bosons adds a few new production processes at the LHC
along with the doublet-like charged Higgs production process. For a doublet-like charged 
Higgs boson the production processes are dominated by the top quark decay for the light charged Higgs boson ($m_{h^\pm_i} < m_t$) or $b g \to t h^\pm_i$ for  ($m_{h^\pm_i} > m_t$) which are governed by the 
corresponding Yukawa coupling and $\tan{\beta}$ viz, in 2HDM, MSSM and NMSSM. In TNMSSM however the charged Higgs bosons can be triplet-like,  and hence they do not couple to fermions. Fermionic channels, including top and bottom and, in general, all the fermions, are then suppressed. The presence of the $h_i^\pm-W^\mp-Z$ vertex generates new production channels and also modifies the known processes for the production of a charged Higgs  boson $h^\pm_i$. In these sections we address the dominant and characteristically different production mechanisms for the light charged Higgs  bosons $h^\pm_1$ at the LHC. For this purpose we select in the parameter space the benchmark points with a discovered Higgs boson around $125$ GeV and with the lightest charged Higgs boson $h^\pm_1$ that is triplet-like ($\geq 90\%$). The cross-sections are calculated at the LHC with a center of mass energy of 14 TeV for such events. We have performed our analysis at leading order, using $\mathtt{CalcHEP\_3.6.25}$ \cite{calchep}, using the CTEQ6L \cite{6teq6l} set of parton distributions and  a renormalization/factorization scale $Q=\sqrt{\hat{s}}$   where $\hat{s}$ denotes the total center of mass energy squared at parton level.

\subsection{Associated $W^\pm$}
The dominant channels are shown in Figure~\ref{prodchW}, which are mediated by
the neutral Higgs bosons, the $Z$ boson and the quarks. Figure~\ref{prodchW}(b) which describe the $Z$ mediation requires the non-zero $h_1^\pm-W^\mp-Z$ vertex which is absent in theories without the $Y=0,\pm2$ triplet-extended Higgs sector. For a  doublet-like charged Higgs, the only contributions comes from the neutral Higgs-mediated diagrams in the s-channel and $t$-quark mediated diagram in the t-channel  (see Figure~\ref{prodchW}(a), (c)). For low $\tan{\beta}$ case, the t-channel contribution in $b\bar{b}$ fusion is really large due to large Yukawa coupling. We will see that this admixture of doublet still affects the production cross-section for low $\tan{\beta}$. 
\begin{figure}[thb]
\begin{center}
\mbox{\subfigure[]{
\includegraphics[width=0.35\linewidth]{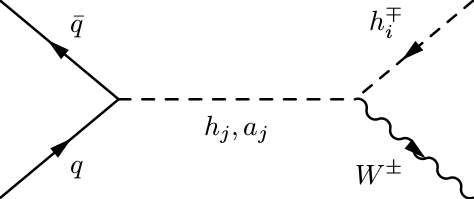}}
\hskip 15pt
\subfigure[]{\includegraphics[width=0.35\linewidth]{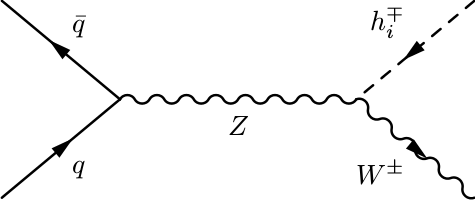}}}
\mbox{\subfigure[]{\includegraphics[width=0.3\linewidth]{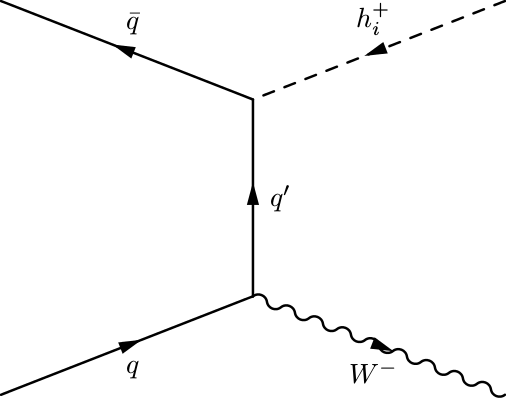}}
}
\caption{Figures (a-c) describe the production of charged Higgs boson in association with $W^\pm$ boson via $h_j/a_j$, $Z$ and q' exchange respectively.}
\end{center}
\end{figure}
\begin{figure}[thb]
\begin{center}
\includegraphics[width=0.7\linewidth]{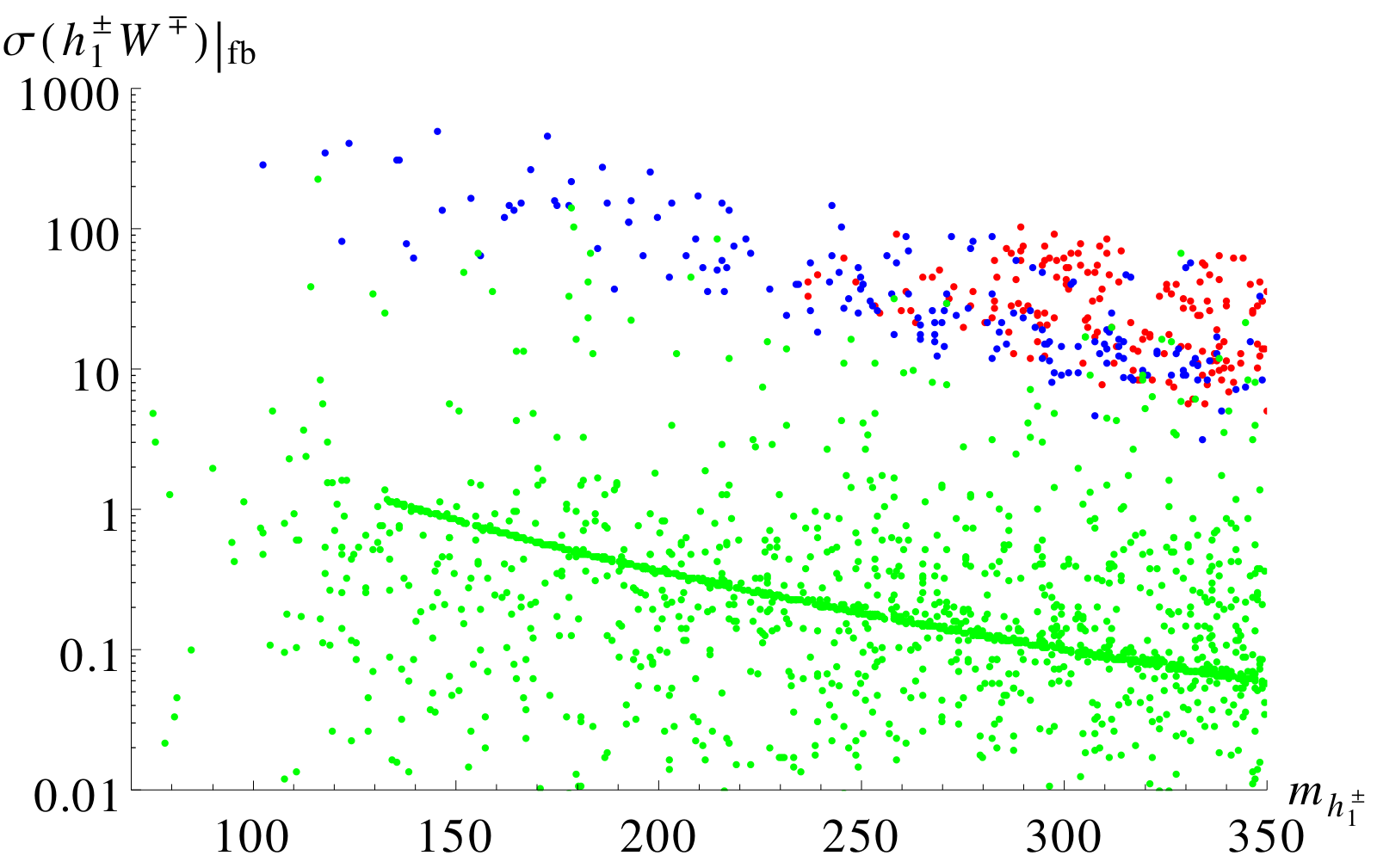}
\caption{The production cross-section of $h^\pm_1W^\mp$ at the LHC versus the lightest charged Higgs boson mass $m_{h^\pm_1}$. The red coloured ones are $\geq 90\%$ doublet-like, green ones are  $\geq 90\%$ triplet-like  and blue ones are mixed type light charged Higgs bosons.}\label{ch1Wcs}
\end{center}
\end{figure}

The contribution of $h_1$ is subdominant because $h_1$ and $h_1^\pm$ are selected to be mostly doublet and triplet respectively, in order to satisfies the LHC data. The coupling of a totally triplet charged Higgs boson with a totally doublet neutral Higgs boson and a $W$ boson is not allowed by gauge invariance. For the lightest triplet-like charged Higgs boson,  one of the degenerate neutral Higgs boson, either $h_2$ or $a_2$, is also triplet-like, and fails to contribute as mediator in $b\bar{b}$ fusion mode (Figure~\ref{prodchW}(a)). The other relevant neutral Higgs boson which is not degenerate with  the lightest charged Higgs boson $h^\pm_1$ contributes to $b\bar{b}$ fusion production process via its doublet mixings. Thus doublet-triplet mixing part plays an important role even when we are trying to produce a light charged Higgs boson which is triplet-like.  This feature also has been observed in Triplet Extended Supersymmetric Standard Model (TESSM) \cite{pbas3}. Even the off-shell doublet type neutral Higgs mediation ($h_{125}$) in s-channel via gluon-gluon fusion fails to give sufficient contribution to $h^\pm_1 W^\mp$ final state. We checked such process at the LHC for the center of mass energy of 14 TeV and a triplet-like charged Higgs of mass $\sim 300$ GeV and $h^\pm_1 W^\mp$ cross-section is below $\mathcal{O}(10^{-3})$ fb. 
 
In Figure~\ref{ch1Wcs} we present the associated production cross-section for a light charged Higgs boson $h^\pm_1$ together with the light charged Higgs boson mass $m_{h^\pm_1}$. 
The red coloured ones are $\geq 90\%$ doublet-like, green ones are  $\geq 90\%$ triplet-like  and blue ones are mixed type light charged Higgs bosons. It can be seen that as the doublet 
the fraction grows, the production cross-section also grows. At $\lambda_T\simeq0$ the lightest
charged Higgs cannot be completely triplet-like, due to the doublet fraction $\frac{v_T}{v}$.
In this limit the cross section follows the line given by the green points in Figure~\ref{ch1Wcs}. As we have seen in the previous section, for $\lambda_T\neq 0$ the coupling $g_{h_1^\pm W^\mp Z}$ is very small even if the lightest charged Higgs is completely triplet-like. This means that the $Z$ propagator (cfr. Figure~\ref{prodchW}(b)) does not give contribution. However, since for $\lambda_T\neq 0$ the triplet fraction of $h_1^\pm$ is not fixed, the cross-section can be enhanced or decreased compared to the $|\lambda_T|\simeq0$ one.

\subsection{Associated $Z$ }
\begin{figure}[thb]
\begin{center}
\mbox{\subfigure[]{
\includegraphics[width=0.35\linewidth]{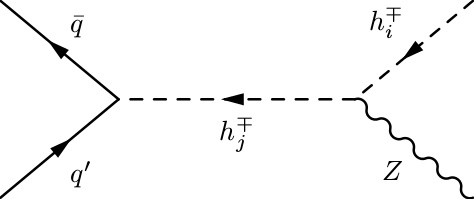}}\hskip 15pt
\subfigure[]{\includegraphics[width=0.35\linewidth]{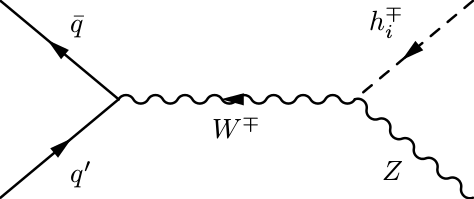}}}
\caption{Figures (a,b) describe the production of the charged Higgs boson in association with $Z$ boson via $h_j^\pm$ and $W^\pm$ boson exchange.}\label{prodch1Z}
\end{center}
\end{figure}
Unlike the previous case, the charged Higgs production in association with $Z$ does not have sizeable contributions from the doublet part of the Higgs boson spectrum. For instance, the doublet nature of the charged Higgs allows its exchange in the s-channel, as shown in  Figure~\ref{prodch1Z}(a), via an annihilation process ($q \bar{q}')$ which requires quarks of different flavours. The contributions from the valence $u/\bar{d}, \bar{u}/d$ distributions, in a $pp$ collision are strongly suppressed by the much lower Yukawa couplings. On the other hand contributions from heavier generations such as $c/\bar{b},\bar{c} /b$ are suppressed by CKM mixing angles and the involvement of sea quarks in the initial state.

\begin{figure}[thb]
\begin{center}
\includegraphics[width=0.7\linewidth]{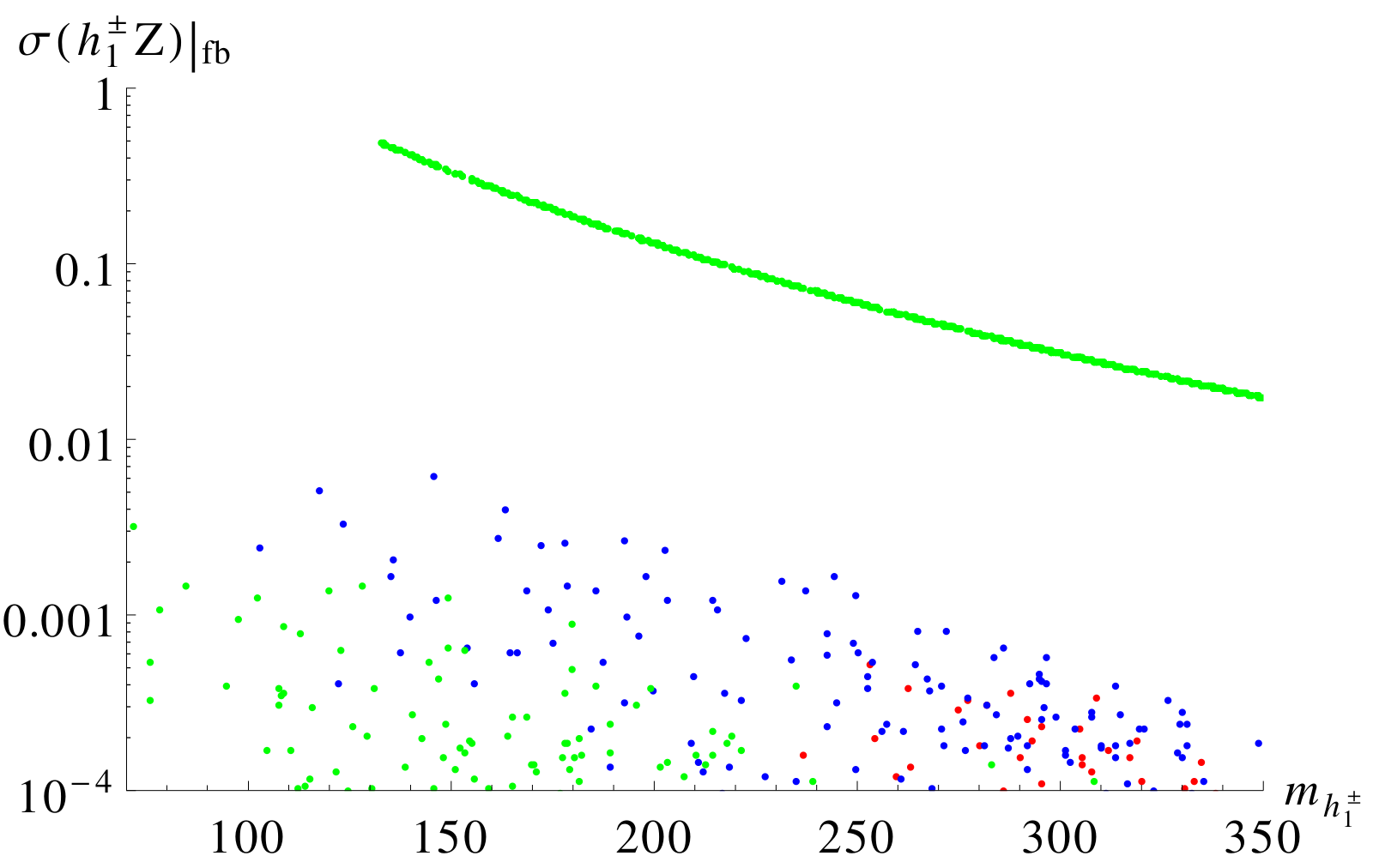}
\caption{The production cross-section of light charged Higgs boson $h^\pm_1$ in association with
$Z$ boson versus the light charged Higgs boson mass $m_{h^\pm_1}$.}\label{ch1Zcs}
\end{center}
\end{figure}

Nevertheless, in the case of the TNMSSM, a non-zero $h_1^\pm-W^\mp-Z$ vertex gives an extra contribution to this production process, which is absent in the case of doublet-like charged Higgs bosons. In fact, for $\lambda_T \simeq 0$, which corresponds to what we have called decoupling limit,  the $T^+_1$ and $T^-_2$ interaction eigenstates contribute additively to the $h_1^\pm-W^\mp-Z$, as can be seen from Eq.~\ref{zwch} and also can be realised from Figure~\ref{ssoslmbda} and Figure~\ref{ghmp1WZ}. However we can see from   Figure~\ref{ch1Zcs} that the $h^\pm_1Z$ production cross-section is smaller than the respective production in association with a $W^\pm$. This is  due to the fact that there are no other efficient contributions beside the channel with the $W^\pm$ in the propagator, as discussed earlier.

\subsection{Associated $h_1$}
\begin{figure}[thb]
\begin{center}
\mbox{\subfigure[]{
\includegraphics[width=0.35\linewidth]{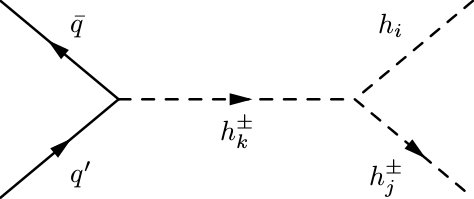}}\hskip 15pt
\subfigure[]{\includegraphics[width=0.35\linewidth]{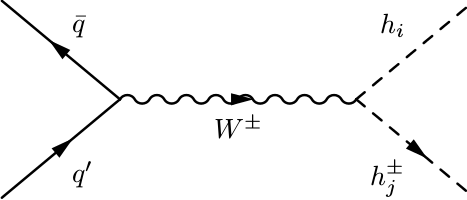}}}
\caption{Figures (a,b) describes the production of charged Higgs boson in association with $h_i$ boson via $h_k^\pm$ and $W^\pm$ boson exchange.}\label{prodchhi}
\end{center}
\end{figure}

We have considered, than, the production of the charged Higgs boson production in association with a scalar Higgs boson, $h_i$. It is clear from Figure~\ref{prodchhi} that there are two contributions to this channel, one via the doublet-type
charged Higgs boson and another mediated by the $W^\pm$ boson.  However the charged Higgs mediated diagrams are suppressed, for the same reasons discussed earlier in the associated $Z$ production. Both the triplet and doublet Higgs bosons couple to $SU(2)$ gauge boson $W^\pm$. However a careful look on the vertex, given in Eq.~\ref{hachW}, shows that their mixing angles can have relative signs. In general their coupling in association with neutral Higgs bosons have to be doublet(triplet) type for doublet(triplet) type charged Higgs bosons. 
 
\begin{figure}[thb]
\begin{center}
\includegraphics[width=0.7\linewidth]{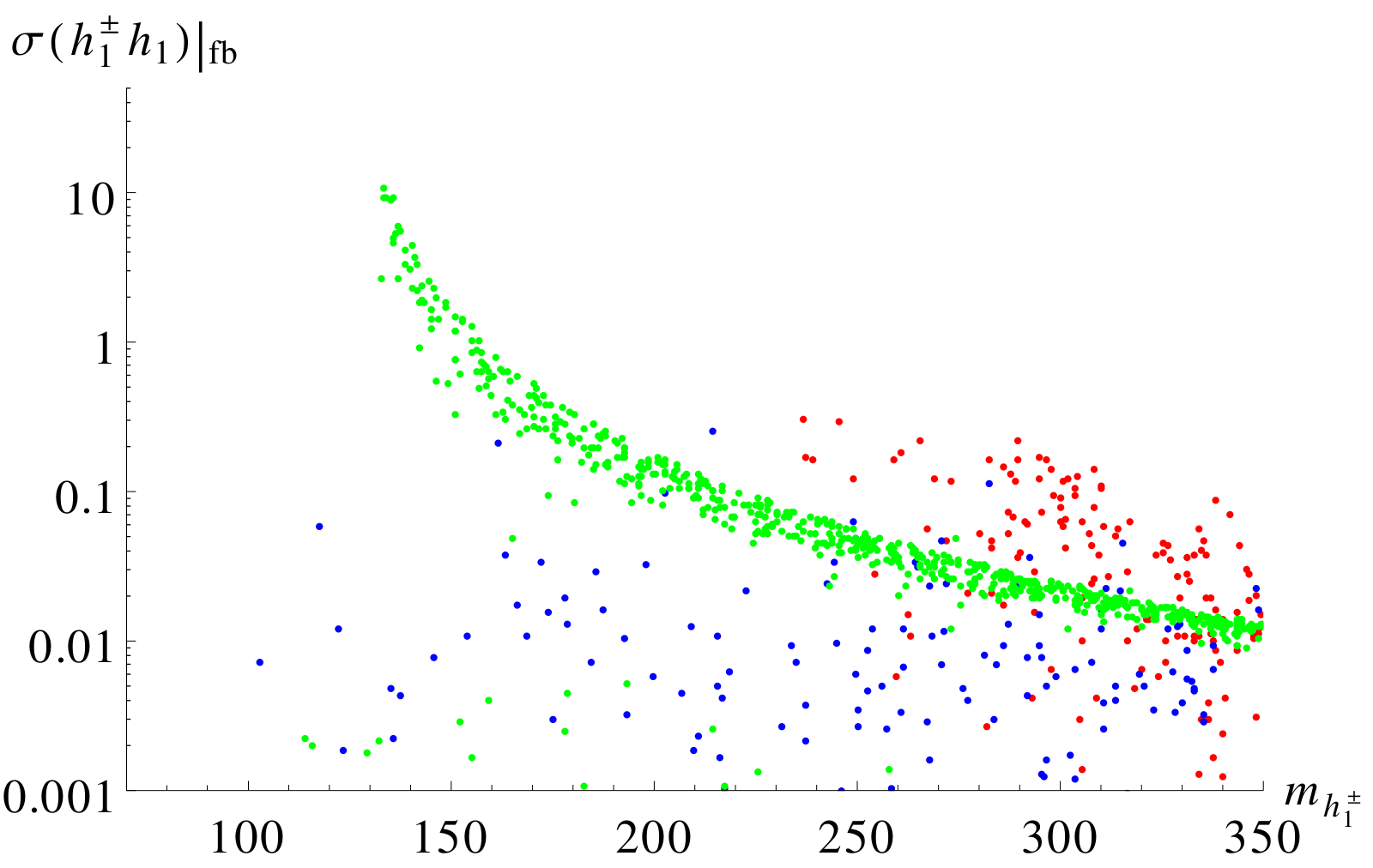}
\caption{The production cross-section of a light charged Higgs boson $h^\pm_1$ in association with
the $h_1$ boson versus the light charged Higgs boson mass $m_{h^\pm_1}$.}\label{ch1h1cs}
\end{center}
\end{figure}

This behaviour can be seen from Figure~\ref{ch1h1cs}, where we plot the production cross-section versus the mass of the lightest charged Higgs boson, $m_{h^\pm_1}$. The colour code for the charged Higgs boson remains as before. It is quite evident that, for a triplet-like charged Higgs boson, the cross-sections in association with $h_1$, which is mostly doublet, are very small, except for the $\lambda_T\simeq0$ points.  We can see the enhanced cross-section for the mostly doublet charged Higgs boson in association with doublet-like $h_1$ (red points). The situation is different for $\lambda_T\simeq0$, where 
 it is easy to produce a mostly triplet charged Higgs boson in this channel due to the enhancement of the $h^\pm_1-W^\mp-h_1$ coupling, given in Eq.~\ref{hachW}. This is due to the fact that  for  $\lambda_T\simeq0$ the rotation angles $\mathcal{R}^C_{12}$ and $\mathcal{R}^C_{14}$ of the triplet sector, which appear in the coupling given in Eq.~\ref{hachW}, take same sign (in the decoupling limit see Figure~\ref{ssoslmbda}).


\subsection{Associated $a_1$}
\begin{figure}[thb]
\begin{center}
\mbox{\subfigure[]{
\includegraphics[width=0.35\linewidth]{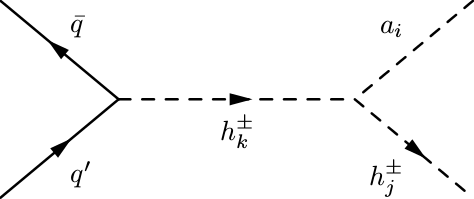}}\hskip 15pt
\subfigure[]{\includegraphics[width=0.35\linewidth]{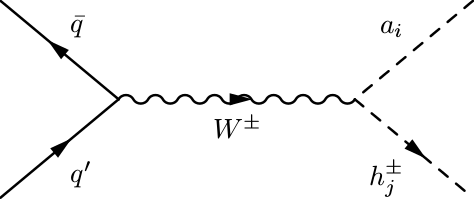}}}
\caption{Figures (a,b) describes the production of charged Higgs boson in association with $a_i$ boson via $h_k^\pm$ and $W^\pm$ boson exchange.}\label{prodchai}
\end{center}
\end{figure}
\begin{figure}[thb]
\begin{center}
\includegraphics[width=0.7\linewidth]{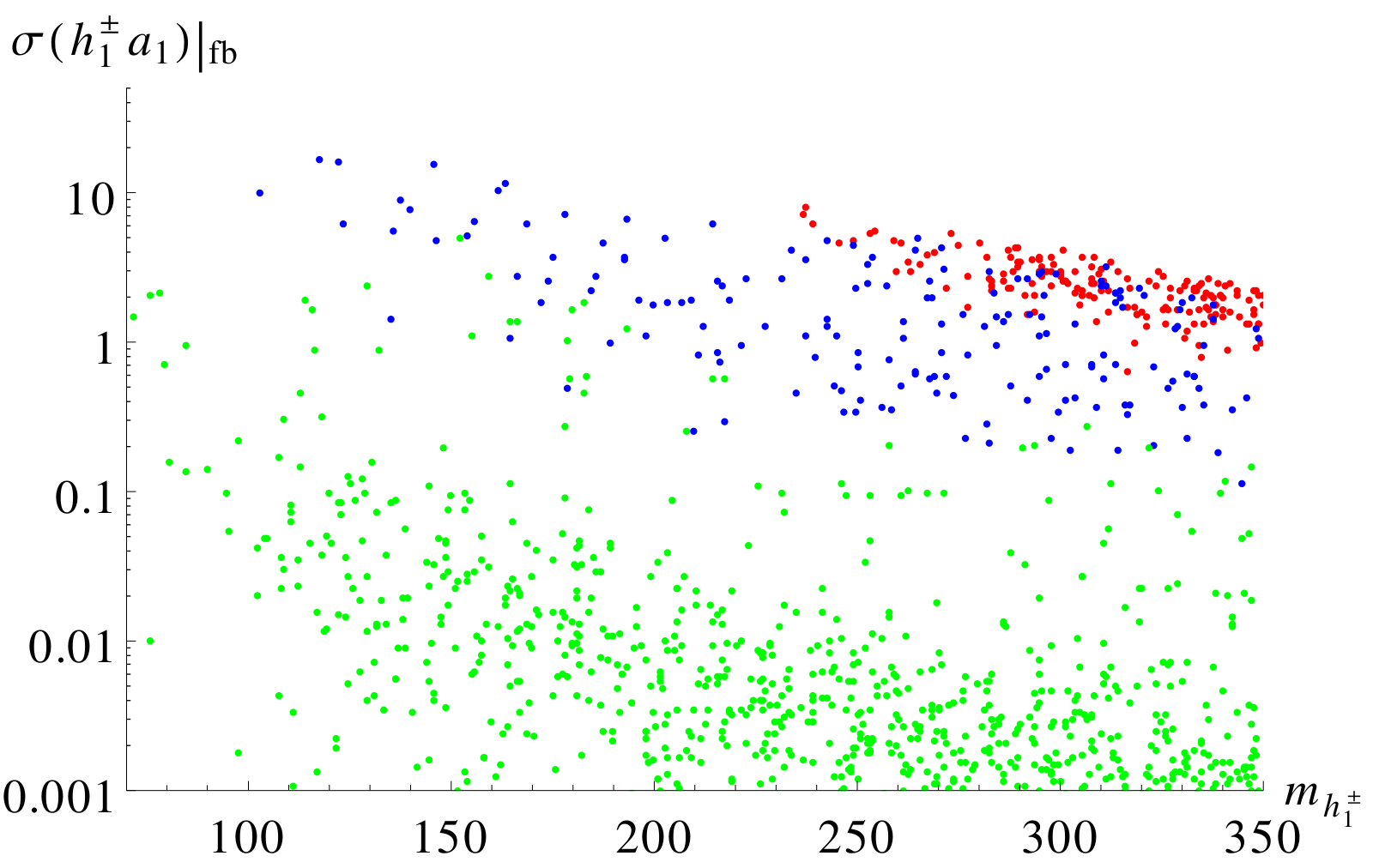}
\caption{The production cross-section of light charged Higgs boson $h^\pm_1$ in association with
the $a_1$ boson versus the light charged Higgs boson mass $m_{h^\pm_1}$.}\label{ch1a1cs}
\end{center}
\end{figure}

Similarly, we can also produce the charged Higgs boson in association with a pseudoscalar Higgs boson, as shown in Figure~\ref{prodchai}. Here we also include the two contributions coming from $h^\pm_i$ and $W^\pm$ respectively even though, as before, the contribution from the charged Higgs propagator is negligible. Figure~\ref{ch1a1cs} presents the variation of the cross-section with the mass of the lightest charged Higgs boson. The cross-section stays very low for the triplet-like points (green ones) and reaches a maximum around 10 fb for doublet- and mixed-like points (red and blue ones). For $\lambda_T\simeq0$ points, the triplets ($T^+_1, T^{-*}_2$) rotation angles $\mathcal{R}^C_{i2, i4}$ appear with a relative sign in the coupling $h^\pm_i-W^\mp-a_j$, as can be seen in Eq.~\ref{hachW}. The $h^\pm_1 a_1$ cross-section thus gets a suppression in the decoupling limit, i.e. for $|\lambda_T|\simeq 0$, unlike the $h_ih^\pm_1$ case, as discussed in the previous section.

\subsection{Charged Higgs pair production }
\begin{figure}[thb]
\begin{center}
\mbox{\subfigure[]{
\includegraphics[width=0.35\linewidth]{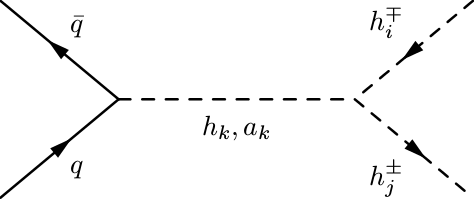}}\hskip 15pt
\subfigure[]{\includegraphics[width=0.35\linewidth]{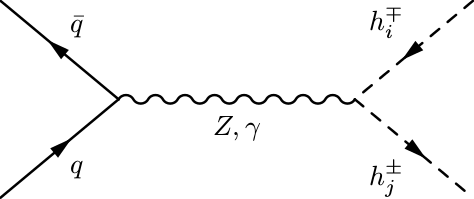}}}
\caption{Figures (a,b) describes the production of charged Higgs boson pair via $h_k/a_k$ and $Z/\gamma$ boson exchange.}\label{prodchij}
\end{center}
\end{figure}

\begin{figure}[thb]
\begin{center}
\subfigure[]{\hspace{-.5cm}
\includegraphics[width=0.7\linewidth]{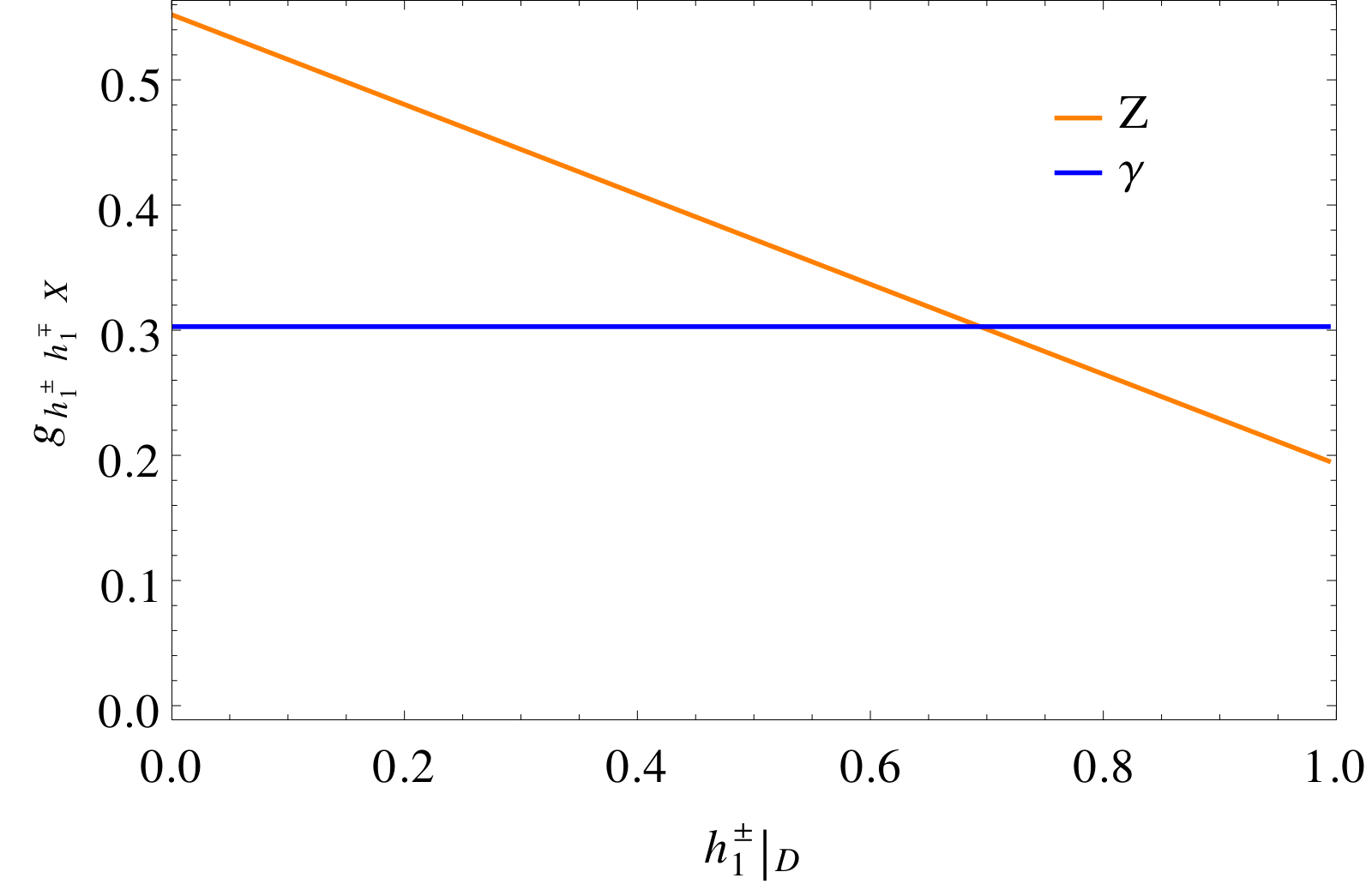}}
\caption{Value of the coupling $g_{h_1^\pm h_1^\mp X}$ as a function of the doublet fraction of the lightest charged Higgs boson. In the case of the photon this coupling is just the value of the electric charge.}\label{ZphoHpmHpm}
\end{center}
\end{figure}

\begin{figure}[thb]
\begin{center}
\includegraphics[width=0.8\linewidth]{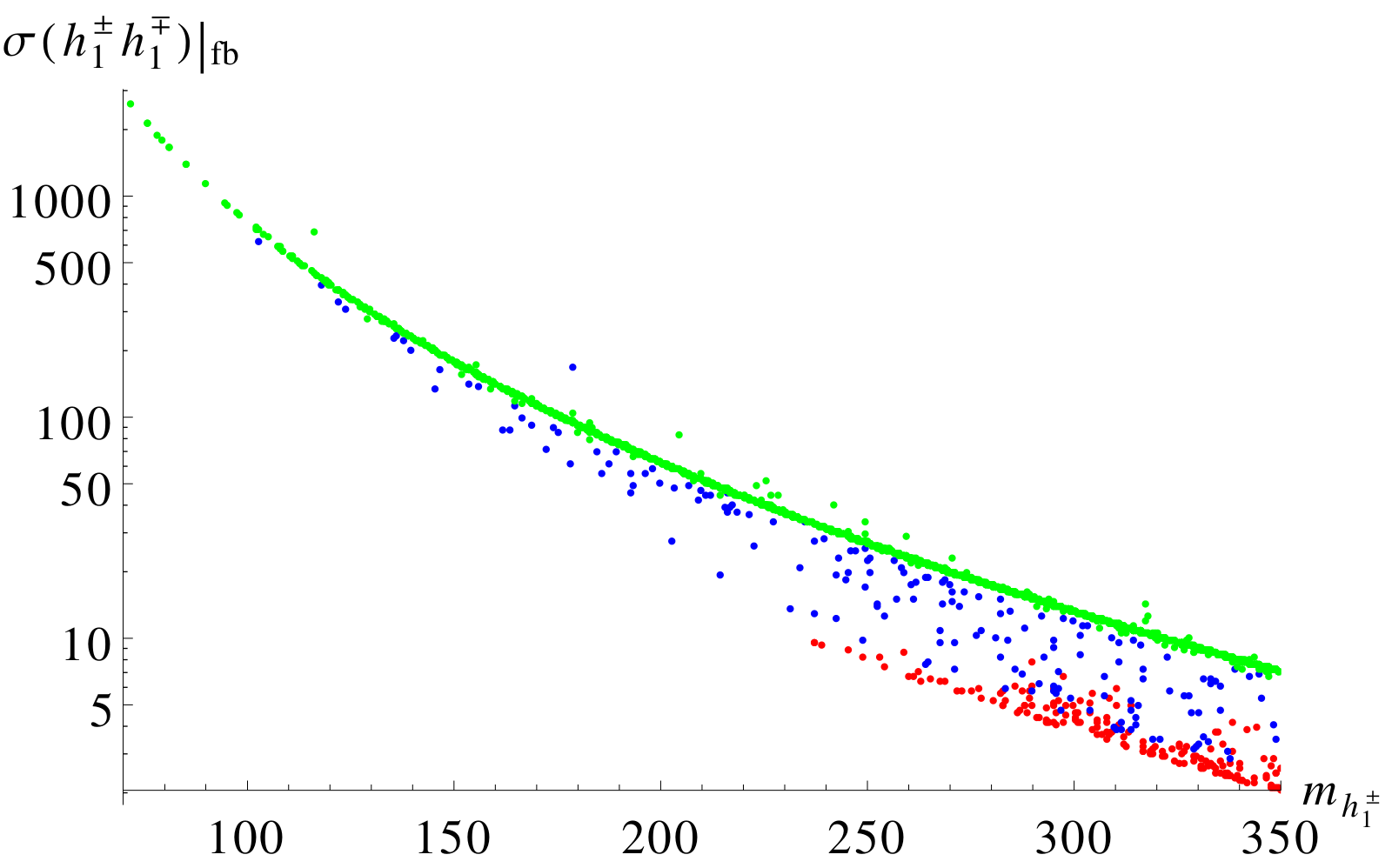}
\caption{The production cross-section of light charged Higgs boson pair  $h^\pm_1h^\mp_1$ versus the light charged Higgs boson mass $m_{h^\pm_1}$.}\label{ch1ch1cs}
\end{center}
\end{figure}

Here we move to the description of the charged Higgs pair production for the lightest charged Higgs boson $h^\pm_1$. The Feynman diagrams for this process are given in Figure~\ref{prodchij}, with the neutral Higgses and $Z,\gamma$ bosons contributing to the process. However, if the lightest charged Higgs boson $h^\pm_1$ is triplet-like, the diagrams of Figure~\ref{prodchij}(a) give less contribution to the cross section. In fact $a_1$ is selected to be singlet-like, so it does not couple to the fermoins, and the diagram with $h_{125}$ in the propagator is subdominant. The reason is that the coupling $g_{h_1^\pm h_1^\mp h_1}$ of a totally doublet scalar Higgs boson with two totally triplet charged Higgs bosons is prevented by gauge invariance. The triplet charged Higgs pair production is more suppressed  than the single triplet-like charged Higgs production via a doublet-like neutral Higgs boson. In that case pair production cross-section via off-shell doublet type neutral Higgs mediation ($h_{125}$) in s-channel via gluon-gluon fusion is below $\mathcal{O}(10^{-6})$ fb. Hence for triplet-like $h_1^\pm$ the diagrams of Figure~\ref{prodchij}(b) are the most relevant ones. The coupling of a pair of $h_1^\pm$ to the $Z$ and the $\gamma$ bosons is shown in Figure~\ref{ZphoHpmHpm} as a function of the doublet fraction. The coupling $g_{h_1^\pm h_1^\mp \gamma}$ is independent of the structure of $h_1^\pm$ as it should be because of the $U(1)_{\rm{em}}$ symmetry. In fact the value of this coupling is just the value of the electric charge. Conversely, the coupling of the $Z$ boson to a pair of charged Higgs depends on the structure of the charged Higgs. When the charged Higgs is totally doublet its coupling approaches the MSSM value $\frac{g_L}{2}\frac{\cos\,2\theta_w}{\cos\theta_w}$. If the charged Higgs is totally triplet the value of the coupling is $g_L\cos\theta_w$, the same of the $W^\pm-W^\mp-Z$ interaction.
In Figure~\ref{ch1ch1cs} we show the variation of the cross-sections with respect to the 
 lightest charged Higgs boson mass $m_{h^\pm_1}$. The colour code of the points are as the previous ones. We can see that for triplet-like points with mass around $\sim 100$ GeV the cross-section reach around a picobarn. This large cross-section makes this production a viable channel to be probed at the LHC for the light triplet type charged Higgs boson. We discuss the corresponding phenomenology in section~\ref{pheno}.

\subsection{Vector boson fusion}
Neutral Higgs boson production via vector boson fusion is second most dominant 
production mode in SM. Even in 2HDM or MSSM this production mode of the neutral Higgs boson is one of the leading ones. However no such channel exist for charged Higgs boson as $h^\pm_i-W^\mp-Z$ vertex is zero at the tree-level, as long as custodial symmetry 
is preserved. The introduction of a $Y=0$ triplet breaks the custodial symmetry at tree-level, giving a non-zero $h^\pm_i-W^\mp-Z$ vertex, as shown in Eq.~\ref{zwch}. This vertex gives rise to the striking production channel of the vector boson fusion into a single charged Higgs boson, which is absent in the MSSM and in the 2-Higgs-doublet model (2HDM) at tree-level. This is a signature of the triplets with $Y=0, \pm 2$ which break custodial symmetry at the tree-level.
\begin{figure}[t]
\begin{center}
\mbox{\subfigure[]{
\includegraphics[width=0.35\linewidth]{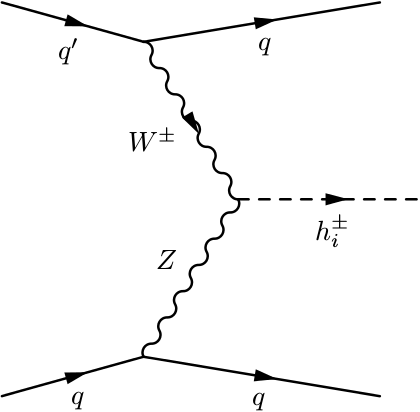}}}
\caption{The Feynman diagram for the charged Higgs production via vector boson fusion at the LHC.}\label{prodvvfch}
\end{center}
\end{figure}

\begin{figure}[thb]
\begin{center}
\includegraphics[width=0.7\linewidth]{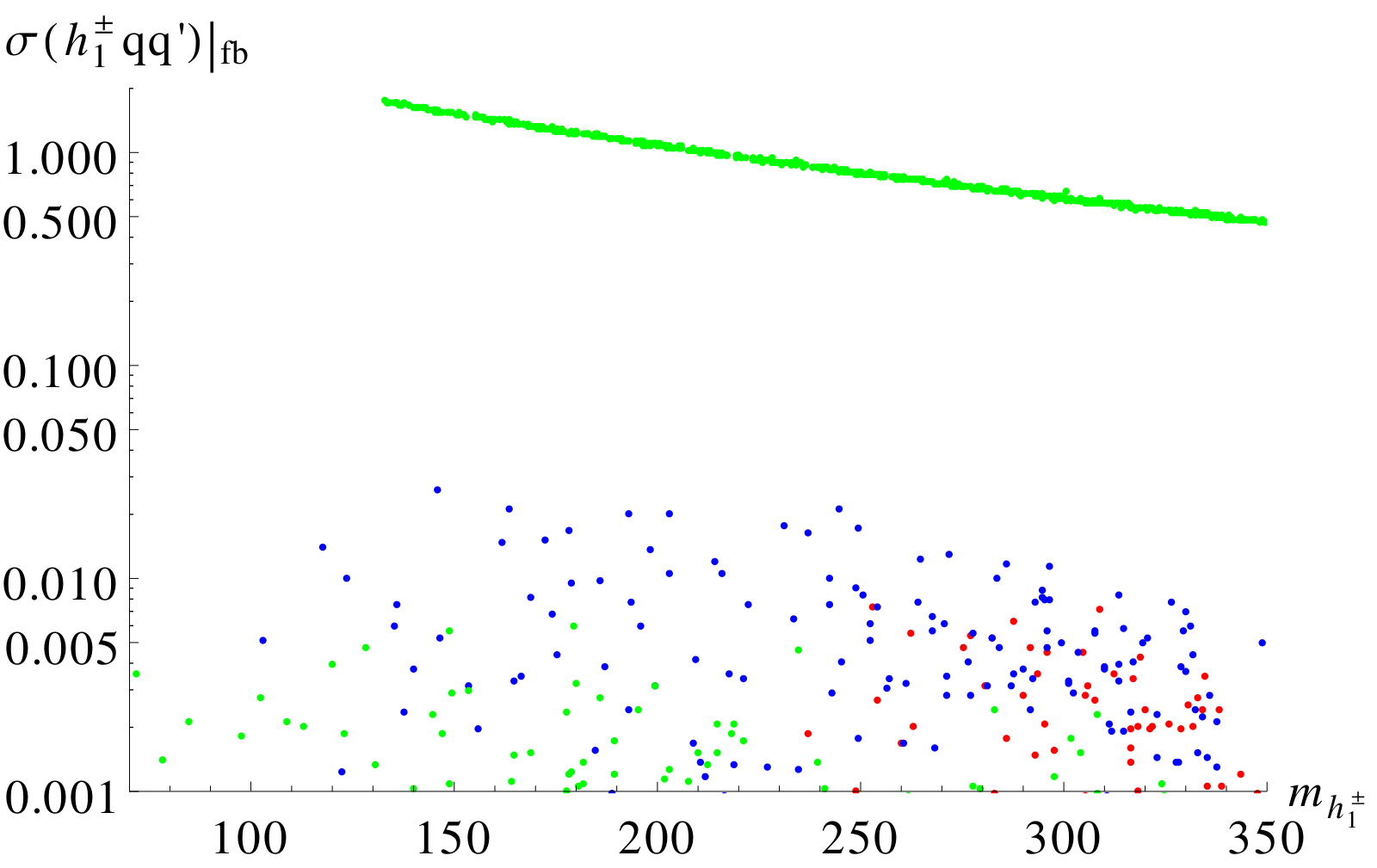}
\caption{The production cross-section of a light charged Higgs boson via  vector boson fusion versus the light charged Higgs boson mass $m_{h^\pm_1}$.}\label{VBFcs}
\end{center}
\end{figure}
Figure~\ref{VBFcs} shows the cross-section variation with respect to the lightest charged Higgs boson mass $m_{h^\pm_1}$. As expected, doublet-like points (in red) have very small cross-sections, and for the mixed points (in blue) we see a little enhancement. Green points describe the cross-sections for the triplet-like points. We see that a triplet-like charged Higgs boson does not necessarily guarantee large values for the cross-section. As one can notice from Eq.~\ref{zwch}, the coupling $g_{h_1^\pm W^\mp Z}$ is a function of $\mathcal{R}^{C}_{12}$ and $\mathcal{R}^{C}_{14}$ and their relative sign plays an important role. From Figure~\ref{ghmp1WZ} we see that only in the decoupling limit, where where $\lambda_T=0$, both $\mathcal{R}^{C}_{12}$ and $\mathcal{R}^{C}_{14}$ take the same sign, thereby enhancing the $h_1^\pm- W^\mp -Z$ coupling and thus the cross-section.  It can been seen that only for lighter masses $\sim 150-200$ GeV the cross-sections is around few femtobarns. Such triplet-like charged Higgs bosons can be probed at the LHC as a single charged Higgs production channel without  the top quark. This channel thus can be used to distinguish from other known single charged Higgs production mode in association with the top quark, which characterises  a doublet-like charged Higgs bosons.

\subsection{Associated top quark}
\begin{figure}[thb]
\begin{center}
\mbox{\subfigure[]{
\includegraphics[width=0.35\linewidth]{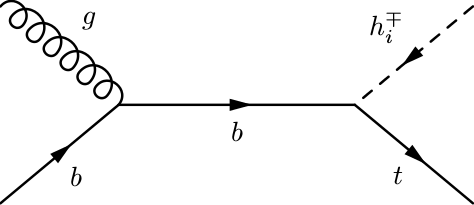}}\hskip 20pt
\subfigure[]{\includegraphics[width=0.3\linewidth]{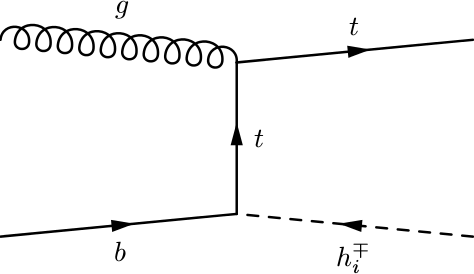}}}
\caption{Figures (a,b) describes the production of charged Higgs boson in association with a top quark via $b$ and $t$ exchange.}\label{prodtch}
\end{center}
\end{figure}
In the TNMSSM the triplet sector does not couple to fermions, which causes a natural suppression of the production of a triplet-like charged Higgs in association with a top quark. The only way for this channel to be allowed is via the mixing with doublets. Figure~\ref{prodtch} shows the Feynman diagrams of such production processes, which are dominant and take place via a $b$ quark and gluon fusion. They are highly dependent on the value of $\tan{\beta}$ \cite{djuadi, moretti}.
\begin{figure}[thb]
\begin{center}
\includegraphics[width=0.7\linewidth]{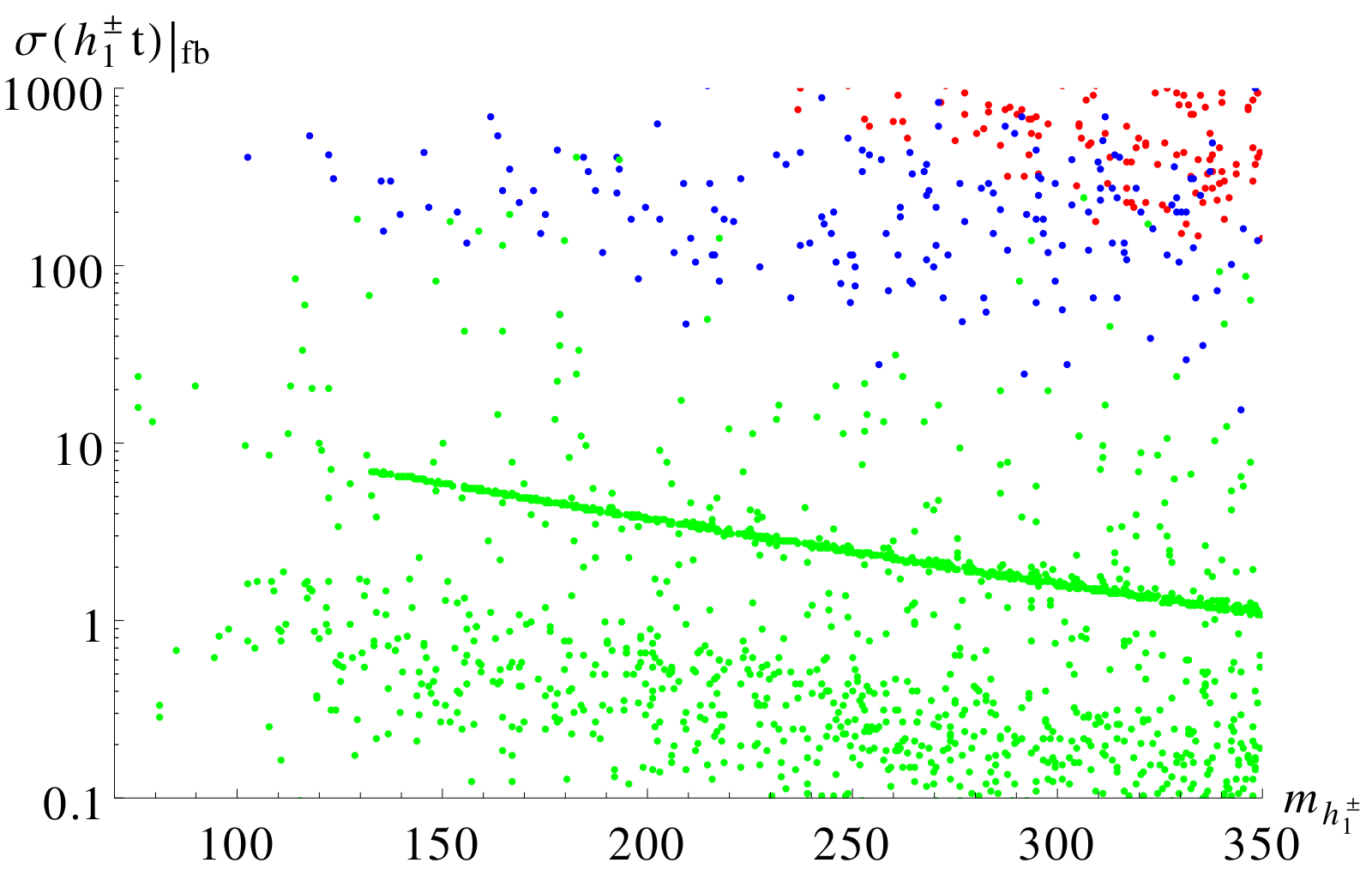}
\caption{The production cross-section of light charged Higgs boson in association with top quark versus the light charged Higgs boson mass $m_{h^\pm_1}$.}\label{tchcs}
\end{center}
\end{figure}
Figure~\ref{tchcs}(b) shows the production cross-section as a function of the lightest charged Higgs boson mass, where the green points correspond to linear combinations which are mostly triplet ($\gsim 90\%$), while red points correspond to those which are mostly of doublet ($\gsim 90\%$) and the blue points are of mixed type. Triplet-like points have a naturally suppressed cross-section whereas the doublet-like points have a large cross-sections, that can be $\sim$ pb. The mixed points lay in between, with cross-sections $\mathcal{O}(100)$ fb. One can also notice the certain enhanced line in the green points which correspond to  $|\lambda_T| \simeq 0$. As already explained in the previous sections, in this limit some portion ($\sim(\frac{v_T}{v})^2$) of the lightest charged Higgs boson $h^\pm_1$ remains doublet type, as shown in Figure~\ref{chpslmbda}, and is responsible for the enhancement of the cross-section.

Thus not finding a charged Higgs boson in this channel does not mean that it is completely ruled out, simply it can come from higher representation of $SU(2)$.

\section{Charged Higgs boson phenomenology}\label{pheno}
As was pointed out before, the TNMSSM with a $Z_3$ symmetry allows for a very light singlet-like pseudoscalar in its spectrum, which turns into a pseudo-NG mode in the limit of small soft parameters $A_i$ \cite{TNMSSM1}. The existence of such a light and still hidden scalar prompts the decay of a light charged Higgs boson $h^\pm_1 \to a_1 W^\pm$. Of course the gauge invariant structure of the vertex further restricts such decay mode, which is only allowed by the mass mixing of the singlet with the doublets or the triplet.  In the extended supersymmetric scenarios with only triplet, one cannot naturally obtain such light triplet-like pseudoscalar, because imposing $Z_3$ symmetry would be impossible due to existence of $\mu$ term, which is necessary to satisfy the lightest chargino mass bound \cite{pbas3}. The existence of a light pseudoscalar mode has been observed and studied in the context of the NMSSM \cite{han, colepa, guchait, pbsnkh}. Unlike NMSSM, in TNMSSM with a $Z_3$ symmetry the decay $h^\pm_1 \to ZW^\pm$ is possible for a triplet-type light charged Higgs boson. Below we discuss the phenomenology of such charged Higgs bosons at the LHC.
 
 {For this phenomenological analysis we have selected three benchmark point, named BP1, BP2 and BP3 given in Table~\ref{BP}.
\begin{table}
\begin{center}
\renewcommand{\arraystretch}{1.4}
\begin{tabular}{||c||c|c|c|c|c||}
\hline
\hline
&$m_{h_1^\pm}$&$m_{a_1}$&$\mathcal{Br}(a_1W^\pm)$&$\mathcal{Br}(Z\,W^\pm)$&$\mathcal{Br}(\tau\nu_\tau)$\\
\hline
BP1&179.69&41.22&$9.7\times10^{-1}$&$2.1\times10^{-2}$&$1.3\times10^{-4}$\\
\hline
BP2&112.75&29.77&$9.9\times10^{-1}$&$6.3\times10^{-5}$&$5.5\times10^{-3}$\\
\hline
BP3&172.55&48.94&$6.3\times10^{-5}$&$9.8\times10^{-1}$&$2.4\times10^{-3}$\\
\hline
\hline
\end{tabular}
\caption{The mass of $h_1^\pm$, the mass of $a_1$ and the relevant branching ratios for the three benchmark points used in the phenomenological analysis.}\label{BP}
\end{center}
\end{table}
All of them are characterised by a triplet-like charged Higgs boson $h_1^\pm$, which make the charged Higgs branching fractions into fermions, e.g. $\mathcal{Br}(h_1^\pm\to\tau\nu_\tau)$ or $\mathcal{Br}(h_1^\pm\to t\,b)$, strongly suppressed. We choose this scenario of triplet-like charged Higgs boson to look for new physics signals that is not there in two Higgs doublet model (2HDM), MSSM and NMSSM. The benchmark points maximize following decay modes;
 \begin{itemize}
   \item BP1:  \\
   $\sigma_{pp\to h_1^\pm h_1^\mp} \times \mathcal{Br}(h_1^\pm \to a_1W^\pm)\mathcal{Br}(h_1^\mp \to  Z\,W^\mp)$ ,
   \item BP2:\\
    $\sigma_{pp\to h_1^\pm h_1^\mp} \times \mathcal{Br}(h_1^\pm \to a_1W^\pm)\mathcal{Br}(h_1^\mp \to   a_1W^\pm)$ 
    
    \item BP3:\\
    $\sigma_{pp\to h_1^\pm h_1^\mp} \times \mathcal{Br}(h_1^\pm \to Z\,W^\mp)\mathcal{Br}(h_1^\mp \to   Z\,W^\mp)$.
\end{itemize}
 We will discuss the final sate searches along with dominant SM backgrounds  below starting for BP1 to BP3. A detailed collider study is in preparation \cite{pbch}. 
 
 If the lightest charged Higgs boson is pair produced, it can have the following decay topologies
\bea\label{fs1}
pp &\to& h^\pm_1h^\mp_1\nn \\
 &\to & a_1 W^\pm Z W^\mp \nn \\
   &\to & 2\tau (2b)+ 2j+ 3\ell \nn +\etmiss \\
   & \to  & 2\tau (2b)+ 4\ell +\etmiss .
\eea
Eq.~\ref{fs1} shows that when one of the charged Higgs bosons decays to $a_1 W^\pm$, which is a signature of the existence of singlet-type pseudoscalar, and the other one decays to $ZW^\pm$, which is the triplet signature. Thus we end up with $a_1 +2 W^\pm +Z$ intermediate state. Depending on the decays of the gauge bosons; hadronic or leptonic, and that of the light pseudoscalar (into $b$ or $\tau$ pairs), we can have final states with multi-lepton plus two $b$- or $\tau$-jets. The tri-lepton and four-lepton backgrounds are generally rather low in SM. In this case they are further tagged with $b$ or $\tau$-jet pair, which make these channels further clean. As mentioned earlier the detailed signal, backgrounds study is in progress as a separate study in \cite{pbch}. However in Table~\ref{finalstates}
we look for $ \geq3\ell+2\tau+\etmiss$ and $\geq3\ell+2b+\etmiss$ final states event numbers at an integrated luminosity of 1000 fb$^{-1}$ for both BP1 and dominant SM backgrounds. The demand $\geq 3\ell$ over $4\ell$ was chosen to enhance the signal numbers. The  kinematical cuts on the momentum and various isolation cuts  and tagging efficiencies for $b$-jets \cite{btag} and $\tau$-jets \cite{tautag} reduce the final state numbers. The $b$-tagging efficiency has been chosen to be $0.5$ and $\tau$-jet tagging efficiency varies a lot with the momentum of the $\tau$-jet ($30-70\%$) are taken into account while giving the final state numbers. 
For $\geq3\ell+2\tau+\etmiss$ and $\geq3\ell+2b+\etmiss$ final states the dominant backgrounds mainly come from triple gauge boson productions $ZZZ$ and $ZWZ$ respectively.  We can see that that  $\geq3\ell+2b+\etmiss$ reaches around $3\sigma$ of signal significance at an integrated luminosity of 1000 fb$^{-1}$. However a point with larger branching to both $aW^\pm$ and $ZW^\pm$ decay modes can be probed with much earlier data.

 In the case of a TESSM \cite{pbas1, pbas3} we have have only the triplet signature of charged Higgs decaying into $ZW^\pm$, which carries a different signature respect to the doublet-like charged Higgs boson. On the other hand, in the NMSSM we
only have $a_1 W^\pm$ decay \cite{han, colepa, guchait, pbsnkh}, which is characterised by a different signature respect to the MSSM \cite{ChCMS,ChATLAS}. In comparison, Eq.~\ref{fs1} provides a golden plated mode in the search of an extended Higgs sector, as predicted by the TNMSSM. Finding out both $a_1 W^\pm$ and $Z W^\pm$ decay modes at the LHC can prove the existence of both a singlet and  a triplet of the model. However, as we can see in Figure~\ref{muCH}, it is very difficult to find out points where both the $\mathcal{Br}( h^\pm_1 \to ZW^\pm)$ and $\mathcal{Br}( h^\pm_1 \to a_1W^\pm)$ are enhanced at the same time. Nevertheless as the final states carry the signatures of both singlet and triplet type Higgs bosons, it is worth exploring for a high luminosity at the LHC or even for higher energy (more than 14 TeV) at the LHC in future.

\begin{figure}[thb]
\begin{center}
\mbox{\subfigure[]{
\includegraphics[width=0.5\linewidth]{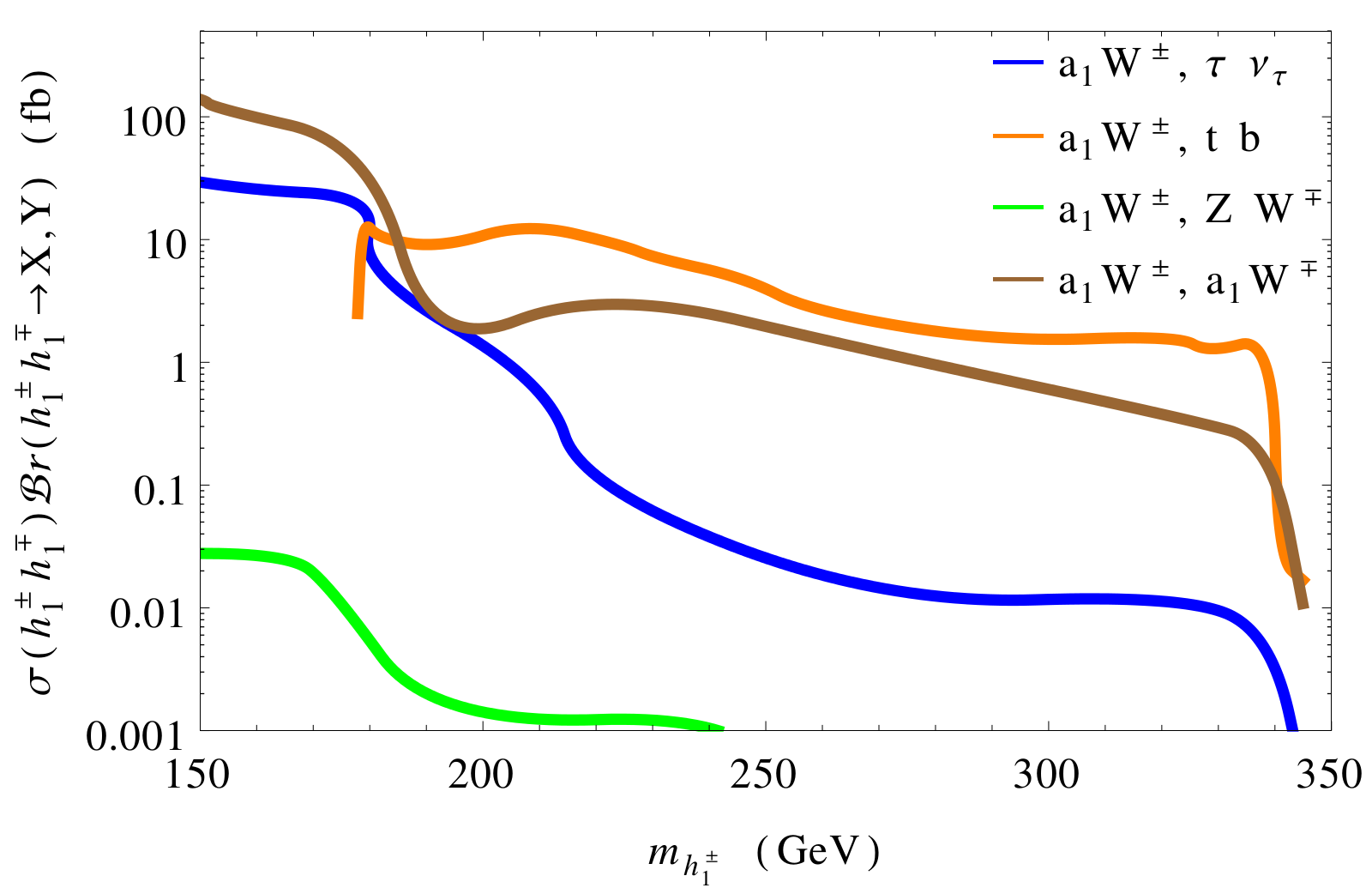}}
\subfigure[]{\includegraphics[width=0.5\linewidth]{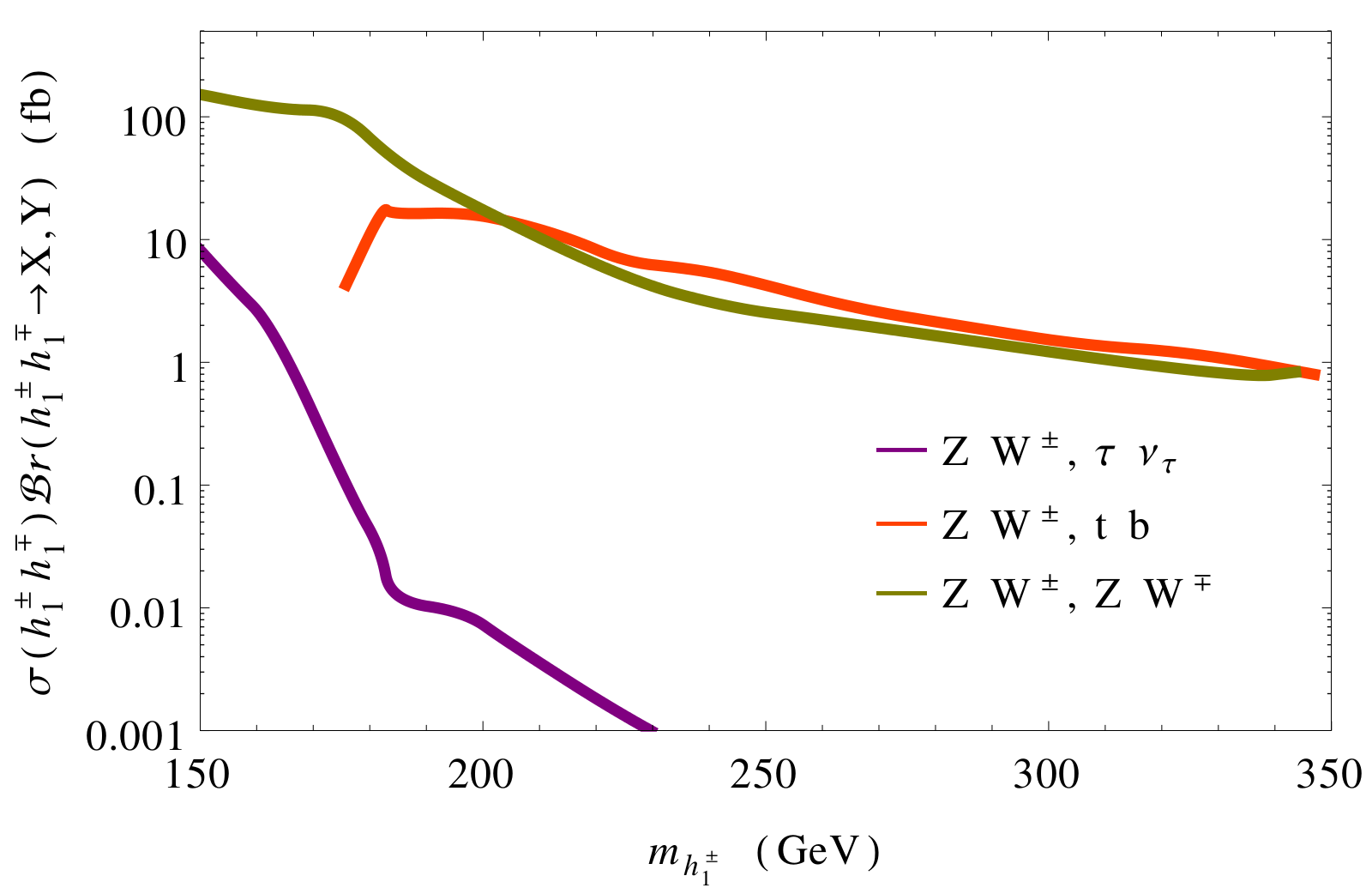}}}

\caption{The signal strength for the pair production of the lightest charged Higgs boson in the intermediate channels of Eq. \ref{fs1}, \ref{fs2}, \ref{fs2b}, \ref{fs2t} (a) and \ref{fs2a}, \ref{fs2c}, \ref{fs2ta} (b) as a function of the mass of the lightest charged Higgs boson.}\label{muCH}
\end{center}
\end{figure}

\begin{table}
\begin{center}
\renewcommand{\arraystretch}{1.5}
\begin{tabular}{||c|c|c||c|c||}
\hline\hline
\multicolumn{3}{||c||}{\multirow{2}{*}{Decay Channels}}&\multicolumn{2}{c||}{$\#$ of Events}\\
\cline{4-5}
\multicolumn{3}{||c||}{}&Signal&Backgrounds\\
\hline
\hline
\multirow{2}{*}{\rotatebox{90}{BP1}}&\multirow{2}{*}{$\;a_1W^\pm\,ZW^\mp\;$}
&$\geq3\ell+2\tau+\etmiss$&1&6\\
\cline{3-5}
&&$\geq3\ell+2b+\etmiss$&21&39\\
\hline
\hline
\multirow{2}{*}{\rotatebox{90}{BP2}}&$\;a_1W^\pm\,\tau\nu_\tau\;$
&$3\tau+1\ell+\etmiss$&13&$<1$\\
\cline{3-5}
\cline{3-5}

&$\;a_1W^\pm\,a_1W^\mp\;$&$\;2b+2\tau+2\ell+\etmiss\;$&164&38\\
\hline
\hline
\multirow{3}{*}{\rotatebox{90}{BP3}}&$\;ZW^\pm\,\tau\nu_\tau\;$
&$1\tau+3\ell+\etmiss$&9&19\\
\cline{3-5}
\cline{3-5}
&\multirow{2}{*}{$\;Z\,W^\pm\,Z\,W^\mp\;$}
&$\geq5\ell+\etmiss$&228&23\\
\cline{3-5}
&&$\;\geq1\ell+2b+2\tau+\etmiss\;$&29&246\\

\hline\hline
\end{tabular}
\caption{The final state numbers for the benchmark points and backgrounds  at an integrated luminosity of 1000 fb$^{-1}$.}\label{finalstates}
\end{center}
\end{table}

}

The light charged Higgs boson can also decay to $\tau\nu$ for $m_{h^\pm_1} < m_t$ and to $tb$ for $m_{h^\pm_1} > m_t$, via its doublet fraction. The charged Higgs pair production then has the signatures given in Eq.~\ref{fs2} and Eq.~\ref{fs2a}, with one of the charged Higgs boson decaying to $\tau\nu$ and the other one to $a_1 W^\pm$ or $Z W^\pm$, respectively. Eq.~\ref{fs2} and Eq.~\ref{fs2a}  probe the existence of singlet, doublet and triplet representations at the same time. The final states with one or more tau-jets along with charged lepton reduce the SM backgrounds but nevertheless $t\bar{t}Z$ and $tZW^\pm$ contribute. 

\bea\label{fs2}
pp &\to & h^\pm_1h^\mp_1 \nn \\
& \to & a_1 W^\pm  \tau\nu \nn \\
 & \to & 3\tau /(2b +1\tau) +1 \ell +\etmiss,  
   \eea

\bea\label{fs2a}
pp &\to & h^\pm_1h^\mp_1 \nn \\
 &\to & Z W^\pm  \tau\nu \nn \\
  & \to & 1(3)\tau + 3(1)\ell  +\etmiss. 
   \eea

Thus these final states would play a very crucial role in determining whether the mechanism of EWSB incorporates a finer structure respect to our current description, with a single Higgs doublet.
{ In Table~\ref{finalstates} we present the number of events in the $3\tau+1\ell+\etmiss$ final state for the channel $a_1W^\pm\,\tau\nu_\tau$ and in the $1\tau+3\ell+\etmiss$ for the channel $ZW^\pm\,\tau\nu_\tau$ at an integrated luminosity of 1000 fb$^{-1}$. As already stated, we chose a triplet-like charged Higgs boson $h_1^\pm$ and hence the branching in $\tau\nu_\tau$ is suppressed, being a signature decay mode for a doublet-type charged Higgs boson. In both the case the dominant backgrounds are the triple gauge bosons $ZZZ$ and $ZWZ$. We can see that that  $3\ell+1\tau+\etmiss$ reaches more than $3\sigma$ of signal significance at an integrated luminosity of 1000 fb$^{-1}$.

}
There are, of course, two other possibilities for the decays of a pair of charged Higgs bosons, that is when both the charged Higgs bosons decays to $a_1W^\pm$ or $ZW^\pm$.
\bea\label{fs2b}
pp &\to& h^\pm_1h^\mp_1\nn \\
 &\to & a_1 W^\pm a_1 W^\mp \nn \\
   &\to & 2\tau +2b+ 2j+ 1\ell \nn +\etmiss \\
   & \to  & 4\tau (4b)+ 2\ell +\etmiss\nn\\
   & \to &  2b+2\tau+2\ell+\etmiss.
\eea
\bea\label{fs2c}
pp &\to& h^\pm_1h^\mp_1\nn \\
 &\to & Z W^\pm Z W^\mp \nn \\
   &\to & 2j+ 4\ell \nn +\etmiss \\
   & \to  & 6\ell +\etmiss \nn\\
   & \to &2b+2\tau+2\ell+\etmiss.
\eea
These channels can prove the existence of singlet and triplet representation separately.
{  For the decay channel $h^\pm_1h^\mp_1\to  a_1 W^\pm a_1 W^\mp$ we have considered the $2b+2\tau+2\ell+\etmiss$ final state for the signal and background analysis. This is because the final states with $\geq1\ell$ have $\bar t t$ as dominant background and hence are strongly suppressed. For $2b+2\tau+2\ell+\etmiss$ the dominant backgrounds are $ZZZ$ and $\bar t tZ$ and we can see from Table~\ref{finalstates} that the signal significance is more than $10\sigma$ for an integrated luminosity of 1000 fb$^{-1}$. A $5\sigma$ of signal significance can be achieved with an integrated luminosity of $\approx$ 200 fb$^{-1}$ at the LHC with 14 TeV center of mass energy.

In the case of $h^\pm_1h^\mp_1\to Z\, W^\pm Z\, W^\mp$ we look into the $\geq5\ell+\etmiss$ and $\geq1\ell+2b+2\tau+\etmiss$ final states where the demand $\geq 1\ell$ over $2\ell$ was chosen to enhance the signal numbers. The $\geq5\ell+\etmiss$ has the triple gauge bosons $ZZZ$ and $ZWZ$ as dominant backgrounds. This is one of cleanest final state and we can see from Table~\ref{finalstates} that it has more than $14\sigma$ of signal significance at an integrated luminosity of 1000 fb$^{-1}$. The integrated luminosity for $5\sigma$ of signal significance is 120 fb$^{-1}$. The dominant backgrounds for the  $\geq1\ell+2b+2\tau+\etmiss$ final state are the triple gauge bosons $ZZZ$ and $ZWZ$ as well as $\bar t tZ$. The $\bar t tZ$ background is the most dominant one in this case and suppress the signal significance, as one can immediately realize looking at Table~\ref{finalstates}.
}

For a charged Higgs bosons heavier than the top quark the channel $h^\pm_1 \to t b$ is kinematically allowed. 
If one of the charged Higgs decays to $tb$ and the other one decays to $ a_1 W^\pm$ we have the final states given by Eq.~\ref{fs2t}. When the other charged Higgs boson decays to $ZW^\pm$, the production of $h^\pm_1h^\mp_1$ results in the final states of  Eq.~\ref{fs2ta} 
\bea\label{fs2t}
pp&\to & h^\pm_1h^\mp_1 \nn \\
 &\to & a_1 W^\pm  t b\nn \\
  & \to & 2\tau + 2b + 2 W \nn \\
     & \to & 2\tau +2b + 2\ell +\etmiss, 
   \eea
   
   \bea\label{fs2ta}
pp&\to & h^\pm_1h^\mp_1 \nn \\
 &\to & Z W^\pm  t b\nn \\
   &\to & 2\tau + 2b + 2 W \nn \\
     & \to & 2\tau +2b + 2\ell +\etmiss\, \nn\\
     & \rm{or}& \,2b+ 4\ell +\etmiss.  
   \eea
The signal related to the intermediate states of the pair production and the decays of the lightest charged Higgs boson in the channels of Eq. \ref{fs1}, \ref{fs2}, \ref{fs2a}, \ref{fs2t} and \ref{fs2ta} is reported in Figure~\ref{muCH}. We can clearly see that for light charged Higgs boson  ($m_{h^\pm_1} \gsim 200$ GeV)  the decay modes in a light pseudoscalar can be probed rather easily at the LHC 
but probing  $a_1W^\pm$ and $ZW^\mp$, i.e., the existence of a light pseudoscalar and the triplet decay modes  together needs higher luminosity.

Another signature of this model could be the existence of the heavier charged Higgs bosons $h^\pm_{2,3}$
which could be produced at the LHC. For our selection points $h^\pm_2$ is triplet-like
and $h^\pm_3$ is doublet-like. Following our discussion in section~\ref{ch1dcy},  
such heavy charged Higgs can decay dominantly to $a_1 h^\pm_1$ or $h_1 h^\pm_1$, as shown in
Eq.~\ref{fs3} and Eq.~\ref{fs4}. The lighter charged Higgs can then decay into final states with  $a_1 W^\pm$
or $Z W^\pm$ giving $2\tau (2b)+  3\ell +\etmiss$ and $4\tau (4b)+  1\ell +\etmiss$ final states

\bea\label{fs3}
pp\to h^\pm_{2,3} +X&  \to & a_1/h_1 h^\mp_1 \nn \\
&  \to & 2\tau (2b)+ Z W^\pm \nn \\
  & \to & 2\tau (2b)+  3\ell + \etmiss,
\eea

\bea\label{fs4}
pp\to h^\pm_{2,3} +X & \to & a_1/h_1 h^\mp_1 \nn \\
 & \to & 2\tau (2b)+ a_1 + W^\pm \nn \\
  & \to & 4\tau (4b)+  1\ell + \etmiss.
\eea

Searching for the above signatures is certainly necessary not only in order to discover a charged Higgs boson but also to determine whether scalars in higher representations of $SU(2)$ are involved in the mechanism of EWSB. 

\section{Discussion }\label{dis}

In this article we have presented a detailed analysis of the charged Higgs sector of the TNMSSM, considering both the doublet- and triplet-like cases, as predicted by the triplet-singlet extension of the MSSM.  { We focus our attention on a typical mass spectrum with a doublet-like CP-even Higgs boson around 125 GeV,  a light triplet-like charged Higgs boson and a light singlet-like pseudoscalar. The existence of light singlet-like pseudoscalar and triplet-like charged Higgs  boson enrich the phenomenology at the LHC and at future colliders. 

In general we expect to have mixing between doublet and triplet type charged Higgs. We find that in the decoupling limit, $\lambda_T \simeq 0$, one should expect two triplet-like and one doublet-like 
massive charged Higgs bosons. However since the Goldstone boson is a linear combination which includes a triplet contribution $\sim {v_T}/{v}$ (see Eq.~\ref{gstn}), one of the massive eigenstates triplet cannot be $100\%$ triplet-like.

 Recent searches by both CMS \cite{ChCMS} and ATLAS  \cite{ChATLAS} are conducted for a 
charged Higgs mainly of doublet-type and coupled to fermions. For this reason such a state can be produced in association with the top
quark and can decay to $\tau\nu$. Clearly, these searches have to be reinvestigated in order to probe the possibility of triplet representations of $SU(2)$ in the Higgs sector.\\
The breaking of the custodial symmetry via a non-zero triplet VEV generates $h_i^\pm-W^\mp-Z$ vertex 
at the tree-level in TNMSSM. This leads to the vector boson fusion channel for the charged Higgs boson, which is not present in the MSSM or the 2HDM.  On top of that the $Z_3$ symmetric 
superpotential of TNMSSM has a  light pseudoscalar $a_1$ as a pseudo NG mode of a global $U(1)$ symmetry, known as the "$R$-axion" in the literature. However the later can also be  found in the context of the $Z_3$ symmetric NMSSM. In this case the light charged Higgs boson can decay to $a_1 W^\pm$ \cite{han, colepa, guchait, pbsnkh} just like in the TNMSSM. In the context of a CP-violating MSSM, such modes can arise due to the possibility of a light Higgs boson $h_1$ and of CP-violating interactions. A charged Higgs boson can decay to $h_1 W^\pm$ \cite{CPVMSSM}, just as in our case. Therefore, one of the challenges at the LHC will be to distinguish among such models once such a mode is discovered. 

Triplet charged Higgs bosons with $Y=0$, however, have some distinctive features because they do not couple to the fermions, while the fusion channel $ZW^\pm$ is allowed.  The phenomenology of such triplet-like charged Higgs boson has already been studied in the context of TESSM \cite{pbas3}. Such charged Higgs bosons also affect the predictions of $B$-observables \cite{pbas1, pbas2} for missing  the coupling to fermions and to the $Z$ boson.  However in TESSM, even though the charged Higgs boson decays to $ZW^\pm$ \cite{pbas3}, the possibility of a light pseudoscalar is not so natural \cite{pbas1, pbas2, DiChiara, pbas3}. Indeed, one way to distinguish between the TESSM and the TNMSSM is to exploit the prediction of a light pseudoscalar in the second model, beside the light triplet type charged Higgs boson. 

We expect that such a Higgs in the TNMSSM will be allowed to decay both to $ZW^\pm$ as well as to $a_1 W^\pm$, the former being a feature of the triplet nature of this state, and the latter of the presence of an $R$-axion in the spectrum of the model. We are currently performing a detailed simulation of both the TESSM and the NMSSM in order to identify specific signatures which can be compared with the TNMSSM \cite{pbch}. A complete simulation of the Standard Model background is also underway.}

\section{Conclusions}\label{concl}
{ Triplet-like charged Higgs bosons do not couple to fermions (see Eq.~\ref{spt}) which makes them hard to be produced at LHC. The non-zero triplet VEV breaks  the custodial symmetry and the consequence can be  seen in  non-zero $h_i^\pm-W^\mp-Z$ coupling. Thus measurement of such coupling  or decay of the charged Higgs boson in $ZW^\pm$ can shed light in determining the role of the triplet in electro-weak symmetry breaking. For this reason we propose few channels which can be probed at the LHC. Specifically if the triplet-like charged Higgs bosons are pair produced at the LHC, it would be interesting to see if both $a_1 W^\pm$ and $ZW^\pm$ decay modes can be probed. Finding these decay modes can surely be a proof of the existence of both the singlet and the triplet in the mass spectrum. This can be a smoking gun signature for TNMSSM at the LHC. General fermiofobic nature however push this settlement at higher luminosity at the LHC.}

\appendix
\section{Mass matrix of the Higgs sector}\label{Higgsm}
The symmetric mass matrices of the Higgs sector are given by
\bea\label{sMM}
\renewcommand{\arraystretch}{1.2}
\mathcal{M}^S=\left(
\begin{array}{cccc}
m^S_{11}&m^S_{12}&m^S_{13}&m^S_{14}\\
 &m^S_{22}&m^S_{23}&m^S_{24}\\
 & &m^S_{33}&m^S_{34}\\
 & & &m^S_{44}
\end{array}
\right),
\eea

\bea\label{pMM}
\renewcommand{\arraystretch}{1.2}
\mathcal{M}^P=\left(
\begin{array}{cccc}
m^P_{11}&m^P_{12}&m^P_{13}&m^P_{14}\\
 &m^P_{22}&m^P_{23}&m^P_{24}\\
 & &m^P_{33}&m^P_{34}\\
 & & &m^P_{44}
\end{array}
\right),
\eea

\bea\label{chMM}
\renewcommand{\arraystretch}{1.2}
\mathcal{M}^C=\left(
\begin{array}{cccc}
m^C_{11}&m^C_{12}&m^C_{13}&m^C_{14}\\
 &m^C_{22}&m^C_{23}&m^C_{24}\\
 & &m^C_{33}&m^C_{34}\\
 & & &m^C_{44}
\end{array}
\right),
\eea
where we have used the following abbreviations
\begin{align}
m^S_{11}&=\frac{1}{4 v_u}\Big(2 v_d \big(\sqrt{2} A_S v_S-v_T \left(A_T+\sqrt{2} v_S \lambda _T \lambda_{TS}\right)+\lambda _S \big(\kappa  v_S^2+v_T^2 \lambda _{TS}\big)\big)+v_u^3
   \big(g_L^2+g_Y^2\big)\Big)\nn\\
m^S_{12}&=\frac{1}{2} \Big(-\sqrt{2} A_S v_S+v_T \left(A_T+\sqrt{2} v_S \lambda _T \lambda _{TS}\right)-\lambda_S \left(\kappa  v_S^2+v_T^2 \lambda _{TS}\right)\Big)-\frac{1}{4} v_d v_u
   \left(g_L^2+g_Y^2-2 \left(2 \lambda _S^2+\lambda _T^2\right)\right)\nn\\
m^S_{13}&=-\frac{A_S v_T}{\sqrt{2}}+v_d \left(\frac{\lambda _T v_T \lambda _{TS}}{\sqrt{2}}-\kappa 
   \lambda_S v_S\right)+\frac{1}{2} v_u \lambda _S \left(2 \lambda _S v_S-\sqrt{2} \lambda _T v_T\right)\nn\\
m^S_{14}&=\frac{1}{2} \Big(v_d \left(A_T-2 \lambda _S v_T \lambda _{TS}\right)+\sqrt{2} v_S \lambda_T \left(v_d \lambda _{TS}-v_u \lambda _S\right)+v_u \lambda _T^2 v_T\Big)\nn\\
m^S_{22}&=\frac{1}{4 v_d}\Big(2 v_u \Big(\sqrt{2} A_S v_S-v_T \left(A_T+\sqrt{2} v_S \lambda _T \lambda _{TS}\right)+\lambda _S \left(\kappa  v_S^2+v_T^2 \lambda _{TS}\right)\Big)+v_d^3
   \left(g_L^2+g_Y^2\right)\Big)\nn\\
m^S_{23}&=-\frac{A_S v_u}{\sqrt{2}}+\frac{1}{2} v_d \lambda_S \left(2 \lambda_S v_S-\sqrt{2} \lambda _T   v_T\right)+v_u \left(\frac{\lambda _T v_T \lambda _{TS}}{\sqrt{2}}-\kappa  \lambda _S   v_S\right)\nn\\
m^S_{24}&=\frac{1}{2} \left(v_u \left(A_T-2 \lambda _S v_T \lambda _{TS}\right)+\sqrt{2} v_S \lambda_T \left(v_u \lambda _{TS}-v_d \lambda _S\right)+v_d \lambda _T^2 v_T\right)\nn\\
m^S_{33}&=\frac{1}{4 v_S}\Big(\sqrt{2} v_T \left(\lambda _T \left(\lambda _S \left(v_d^2+v_u^2\right)-2 v_d
   v_u \lambda _{TS}\right)-2 A_{TS} v_T\right)+2 \sqrt{2} A_S v_d v_u+2
   \sqrt{2} A_{\kappa} v_S^2+8 \kappa ^2 v_S^3\Big)\nn\\
m^S_{34}&=\frac{1}{4} \Big(4 \sqrt{2} A_{TS} v_T-\sqrt{2} \lambda _S \lambda _T
   \left(v_d^2+v_u^2\right)+2 \lambda _{TS} \left(\sqrt{2} v_d v_u \lambda _T+4
   v_S v_T \left(\kappa +2 \lambda _{TS}\right)\right)\Big)\nn\\
m^S_{44}&=\frac{1}{4 v_T}\Big(-2 v_d v_u \left(A_T+\sqrt{2} v_S \lambda _T \lambda _{TS}\right)+\sqrt{2}
   v_d^2 \lambda _S v_S \lambda _T+\sqrt{2} v_u^2 \lambda _S v_S \lambda _T+8 v_T^3 \lambda
   _{TS}^2\Big)\nn
\end{align}

\begin{align}
m^P_{11}&=\frac{v_d}{2 v_u}\Big( \left(\sqrt{2} A_S v_S-v_T \left(A_T+\sqrt{2} v_S \lambda _T \lambda
   _{TS}\right)+\lambda _S \left(\kappa  v_S^2+v_T^2 \lambda _{TS}\right)\right)\Big)\nn\\
m^P_{12}&=\frac{1}{2} \left(\sqrt{2} A_S v_S-v_T \left(A_T+\sqrt{2} v_S \lambda _T \lambda _{TS}\right)+\lambda
   _S \left(\kappa  v_S^2+v_T^2 \lambda _{TS}\right)\right)\nn\\
m^P_{13}&=\frac{1}{2} v_d \left(\sqrt{2} A_S-2 \kappa  \lambda _S v_S+\sqrt{2} \lambda _T v_T \lambda
   _{TS}\right)\nn\\
m^P_{14}&=-\frac{1}{2} v_d \left(A_T+\lambda _{TS} \left(2 \lambda _S v_T-\sqrt{2} v_S \lambda
   _T\right)\right)\nn\\
m^P_{22}&=\frac{v_u}{2 v_d}\Big( \left(\sqrt{2} A_S v_S-v_T \left(A_T+\sqrt{2} v_S \lambda _T \lambda
   _{TS}\right)+\lambda _S \left(\kappa  v_S^2+v_T^2 \lambda _{TS}\right)\right)\Big)\nn\\
m^P_{23}&=\frac{1}{2} v_u \left(\sqrt{2} A_S-2 \kappa  \lambda _S v_S+\sqrt{2} \lambda _T v_T \lambda
   _{TS}\right)\nn\\
m^P_{24}&=-\frac{1}{2} v_u \left(A_T+\lambda _{TS} \left(2 \lambda _S v_T-\sqrt{2} v_S \lambda
   _T\right)\right)\nn\\
m^P_{33}&=\frac{v_T}{4 v_S} \Big(\Big(\sqrt{2} \lambda _T \left(\lambda _S \left(v_d^2+v_u^2\right)-2 v_d
   v_u \lambda _{TS}\right)-2 v_T \left(\sqrt{2} A_{TS}+4 \kappa  v_S \lambda
   _{TS}\right)\Big)+2 \sqrt{2} A_S v_d v_u-6 \sqrt{2} A_\kappa v_S^2\nn\\
   &+8 \kappa 
   v_d v_u \lambda _S v_S\Big)\nn\\
m^P_{34}&=\frac{1}{4} \left(-4 \sqrt{2} A_{TS} v_T-\sqrt{2} \lambda _T \left(\lambda _S
   \left(v_d^2+v_u^2\right)+2 v_d v_u \lambda _{TS}\right)+8 \kappa  v_S
   v_T \lambda _{TS}\right)\nn\\
m^P_{44}&=\frac{-2 v_d v_u}{4
   v_T}\Big( \left(A_T+\lambda _{TS} \left(\sqrt{2} v_S \lambda _T-4 \lambda _S
   v_T\right)\right)+v_S \left(\sqrt{2} v_d^2 \lambda _S \lambda _T-8 v_T \left(\sqrt{2}
   A_{TS}+\kappa  v_S \lambda _{TS}\right)\right)+\sqrt{2} v_u^2 \lambda _S v_S \lambda _T\Big)\nn
\end{align}

\begin{align}
m^C_{11}&=\frac{1}{4} \Big(2 \Big(\sqrt{2} v_S \left(A_S \cot\beta+\lambda _T v_T \left(2 \lambda_{S}-\cot\beta \lambda
   _{TS}\right)\right)+\cot\beta v_T \left(\lambda_{S} v_T \lambda
   _{TS}-A_{T}\right)+\kappa  \cot\beta \lambda_{S} v_S^2\Big)\nn\\
   &+\cos ^2\beta\, v^2 \left(g_L^2-2
   \lambda_{S}^2+\lambda _T^2\right)\Big)\nn\\
m^C_{12}&=\frac{1}{4} v \Big(\lambda _T \left(2 v_S \left(\sin\beta \lambda_{S}-2 \cos\beta \lambda _{TS}\right)+\sqrt{2} \sin\beta \lambda _T v_T\right)-\sqrt{2} \sin\beta g_L^2 v_T\Big)\nn\\
m^C_{13}&=\frac{1}{4} \Big(2 \Big(v_T \left(A_{T}+\lambda_{TS} \left(\lambda_{S} v_T+\sqrt{2} v_S
   \lambda _T\right)\right)+\sqrt{2} A_S v_S+\kappa  \lambda_{S} v_S^2\Big)+\sin\beta \cos\beta v^2
   \left(g_L^2-2 \lambda_{S}^2+\lambda _T^2\right)\Big)\nn\\
m^C_{14}&=\frac{v}{4}\left(\sin\beta \left(\sqrt{2} v_T \left(g_L^2-\lambda _T^2\right)+2 \lambda_{S} v_S
   \lambda _T\right)-2 \sqrt{2} A_{T} \cos\beta\right)\nn\\
m^C_{22}&=\frac{1}{4 v_T}\Big(v_T \Big(v^2 \left(\cos (2 \beta ) \left(g_L^2-\lambda _T^2\right)+2 \sin (2 \beta ) \lambda_{S}
   \lambda_{TS}\right)-4 v_S \left(\sqrt{2} A_{TS}+\kappa  v_S \lambda_{TS}\right)\Big)-A_{T}
   \sin (2 \beta ) v^2\nn\\
   &+2 v_T^3 \left(g_L^2-2 \lambda_{TS}^2\right)+\sqrt{2} v^2 v_S
   \lambda _T \left(\lambda_{S}-\sin (2 \beta ) \lambda_{TS}\right)\Big)\nn\\
m^C_{23}&=\frac{v}{4}\left(2 \sqrt{2} A_{T} \sin\beta+\cos\beta \left(\sqrt{2} v_T \left(\lambda
   _T^2-g_L^2\right)-2 \lambda_{S} v_S \lambda _T\right)\right)\nn\\
m^C_{24}&=\sqrt{2} A_{TS} v_S-\frac{1}{2} g_L^2 v_T^2+\lambda_{TS} \big(\kappa  v_S^2+v_T^2 \lambda_{TS}- \sin\beta \cos\beta
   v^2 \lambda_{S}\big)\nn\\
m^C_{33}&=\frac{1}{4} \Big(2 \Big(\sqrt{2} v_S \left(A_S \tan\beta+\lambda _t v_T \left(2 \lambda_{S}-\tan\beta \lambda_{TS}\right)\right)+\tan\beta v_T \left(\lambda_{S} v_T \lambda
   _{TS}-A_{T}\right)+\kappa  \tan\beta \lambda_{S} v_S^2\Big)\nn\\
   &+\sin ^2\beta g_L^2 v^2+\sin^2\beta v^2 \left(\lambda _T^2-2 \lambda_{S}^2\right)\Big)\nn\\
m^C_{34}&=\frac{v}{4}\Big(\cos\beta \left(\sqrt{2} v_T \left(g_L^2-\lambda _T^2\right)-2 \lambda_{S} v_S
   \lambda _T\right)+4 \sin\beta v_S \lambda _T \lambda_{TS}\Big)\nn\\
m^C_{44}&=\frac{1}{4 v_T}\Big(v_T \Big(v^2 \left(\cos (2 \beta ) \left(\lambda _T^2-g_L^2\right)+2 \sin (2 \beta ) \lambda_{S}
   \lambda_{TS}\right)-4 v_S \left(\sqrt{2} A_{TS}+\kappa  v_S \lambda_{TS}\right)\Big)-A_{T}
   \sin (2 \beta ) v^2\nn\\
   &+2 v_T^3 \left(g_L^2-2 \lambda_{TS}^2\right)+\sqrt{2} v^2 v_S
   \lambda _T \left(\lambda_{S}-\sin (2 \beta ) \lambda_{TS}\right)\Big)\nn
\end{align}
\normalsize
As already explained, the massive eigenvectors of the charged mass matrix are function of all the parameters of the model, including the parameters that are related to the singlet, e.g. $v_S$, $\lambda_S$, $\kappa$, whereas the Goldstone eigenvector is a function of the doublets and triplet VEV only.
This is also true for for the eigenvectors of the pseudoscalar mass matrix. In this case the Goldstone eigenvector is a function of the doublets VEV only.



\begin{thebibliography}{99}

\bibitem{CMS}
  G.~Aad {\it et al.} [ATLAS and CMS Collaborations],
  Phys.\ Rev.\ Lett.\  {\bf 114} (2015) 191803
  doi:10.1103/PhysRevLett.114.191803
  [arXiv:1503.07589 [hep-ex]].
  S.~Chatrchyan {\it et al.}  [CMS Collaboration],
  JHEP {\bf 1401} (2014) 096
  [arXiv:1312.1129 [hep-ex]].
  S.~Chatrchyan {\it et al.}  [CMS Collaboration],
  Phys.\ Rev.\ D {\bf 89} (2014) 9,  092007
  [arXiv:1312.5353 [hep-ex]].
\bibitem{CMS2}
  V.~Khachatryan {\it et al.} [CMS Collaboration],
  Eur.\ Phys.\ J.\ C {\bf 75} (2015) no.5,  212
  doi:10.1140/epjc/s10052-015-3351-7
  [arXiv:1412.8662 [hep-ex]].
  [CMS Collaboration],
  CMS-PAS-HIG-13-002.

\bibitem{ATLAS}
ATLAS-CONF-2015-007
  G.~Aad {\it et al.}  [ATLAS Collaboration],
  Phys.\ Rev.\ D {\bf 91} (2015) 1,  012006
  [arXiv:1408.5191 [hep-ex]].
  G.~Aad {\it et al.} [ATLAS Collaboration],
  Phys.\ Rev.\ D {\bf 92} (2015) no.1,  012006
  doi:10.1103/PhysRevD.92.012006
  [arXiv:1412.2641 [hep-ex]].
  G.~Aad {\it et al.}  [ATLAS Collaboration],
  Phys.\ Rev.\ D {\bf 90} (2014) 5,  052004
  [arXiv:1406.3827 [hep-ex]].
 \bibitem{rparity}
  R.~Barbier, C.~Berat, M.~Besancon, M.~Chemtob, A.~Deandrea, E.~Dudas, P.~Fayet and S.~Lavignac {\it et al.},
  Phys.\ Rept.\  {\bf 420} (2005) 1
  [hep-ph/0406039].
  \bibitem{ChCMS}
  CMS Collaboration [CMS Collaboration],
  CMS-PAS-HIG-14-020.
  CMS Collaboration [CMS Collaboration],
  CMS-PAS-HIG-13-026.
  \bibitem{ChATLAS}  
  G.~Aad {\it et al.} [ATLAS Collaboration],
  JHEP {\bf 1503} (2015) 088
  [arXiv:1412.6663 [hep-ex]].

  
\bibitem{Ellwanger}
  U.~Ellwanger, C.~Hugonie and A.~M.~Teixeira,
  Phys.\ Rept.\  {\bf 496} (2010) 1
  doi:10.1016/j.physrep.2010.07.001
  [arXiv:0910.1785 [hep-ph]].


 \bibitem{pbas1}
  P.~Bandyopadhyay, K.~Huitu and A.~Sabanci,
  JHEP {\bf 1310} (2013) 091
  [arXiv:1306.4530 [hep-ph]].
 
\bibitem{pbas2}
  P.~Bandyopadhyay, S.~Di Chiara, K.~Huitu and A.~S.~Keceli,
  JHEP {\bf 1411} (2014) 062
  [arXiv:1407.4836 [hep-ph]].
\bibitem{DiChiara}
  S.~Di Chiara and K.~Hsieh,
  Phys.\ Rev.\ D {\bf 78} (2008) 055016
  doi:10.1103/PhysRevD.78.055016
  [arXiv:0805.2623 [hep-ph]].
\bibitem{pbas3}
  P.~Bandyopadhyay, K.~Huitu and A.~S.~Keceli,
  JHEP {\bf 1505} (2015) 026
  [arXiv:1412.7359 [hep-ph]].
\bibitem{EspinosaQuiros}
  J.~R.~Espinosa and M.~Quiros,
  hep-ph/9208226.
  J.~R.~Espinosa and M.~Quiros,
  Nucl.\ Phys.\ B {\bf 384} (1992) 113.
  doi:10.1016/0550-3213(92)90464-M



\bibitem{TNMSSM1}
  P.~Bandyopadhyay, C.~Corian\`o and A.~Costantini,
  JHEP {\bf 1509} (2015) 045
  doi:10.1007/JHEP09(2015)045
  [arXiv:1506.03634 [hep-ph]].

\bibitem{TNMSSM2}
  P.~Bandyopadhyay, C.~Corian\`o and A.~Costantini,
  JHEP {\bf 1512} (2015) 127
  doi:10.1007/JHEP12(2015)127
  [arXiv:1510.06309 [hep-ph]].
 





  
     
     
       

     
\bibitem{tnssm}
  T.~Basak and S.~Mohanty,
  Phys.\ Rev.\ D {\bf 86}, 075031 (2012)
  [arXiv:1204.6592 [hep-ph]].

  T.~Basak and S.~Mohanty,
  JHEP {\bf 1308} (2013) 020
  [arXiv:1304.6856 [hep-ph]].
  
  

  L.~Basso, O.~Fischer and J.~J.~van der Bij,
  Europhys.\ Lett.\  {\bf 101} (2013) 51004
  [arXiv:1212.5560].
  
  O.~Fischer and J.~J.~van der Bij,
  JCAP {\bf 1401} (2014) 01,  032
  [arXiv:1311.1077 [hep-ph]].
  
 \bibitem{tnssma}
  K.~Agashe, A.~Azatov, A.~Katz and D.~Kim,
  Phys.\ Rev.\ D {\bf 84} (2011) 115024
  [arXiv:1109.2842 [hep-ph]].
     
     \bibitem{LEPb} 
  R.~Barate {\it et al.}  [LEP Working Group for Higgs boson searches and ALEPH and DELPHI and L3 and OPAL Collaborations],
  Phys.\ Lett.\ B {\bf 565} (2003) 61
  [hep-ex/0306033].
  S.~Schael {\it et al.}  [ALEPH and DELPHI and L3 and OPAL and LEP Working Group for Higgs Boson Searches Collaborations],
  Eur.\ Phys.\ J.\ C {\bf 47} (2006) 547
  [hep-ex/0602042].

     
  \bibitem{rho}
  J.~Beringer {\it et al.}  [Particle Data Group Collaboration],
  Phys.\ Rev.\ D {\bf 86} (2012) 010001.
   \bibitem{bottomonium1}
  P.~Franzini, D.~Son, P.~M.~Tuts, S.~Youssef, T.~Zhao, J.~Lee-Franzini, J.~Horstkotte and C.~Klopfenstein {\it et al.},
  Phys.\ Rev.\ D {\bf 35} (1987) 2883.
   \bibitem{bottomonium}
  J.~S.~Lee and S.~Scopel,
  Phys.\ Rev.\ D {\bf 75} (2007) 075001
  [hep-ph/0701221 [HEP-PH]].

  \bibitem{lhcb}
  V.~Khachatryan {\it et al.} [CMS and LHCb Collaborations],
  Nature {\bf 522}, 68 (2015)
  doi:10.1038/nature14474
  [arXiv:1411.4413 [hep-ex]].


\bibitem{Asakawa}
  E.~Asakawa and S.~Kanemura,
  Phys.\ Lett.\ B {\bf 626} (2005) 111
  doi:10.1016/j.physletb.2005.08.091
  [hep-ph/0506310].
 
  \bibitem{anatomy2}
  A.~Djouadi,
  Phys.\ Rept.\  {\bf 459} (2008) 1
  [hep-ph/0503173].
  
  \bibitem{sarah}
  F.~Staub,
  Comput.\ Phys.\ Commun.\  {\bf 184} (2013) pp. 1792
   [Comput.\ Phys.\ Commun.\  {\bf 184} (2013) 1792]
  [arXiv:1207.0906 [hep-ph]].
  
\bibitem{calchep}
  A.~Pukhov,
``CalcHEP 3.2: MSSM, structure functions, event generation, batchs, and
generation of matrix elements for other packages'',
  [arXiv:hep-ph/0412191].
  

\bibitem{6teq6l}
  J.~Pumplin, D.~R.~Stump, J.~Huston, H.~L.~Lai, P.~Nadolsky and W.~K.~Tung,
  JHEP {\bf 0207}, 012 (2002)
  [arXiv:hep-ph/0201195].,
  see also \url{http://www.physics.smu.edu/scalise/cteq/}
  
\bibitem{djuadi}
  F.~Borzumati and A.~Djouadi,
  Phys.\ Lett.\ B {\bf 549} (2002) 170
  doi:10.1016/S0370-2693(02)02889-7
  [hep-ph/9806301].

\bibitem{moretti}
  D.~J.~Miller, S.~Moretti, D.~P.~Roy and W.~J.~Stirling,
  Phys.\ Rev.\ D {\bf 61} (2000) 055011
  doi:10.1103/PhysRevD.61.055011
  [hep-ph/9906230].
  
\bibitem{han}
  N.~D.~Christensen, T.~Han, Z.~Liu and S.~Su,
  JHEP {\bf 1308} (2013) 019
  doi:10.1007/JHEP08(2013)019
  [arXiv:1303.2113 [hep-ph]].

\bibitem{colepa}
  B.~Coleppa, F.~Kling and S.~Su,
  JHEP {\bf 1412} (2014) 148
  doi:10.1007/JHEP12(2014)148
  [arXiv:1408.4119 [hep-ph]].

\bibitem{guchait}
  M.~Drees, M.~Guchait and D.~P.~Roy,
  Phys.\ Lett.\ B {\bf 471} (1999) 39
  doi:10.1016/S0370-2693(99)01329-5
  [hep-ph/9909266].

  
\bibitem{pbsnkh}
  P.~Bandyopadhyay, K.~Huitu and S.~Niyogi,
  JHEP {\bf 1607} (2016) 015
  doi:10.1007/JHEP07(2016)015
  [arXiv:1512.09241 [hep-ph]].
 
 \bibitem{CPVMSSM}
  P.~Bandyopadhyay and K.~Huitu,
  JHEP {\bf 1311} (2013) 058
  [arXiv:1106.5108 [hep-ph]].
  P.~Bandyopadhyay,
  JHEP {\bf 1108} (2011) 016
  [arXiv:1008.3339 [hep-ph]].
  P.~Bandyopadhyay, A.~Datta, A.~Datta and B.~Mukhopadhyaya,
  Phys.\ Rev.\ D {\bf 78} (2008) 015017
  [arXiv:0710.3016 [hep-ph]].
  D.~K.~Ghosh and S.~Moretti,
  Eur.\ Phys.\ J.\ C {\bf 42} (2005) 341
  [hep-ph/0412365].
  D.~K.~Ghosh, R.~M.~Godbole and D.~P.~Roy,
  Phys.\ Lett.\ B {\bf 628} (2005) 131
  [hep-ph/0412193].
  A.~Pilaftsis and C.~E.~M.~Wagner,
  Nucl.\ Phys.\ B {\bf 553} (1999) 3
  [hep-ph/9902371].
 
\bibitem{pbch}
P.~Bandyopadhyay, C.~Corian\`o and A.~Costantini 

\bibitem{btag}
  I.~R.~Tomalin [CMS Collaboration],
  J.\ Phys.\ Conf.\ Ser.\  {\bf 110} (2008) 092033.
\bibitem{tautag}
  G.~Bagliesi,
  arXiv:0707.0928 [hep-ex].
  G.~L.~Bayatian {\it et al.}  [CMS Collaboration],
  J.\ Phys.\ G {\bf 34} (2007) 995.
  
\end{thebibliography}
\end{document}